\documentclass[preprintnumbers,floatfix,letterpaper,aps,prd,nofootinbib,onecolumn]{revtex4-2}
\usepackage{amsmath,amsfonts,latexsym,amssymb,graphicx,graphics,epsfig,subfigure,color,makeidx,array}
\usepackage{xcolor,diagbox}
\usepackage{multirow}
\usepackage[colorlinks,linkcolor=blue,anchorcolor=blue,citecolor=blue,urlcolor=blue]{hyperref}
\usepackage{mathrsfs}
\usepackage{amsmath}
\usepackage{bm,graphicx,dcolumn,epstopdf,epsf,latexsym,mathbbol, amssymb,amsmath,color,slashed, mathrsfs,mathcomp,simplewick}
\usepackage{makecell}

\newcommand{\bq}{\begin{equation}}
\newcommand{\eq}{\end{equation}}
\newcommand{\bqn}{\begin{eqnarray}}
\newcommand{\eqn}{\end{eqnarray}}
\newcommand{\nb}{\nonumber}
\newcommand{\lb}{\label}
\newcommand{\f}{\frac} 
\newcommand{\p}{\partial}
\newcommand{\tx}{\text}

\newcommand{\lf}{\left}
\newcommand{\rt}{\right}

\begin{document}

\title{Gravitational waveforms from periodic orbits around a quantum-corrected black hole}

\author{Sen Yang${}^{a, b}$}
\email{120220908881@lzu.edu.cn}

\author{Yu-Peng Zhang${}^{a, b}$}
\email{zhangyupeng@lzu.edu.cn}

\author{Tao Zhu${}^{c, d}$}
\email{zhut05@zjut.edu.cn}

\author{Li Zhao${}^{a, b}$}
\email{lizhao@lzu.edu.cn, corresponding author}

\author{Yu-Xiao Liu${}^{a, b}$}
\email{liuyx@lzu.edu.cn, corresponding author}

\affiliation{
${}^{a}$ Lanzhou Center for Theoretical Physics, \\ Key Laboratory for Quantum Theory and Applications of the Ministry of Education,\\
Key Laboratory of Theoretical Physics of Gansu Province,\\
School of Physical Science and Technology,\\
Lanzhou University, Lanzhou 730000, China\\
${}^{b}$ Institute of Theoretical Physics $\&$ Research Center of Gravitation, Lanzhou University, Lanzhou 730000, China\\
${}^{c}$ Institute for Theoretical Physics and Cosmology, Zhejiang University of Technology, Hangzhou, 310023, China\\
${}^{d}$ United Center for Gravitational Wave Physics (UCGWP), Zhejiang University of Technology, Hangzhou, 310023, China\\}
\date{\today}

\begin{abstract}

Extreme mass-ratio inspirals are crucial sources for future space-based gravitational wave detections. Gravitational waveforms emitted by extreme mass-ratio inspirals are closely related to the orbital dynamics of small celestial objects, which vary with the underlying spacetime geometry. Despite the tremendous success of general relativity, there are unsolved issues such as singularities in both black holes and cosmology. Loop quantum gravity, a theory addressing these singularity problems, offers a framework for regular black holes. In this paper, we focus on periodic orbits of a small celestial object around a supermassive quantum-corrected black hole in loop quantum gravity and compute the corresponding gravitational waveforms. We view the small celestial object as a massive test particle and obtain its four-velocity and effective potential. We explore the effects of quantum corrections on marginally bound orbits, innermost stable circular orbits, and other periodic orbits. Using the numerical kludge scheme, we further explore the gravitational waveforms of the small celestial object along different periodic orbits. The waveforms exhibit distinct zoom and whirl phases in a complete orbital period, closely tied to the quantum parameter $\hat \alpha$. We also perform a spectral analysis of the gravitational waves from these periodic orbits and assess their detectability. With the steady progress of space-based gravitational wave detection programs, our findings will contribute to utilizing extreme mass-ratio inspirals to test and understand the properties of quantum-corrected black holes.

\end{abstract}

%\pacs{ 04.50.-h, 11.27.+d}

\maketitle

\section{Introduction}
\label{Introduction}
\renewcommand{\theequation}{1.\arabic{equation}} 
\setcounter{equation}{0}

The successful detection of binary black hole mergers opens the era of gravitational wave astronomy, signifying our ability to study the strong-field dynamics of celestial bodies like black holes using gravitational wave signals \cite{LIGOScientific:2016aoc, LIGOScientific:2016vlm}. As significant sources of gravitational waves, binary black holes have their dynamical properties mainly determined by mass ratio and spins. Due to the varying mass ratios of binary black hole systems, the corresponding gravitational wave frequency bands and amplitudes vary accordingly. Currently, ground-based gravitational wave detectors primarily focus on stellar-mass binary systems with similar masses \cite{LIGOScientific:2018mvr, LIGOScientific:2020ibl, LIGOScientific:2021usb, KAGRA:2021vkt}. However, apart from binaries with similar masses, extreme mass-ratio inspirals (EMRIs) also represent a crucial class of gravitational wave sources \cite{Hughes:2000ssa}. An EMRI system is formed by a stellar-mass object orbiting around a central supermassive black hole \cite{Glampedakis:2005hs}. Such systems are the most exciting potential sources of gravitational waves for low-frequency, space-based gravitational wave detectors such as Laser Interferometer Space Antenna (LISA) \cite{LISA:2017pwj}, Taiji \cite{Hu:2017mde}, TianQin \cite{TianQin:2015yph}, and DECi-heltz Gravitational-wave Observatory (DECIGO) \cite{Musha:2017usi}. The physics of low-frequency gravitational waves, the targets of space-based gravitational wave detection, differs significantly from that of high-frequency gravitational waves \cite{Barausse:2020rsu, LISA:2022yao, Wang:2021srv, Gao:2022hho, Jin:2023sfc, Zhong:2023pjz,GWSTQLISA}. The gravitational waveforms from an EMRI system take accurate information about the orbital dynamics of the small object and the spacetime of the supermassive black hole \cite{Babak:2017tow, dirty, Ghosh:2024arw}. 

The gravitational waveforms of an EMRI system are closely dependent on the orbital dynamics of the small celestial object \cite{Glampedakis:2005hs}. Therefore, studying special orbits in a black hole background will help us identify characteristic gravitational wave signals within an EMRI system. It has been proved that periodic geodesic orbits exist in the background of a Schwarzschild or Kerr black hole \cite{Levin:2008mq, Levin:2008ci, Grossman:2008yk}. A test particle along a periodic orbit must return to its initial location within a finite time and it should have zoom-whirl behavior. The shape of a periodic orbit is indexed with three integers: the zoom number $z$, the whirl number $w$, and the vertex number $v$. The zoom number $z$ describes the number of leaves of a complete periodic orbit. The whirl number $w$ accounts for the number of additional circles that the particle whirls around the central black hole before zooming out to apastron again. The vertex number $v$ depicts the next apastron that the particle will arrive after departing from the starting apastron, labeled as $v = 0$, of the orbit. For every periodic orbit,  radial frequency is associated with angular frequency by a rational number $q$. Periodic orbits possess special values of energy and orbital angular momentum, and they only exist at specific locations in space. When a small perturbation is applied to a periodic orbit, it typically transitions into a non-periodic bound orbit. Over time, this non-periodic orbit will densely fill the torus it lives in. Essentially, while periodic orbits are isolated points in phase space, most bound orbits are aperiodic and represent slight deviations from these periodic solutions. This close relationship between periodic and aperiodic orbits highlights that a generic bound orbit can be viewed as a perturbation of a nearby periodic orbit. \cite{Levin:2008mq}. Periodic orbits are associated with orbital resonances~\cite{Flanagan:2010cd, Berry:2016bit, Speri:2021psr}. When a stellar-mass object approaches or crosses a resonance, it experiences a sharp ``kick", which results in a sudden change in the rate of orbital evolution. This kick leads to substantial modifications in the gravitational wave emission, particularly in the frequency spectrum, and can greatly affect the overall waveform. The sharp changes in the radiation flux during resonance crossings make periodic orbits crucial for the analysis of gravitational wave signals. Because of their importance and interesting properties, the periodic orbits have been studied in many spacetimes, such as charged black holes \cite{Misra:2010pu}, spherically symmetric naked singularities \cite{Babar:2017gsg}, Kerr-Sen black holes \cite{Liu:2018vea}, Einstein–Lovelock black holes \cite{Lin:2021noq}, and so on \cite{ Wei:2019zdf, Deng:2020yfm, Azreg-Ainou:2020bfl, Wang:2022tfo, Tu:2023xab, Li:2024tld, Lin:2023rmo, Zhang:2022psr, Gao:2021arw, Zhou:2020zys, Zhou:2020zys, Deng:2020hxw}. Focusing on the gravitational radiation from these interesting periodic orbits, Refs. \cite{Tu:2023xab} and \cite{Li:2024tld} studied the gravitational waveforms from a test object along the periodic orbits around a supermassive polymer black hole and a supermassive asymptotically flat black hole in Einsteinian cubic gravity, respectively. The gravitational waveforms of periodic orbits from EMRIs were obtained with circular orbit approximation. It was proved that there exist zoom and whirl phases in the gravitational waveforms.

We should note that the properties of geodesic orbits are closely related to the black hole background. Loop quantum gravity is one of the candidates attempting to develop a quantum theory of gravity \cite{Rovelli:1997yv}. Recently, a new model of the regular black hole has been proposed in loop quantum gravity for addressing singularity problems \cite{Lewandowski:2022zce}. The mass of this quantum-corrected black hole exhibits a lower bound from the quantum effects.  Compared with the singular black holes derived from general relativity, the geometry of the regular black hole will be different. Therefore, it is natural to expect that periodic geodesic orbits will be changed by the corrections from loop quantum gravity. This quantum-corrected black hole has attracted significant attention, and its various properties have been studied, such as its shadow and photon ring \cite{Yang:2022btw, Zhang:2023okw, Ye:2023qks}, scalar and vector perturbations \cite{Yang:2022btw, Zhang:2023okw, Gong:2023ghh, Cao:2024oud}, strong cosmic censorship \cite{Shao:2023qlt}, gravitational lensing effects \cite{Zhao:2024elr}, accretion disk \cite{You:2024jeu}, and so on. The stability of this quantum-corrected black hole at huge masses is an important and open question. While gravitational perturbations of the quantum-corrected black hole have not yet been systematically studied, Refs.~\cite{Yang:2022btw, Zhang:2023okw, Gong:2023ghh, Cao:2024oud} show that the quantum-corrected black hole remains stable under scalar and vector field perturbations. Although no definitive conclusions were drawn about the quantum-corrected black hole's stability at huge masses, the existing studies on scalar and vector field perturbations provide a promising foundation for future research.

In this work, we focus on the timelike periodic geodesic orbits around the quantum-corrected black hole. We derive the equations of motion for the massive test particle and obtain the effective potential from the Lagrangian of the test particle \cite{Chandrasekhar:1985kt}. We investigate how the quantum correction affects the effective potential and study the marginally bound orbits (MBOs) and the innermost stable circular orbits (ISCOs) \cite{Misner:1973prb}. We also explore the allowed parameter space of the orbital angular momentum and the energy for the bound orbits around the quantum-corrected black hole. Then, we investigate the periodic orbits around the quantum-corrected black hole. We numerically study the relation between the rational number $q$ and the energy or the orbital angular momentum of periodic orbits. For the periodic orbits indexed with a set of ($z$, $w$, $v$), we calculate the energy and the orbital angular momentum of the periodic orbits, to explore how the quantum correction affects the periodic orbits. Finally, using the numerical kludge scheme \cite{Babak:2006uv}, we investigate the gravitational waveforms from a stellar-mass object moving along different periodic orbits around a supermassive quantum-corrected black hole. The effects of the quantum correction on the waveforms are discussed. We then perform discrete Fourier transforms on the time-domain gravitational waveforms to obtain the corresponding frequency spectra. We also calculate the characteristic strains of gravitational waves from the frequency spectra. To assess the detectability of gravitational waves from the EMRI with periodic orbits, we compare the characteristic strains with the sensitivity curve of LISA \cite{Robson:2018ifk}.

This paper is organized as follows. In Sec.~\ref{sec2}, we analyze the geodesic of a massive test particle on the equatorial orbit around the quantum-corrected black hole, derive the effective potential, and investigate the MBOs and the ISCOs.  In Sec.~\ref{sec3}, we study the rational number $q$ of the periodic orbits around the quantum-corrected black hole and explore the effects of the quantum correction on the energy and the orbital angular momentum of periodic orbits. Then we study the gravitational waveforms of the test object along different periodic orbits around the supermassive quantum-corrected black hole and conduct spectral analysis on the waveforms in Sec.~\ref{sec4}. Finally, the conclusions and discussions of this work are given in Sec.~\ref{sec5}. Throughout the paper, we use the geometrized unit system with $G = c = 1$.

\section{Timelike geodesics} 
\label{sec2}
\renewcommand{\theequation}{2.\arabic{equation}} 
\setcounter{equation}{0}

In this section, we briefly review the geodesic orbits around the quantum-corrected black hole proposed in Ref. \cite{Lewandowski:2022zce}. The corresponding metric is 
\bqn\lb{metric}
ds^2 = - \lf( 1 - \f{2 M}{r} + \f{\alpha M^2}{r^4} \rt) dt^2 + \lf( 1 - \f{2 M}{r} + \f{\alpha M^2}{r^4} \rt)^{-1} dr^2 + r^2 d \theta^2 + r^2 \sin^2 \theta d \phi^2 ,
\eqn 
where $M$ is the Arnowitt–Deser–Misner mass and $\alpha = 16 \sqrt{3} \pi \gamma^3 l_{\tx{P}}^2$ is the quantum-corrected parameter. Here $\gamma$ is the Immirzi parameter and $l_\tx{P}$ is the Planck length. The quantum effect leads to a lower bound $M_{\tx{min}}$ on the mass of the quantum-corrected black hole. For convenience, we use the dimensionless parameter $\hat{\alpha} = \alpha/M^2$ instead of $\alpha$ in this work. Equation \eqref{metric} goes back to the line element of a Schwarzschild black hole when $\hat{\alpha} = 0$.

\begin{figure}[!t]
	\centering
	\subfigure[$\hat{\alpha} =1$]{\includegraphics[scale =0.28]{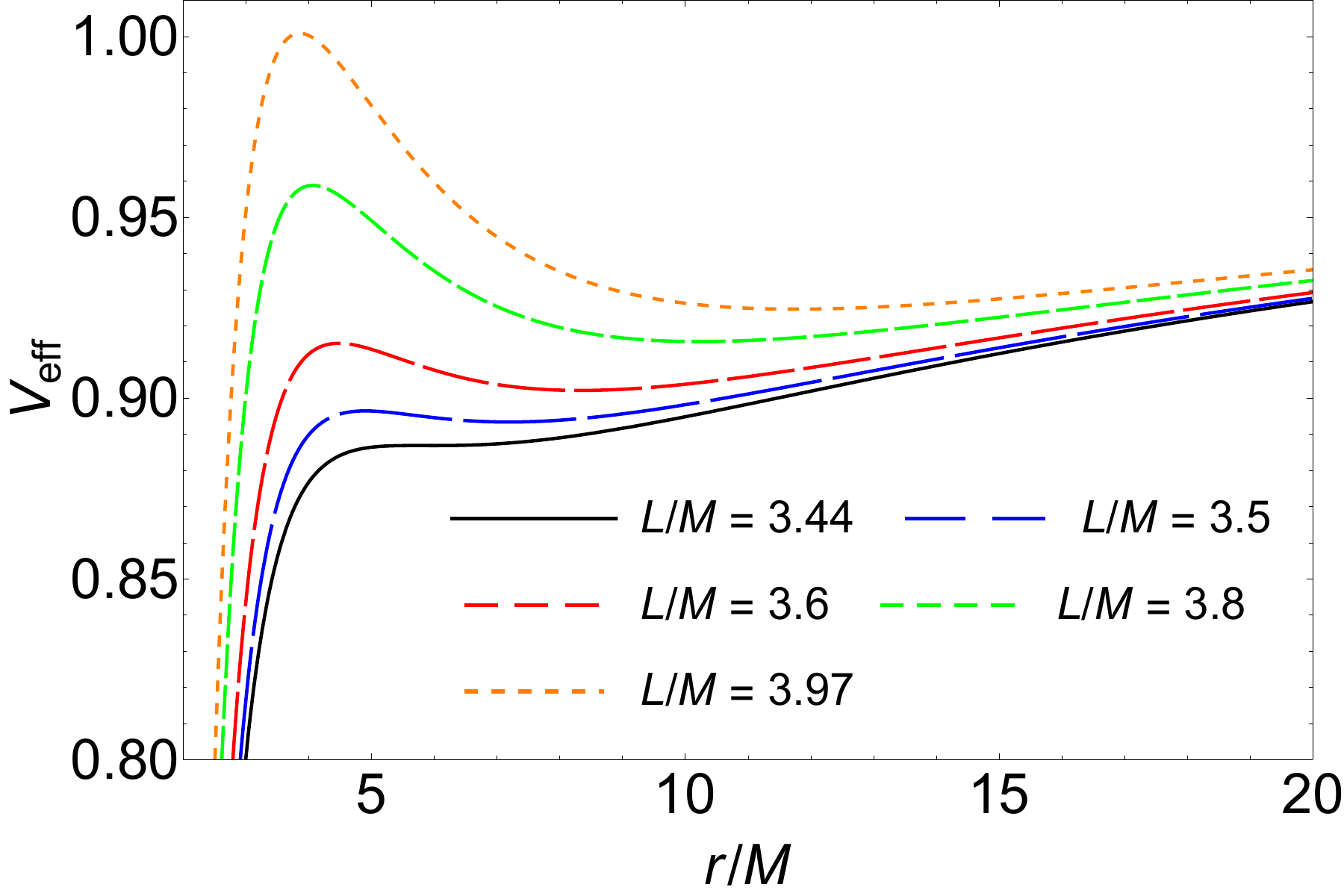}\lb{plot-V-1}}
	\subfigure[$L/M = 3.8$]{\includegraphics[scale =0.28]{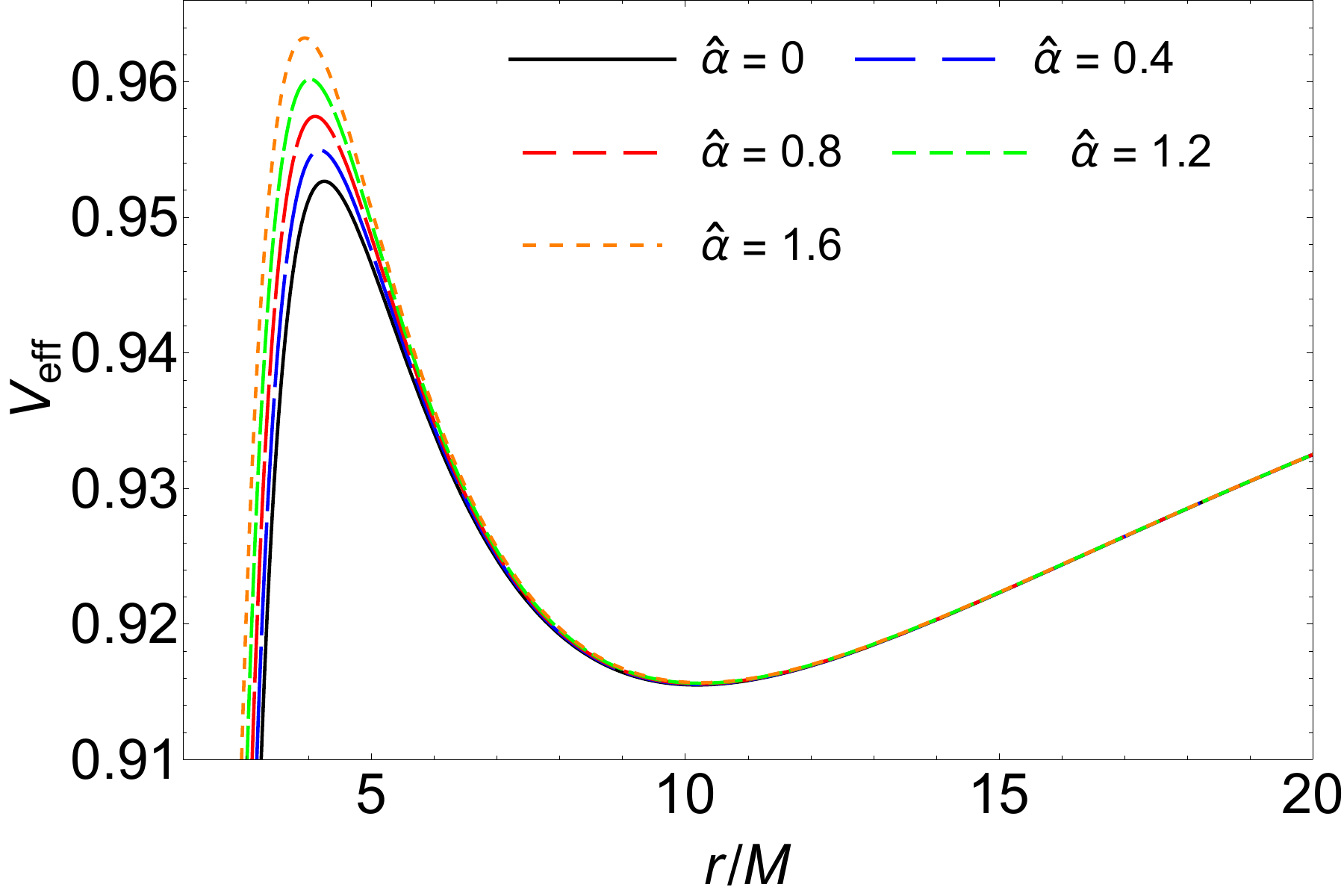}
    \lb{plot-V-2}}
	\caption{Plot of the effective potential \eqref{Veff} of the test particle around the quantum-corrected black hole. (a) The effective potential for different values of the orbital angular momentum and fixed quantum-corrected parameter $\hat{\alpha}$. (b) The effective potential for different values of the quantum-corrected parameter $\hat{\alpha}$ and fixed orbital angular momentum.}
	\label{plot-V}
\end{figure}

Consider a massive test particle moving around the quantum-corrected black hole. The Lagrangian of the particle is \cite{Chandrasekhar:1985kt}
\bqn\lb{Lagrangian}
\mathscr{L} = \f{m}{2} g_{\mu \nu} \f{d x^\mu}{d \tau} \f{d x^\nu}{d \tau},
\eqn 
where $\tau$ and $m$ are the proper time and the mass of the test particle, respectively. For the massive test particle, the corresponding Lagrangian satisfies $2\mathscr{L}=-m$. For simplicity, we let the mass of the test particle $m=1$. Then one can obtain the generalized momentum of the particle:
\bqn\lb{momentum}
p_{\mu} = \f{\p \mathscr{L}}{\p \dot{x}^\mu} = g_{\mu \nu} \dot{x}^\nu,
\eqn
where the dot denotes the derivative with respect to the proper time. Substituting Eqs. \eqref{metric} and \eqref{Lagrangian} into Eq. \eqref{momentum}, one can derive the equations of motion for the particle:
\bqn
p_{t} &=&  - \lf( 1 - \f{2 M}{r} + \f{\alpha M^2}{r^4} \rt) \dot{t} = - E, \lb{momentum-1} \\
p_{\phi} &=& r^2 \sin^2 \theta \dot{\phi} = L, \lb{momentum-2} \\
p_{r} &=& \lf( 1 - \f{2 M}{r} + \f{\alpha M^2}{r^4} \rt)^{-1} \dot{r} , \lb{momentum-3} \\
p_{\theta} &=& r^2 \dot{\theta} , \lb{momentum-4}
\eqn 
where $E$ and $L$ represent the energy and the orbital angular momentum of the particle per unit mass, respectively. From Eqs. \eqref{momentum-1} and \eqref{momentum-2}, one can get
\bqn
\dot{t} &=& \f{E}{ 1 - \f{2 M}{r} + \f{\alpha M^2}{r^4} }, \lb{dott}\\
\dot{\phi} &=& \f{L}{r^2 \sin^2 \theta}. \lb{dotphi}
\eqn 
Because we only consider the spherical quantum-corrected black hole spacetime, we set the orbit of the particle in the equatorial plane, which means $\theta = \pi/2$ and $\dot{\theta} = 0$. Then the Lagrangian \eqref{Lagrangian} simplifies to 
\bqn\lb{Lagrangian-2}
\f{\dot{r}^2}{ 1 - \f{2 M}{r} + \f{\alpha M^2}{r^4} } + \f{L^2}{r^2} - \f{E^2}{ 1 - \f{2 M}{r} + \f{\alpha M^2}{r^4} } = -1.
\eqn 
One can rewrite Eq. \eqref{Lagrangian-2} in the form as follows
\bqn
\dot{r}^2 + V_{\tx{eff}} = E^2,
\eqn
where
\bqn\lb{Veff}
V_{\tx{eff}} = \lf( 1 - \f{2 M}{r} + \f{\alpha M^2}{r^4}  \rt) \lf( 1+\f{L^2}{r^2} \rt) 
\eqn
is defined as the radial effective potential for determining the radial motion of the particle. To understand the properties of the effective potential \eqref{Veff}, we plot its shape in Fig. \ref{plot-V}. Figure \ref{plot-V-1} shows the effective potential will possess extremum with the increase of the orbital angular momentum. Here, the minimum and maximum of the effective potential correspond to the stable and unstable circular orbits. Figure \ref{plot-V-2} shows the value of the maximum of the effective potential increases with the parameter $\hat{\alpha}$. Note that the asymptotic behavior of the effective potential \eqref{Veff} is $V_\tx{eff} \rightarrow 1$ as $r \rightarrow + \infty$. So $E = 1$ is the critical energy of the particle moving along bound or unbound orbits, and the energy of the particle moving on bound orbits must satisfy $E \leq 1$.

\begin{figure}[!t]
	\centering
	\subfigure[]{\includegraphics[scale =0.28]{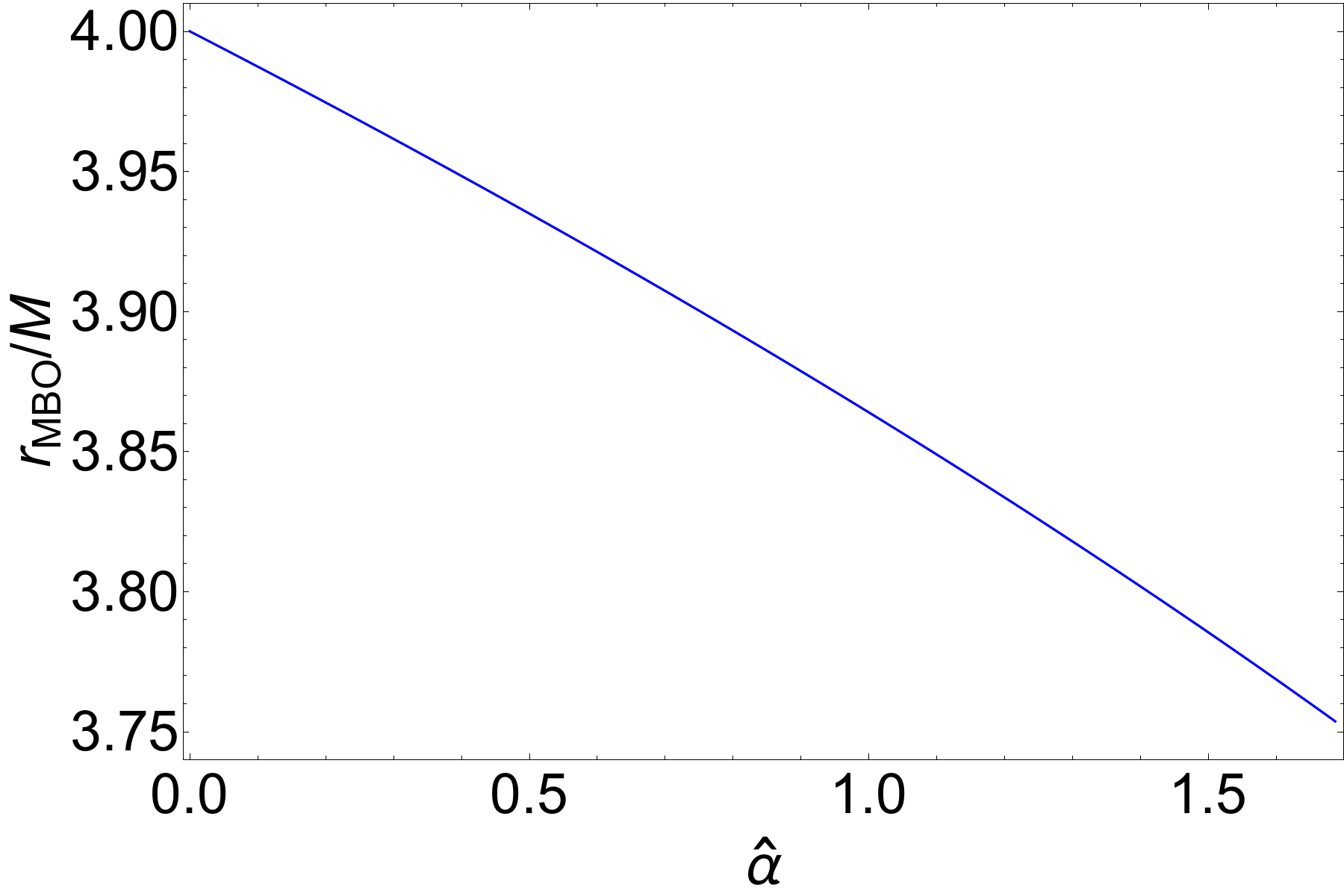}}
	\subfigure[]{\includegraphics[scale =0.28]{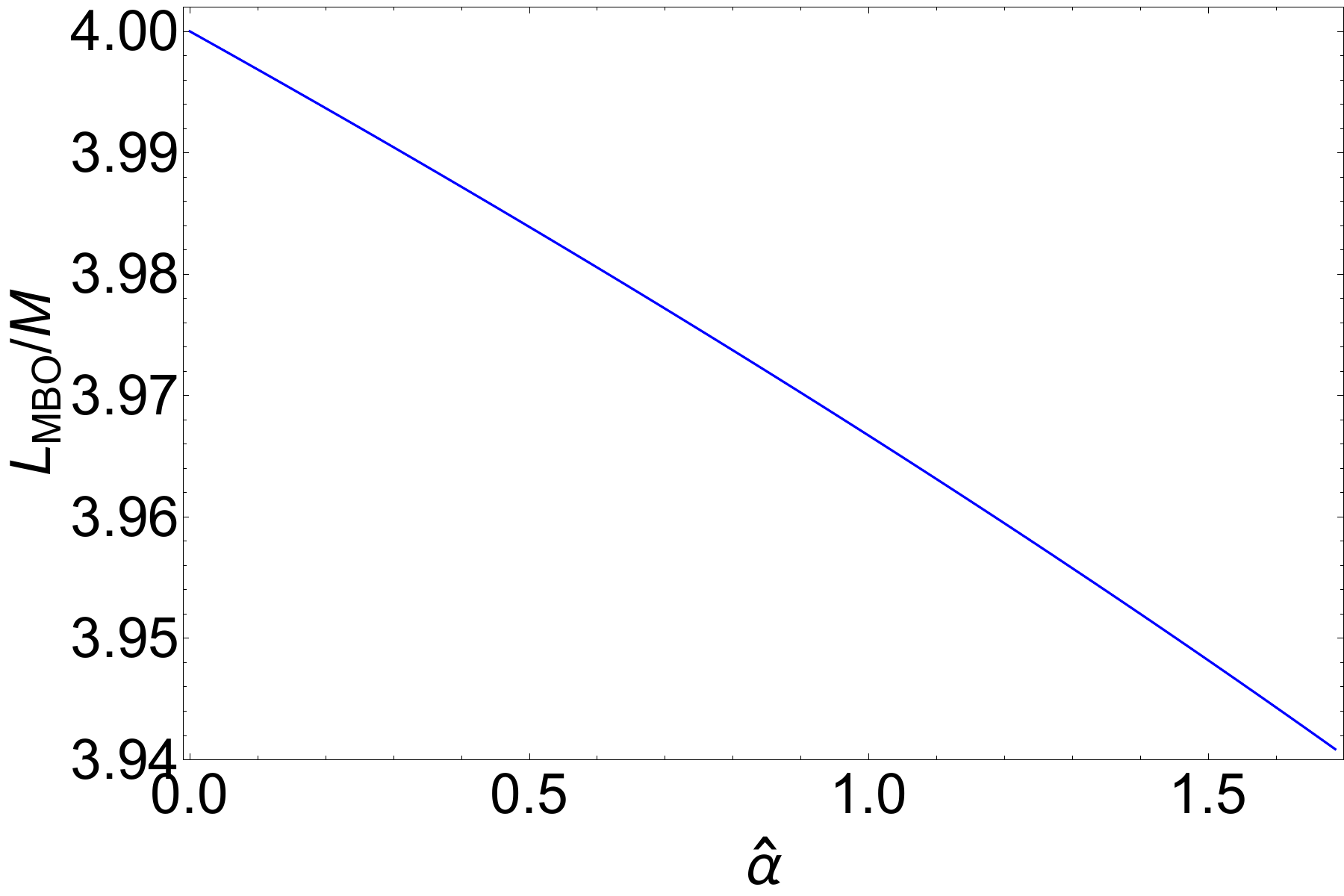}}
	\caption{The properties of the MBO around the quantum-corrected black hole. (a) The radius of the MBO as a function of the parameter $\hat{\alpha}$. (b) The angular momentum of the MBO as a function of the parameter $\hat{\alpha}$.}
	\label{plot-MBO}
\end{figure}

\begin{figure}[!t]
	\centering
	\subfigure[]{\includegraphics[scale =0.28]{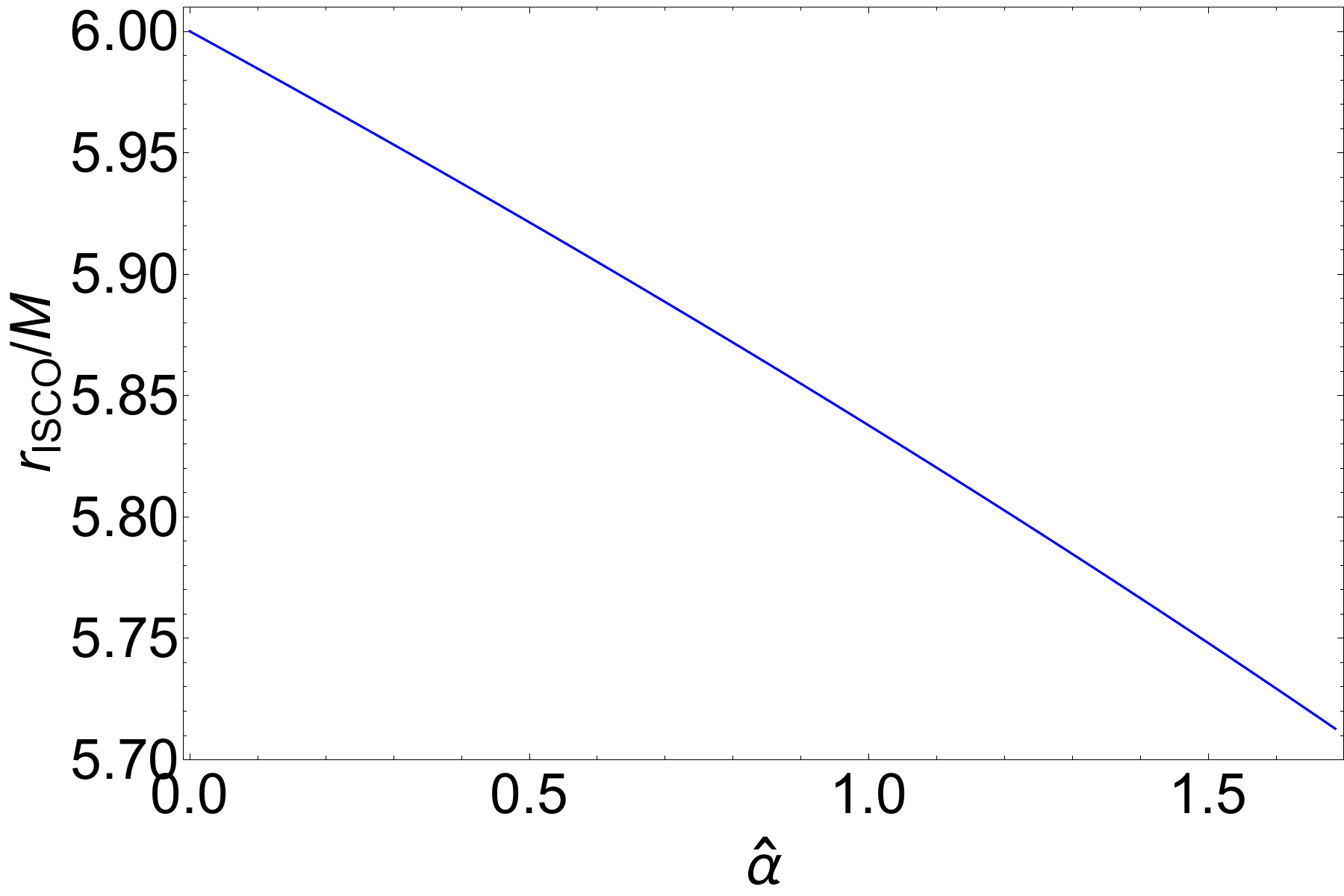}}
	\subfigure[]{\includegraphics[scale =0.28]{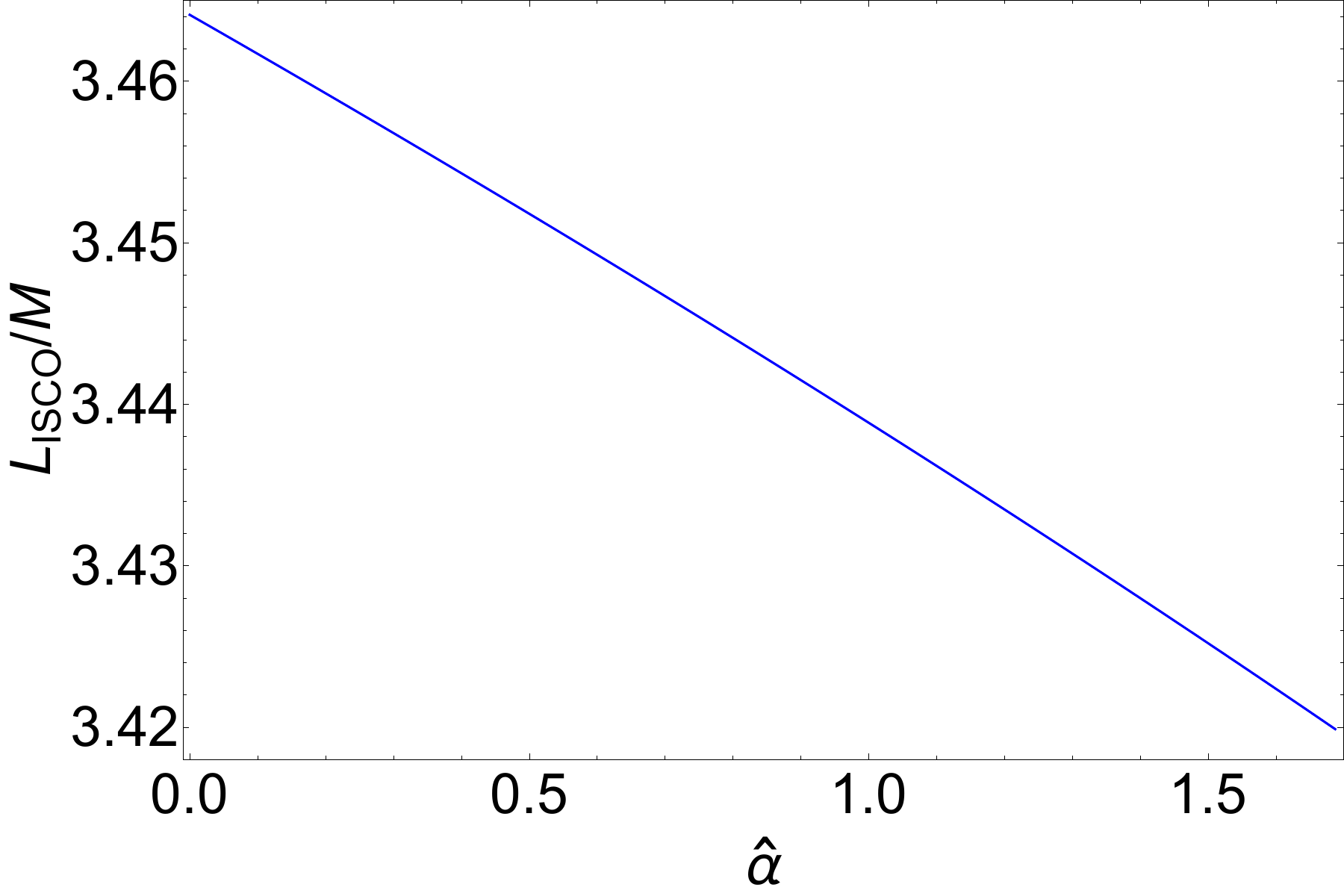}}
    \subfigure[]{\includegraphics[scale =0.28]{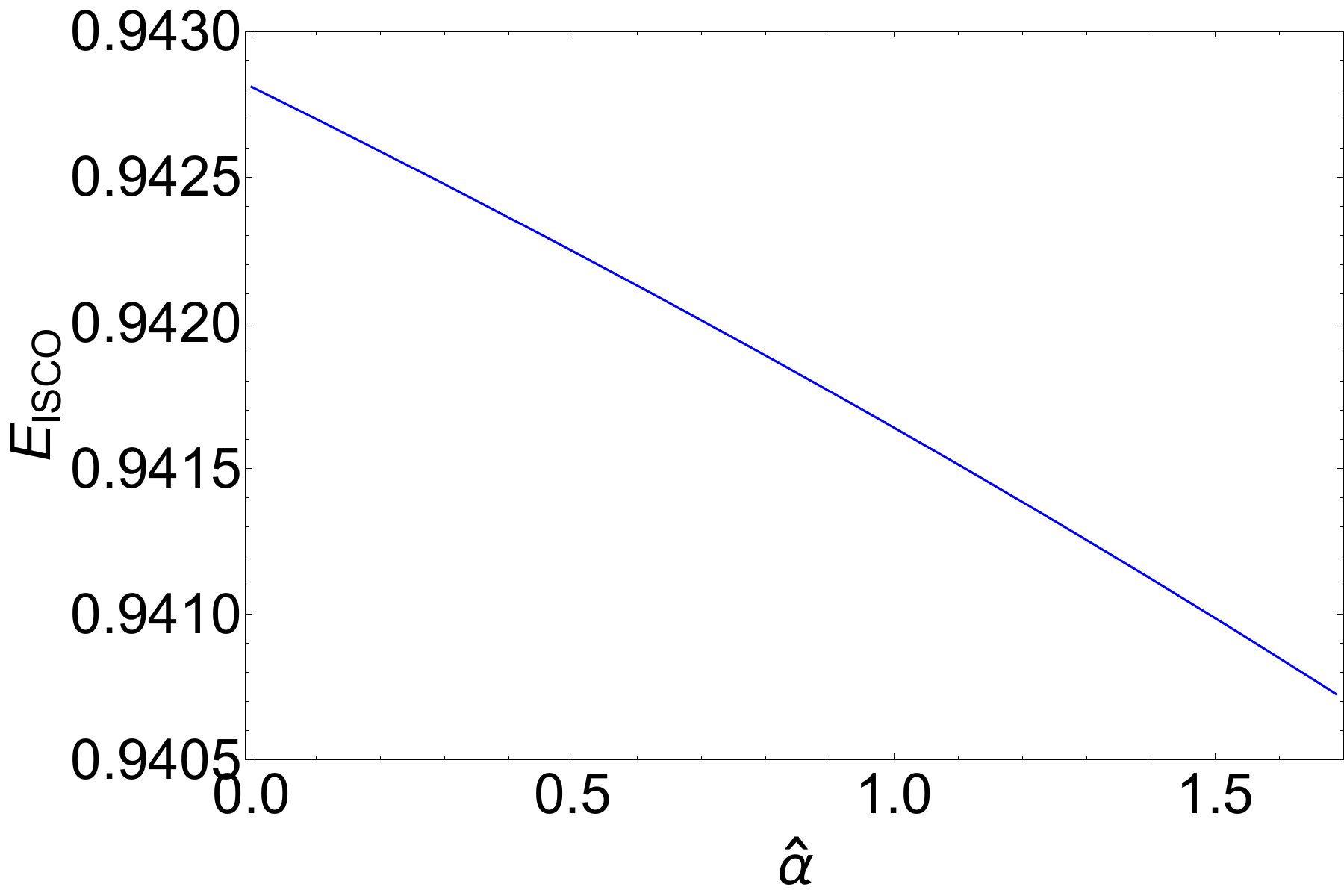}}
	\caption{The properties of the ISCO around the quantum-corrected black hole. (a) The radius of the ISCO as a function of the parameter $\hat{\alpha}$. (b) The orbital angular momentum of the ISCO as a function of the parameter $\hat{\alpha}$. (c) The energy of the ISCO as a function of the parameter $\hat{\alpha}$.}
	\label{plot-ISCO}
\end{figure}

\begin{figure}[!t]
	\centering
	\includegraphics[scale =0.28]{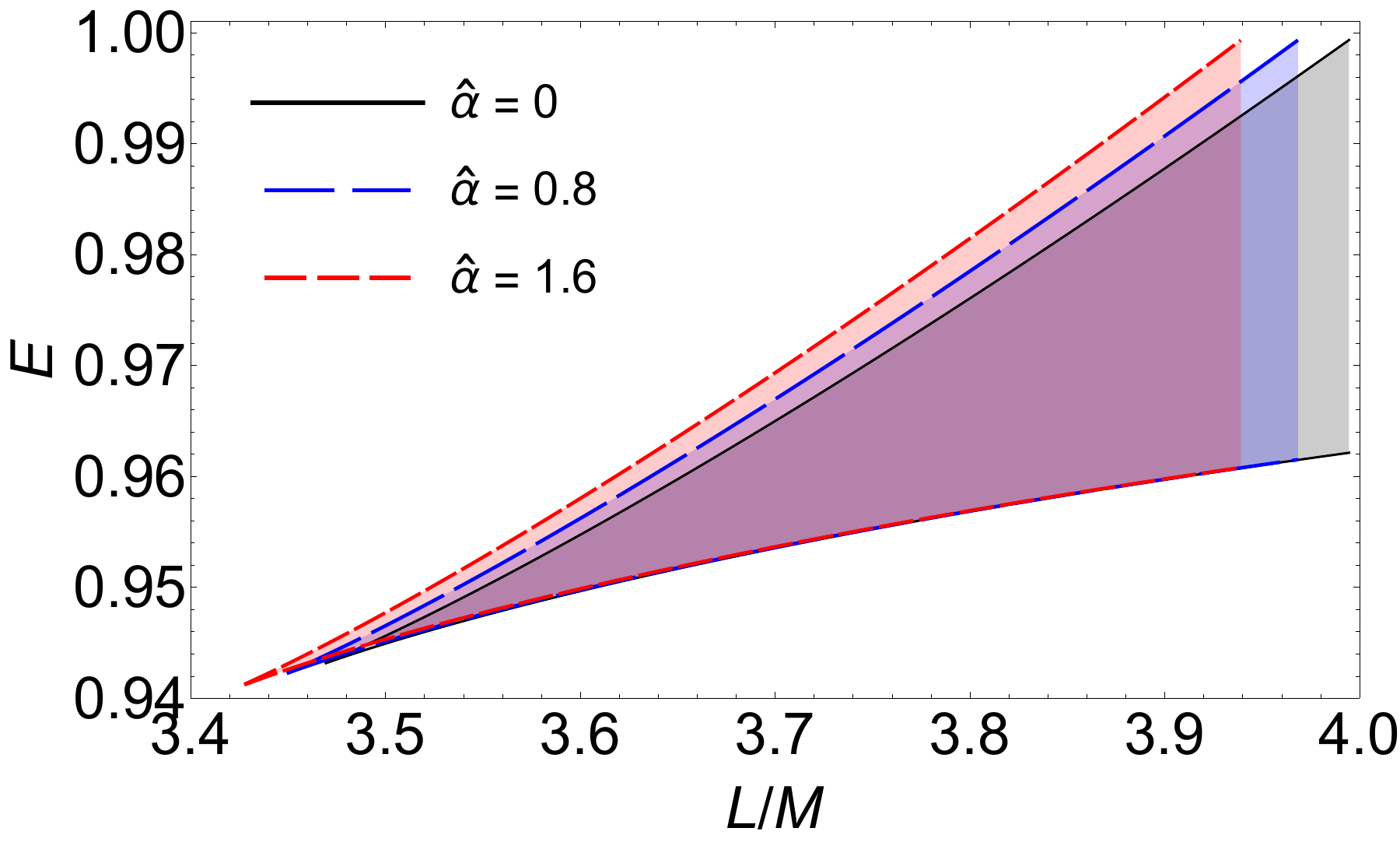}
	\caption{The allowed parameter space of the orbital angular momentum and the energy for the bound orbits around the quantum-corrected black hole with different values of the parameter $\hat{\alpha}$.}
	\label{plot-EvsL}
\end{figure}

In this paper, we mainly focus on the properties of the periodic orbits around the quantum-corrected black hole. We should note that periodic orbits are a kind of bound orbits. For a particle moving along a general bound orbit, its orbital angular momentum and energy should satisfy 
\bqn\lb{range} 
L_{\tx{ISCO}} \leq L ~~~~\tx{and}~~~~E_{\tx{ISCO}} \leq E \leq E_\tx{MBO} = 1,
\eqn
where $L_\tx{ISCO}$ and $E_\tx{ISCO}$ are the orbital angular momentum and the energy of the particle along the ISCO, and $E_\tx{MBO}$ is the energy of the particle along the MBO. The particle along the orbit with $E > 1$ will escape to infinity and the particle along the orbit with $E < E_{\tx{ISCO}}$ will fall into the black hole. To clarify how quantum correction affects the properties of the periodic orbits, we first investigate the properties of the MBO and ISCO in the quantum-corrected black hole. The MBO is the circular bound orbit with the minimum radius and the energy $E_{\tx{MBO}} =1$. Given the effective potential \eqref{Veff} of the motion of the particle, the MBO satisfies the following conditions
\bqn\lb{MBO-condition}
V_{\tx{eff}} = 1,~~~~\f{d V_{\tx{eff}} }{d r} = 0.
\eqn
One can obtain the radius and the orbital angular momentum of the MBO around the quantum-corrected black hole in terms of Eq. \eqref{MBO-condition}. Indeed, in addressing this issue, we encounter high-order equations for which obtaining analytical solutions is not feasible. We numerically solve Eq. \eqref{MBO-condition}, then plot the relations between the radius of the MBO and the parameter $\hat{\alpha}$, and between the orbital angular momentum of the MBO and the parameter $\hat{\alpha}$ in Fig. \ref{plot-MBO}. One can find that both the radius and the orbital angular momentum of the MBO decrease with the parameter $\hat{\alpha}$.
The ISCO is another important kind of bound orbit, which is defined by the conditions 
\bqn\lb{ISCO-condition}
\dot{r} = 0,~~~~\f{d V_{\tx{eff}} }{d r} = 0,~~~~\f{d^2 V_{\tx{eff}} }{d r^2} = 0.
\eqn
We numerically solve Eq.~\eqref{ISCO-condition}, and plot the relations between the radius of the ISCO and the parameter $\hat{\alpha}$, between the orbital angular momentum of the ISCO and the parameter $\hat{\alpha}$, and between the energy of the ISCO and the parameter $\hat{\alpha}$ in Fig.~\ref{plot-ISCO}. One can find that all of the radius, the orbital angular momentum, and the energy of the ISCO decrease with the parameter $\hat{\alpha}$.
From Eq. \eqref{range}, we plot the allowed parameter space of the orbital angular momentum and the energy for the bound orbits around the quantum-corrected black hole with different values of the parameter $\hat{\alpha}$ in Fig. \ref{plot-EvsL}. It shows that the bound orbit, with fixed orbital angular momentum, around the quantum-corrected black hole with a larger value of the parameter $\hat{\alpha}$ has a higher upper energy boundary.

\section{Periodic orbits}\lb{sec3}
\renewcommand{\theequation}{3.\arabic{equation}}
\setcounter{equation}{0} 

\begin{figure}[!t]
	\centering
	\subfigure[]{\includegraphics[scale =0.28]{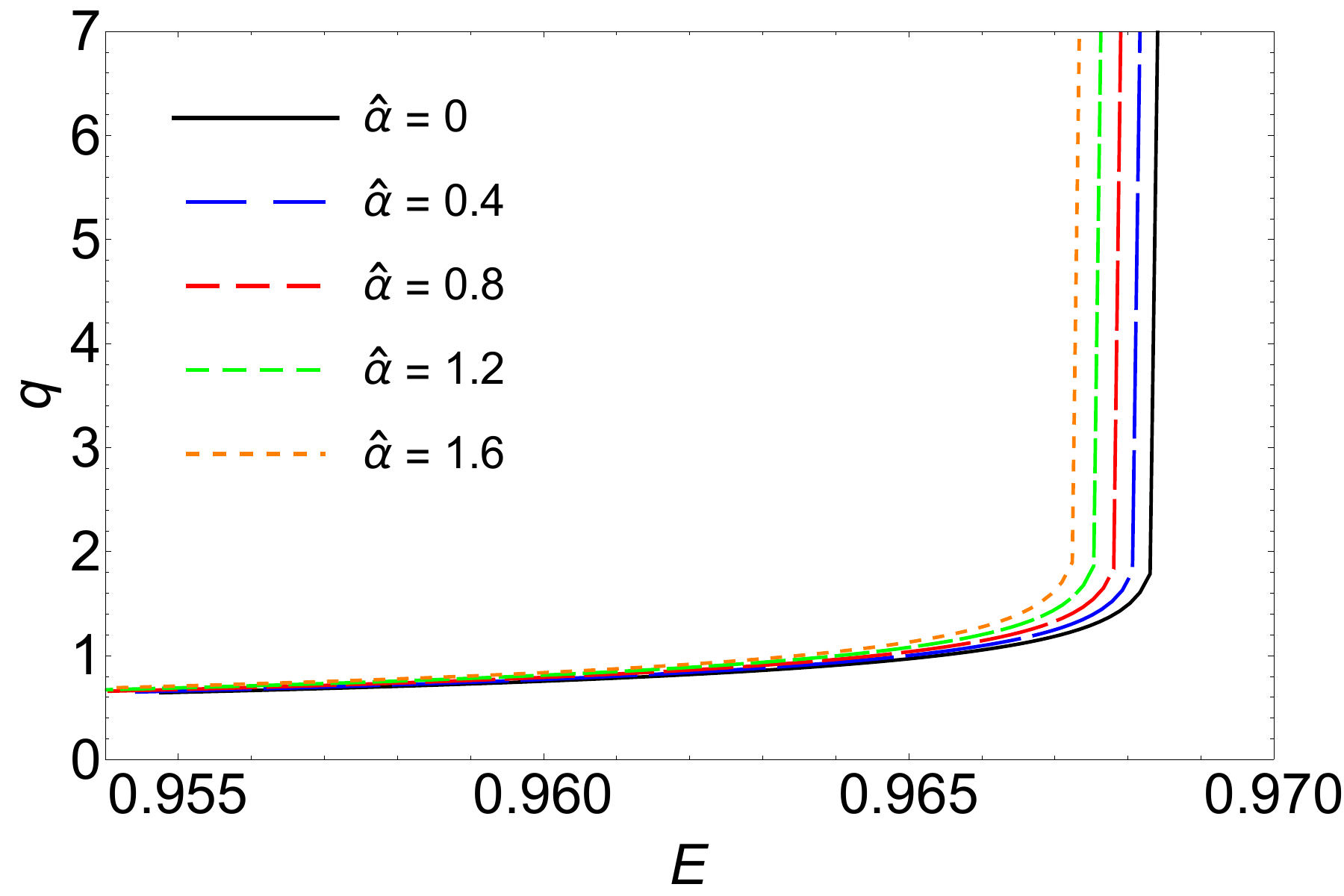}\label{plot-q-E}}
	\subfigure[]{\includegraphics[scale =0.28]{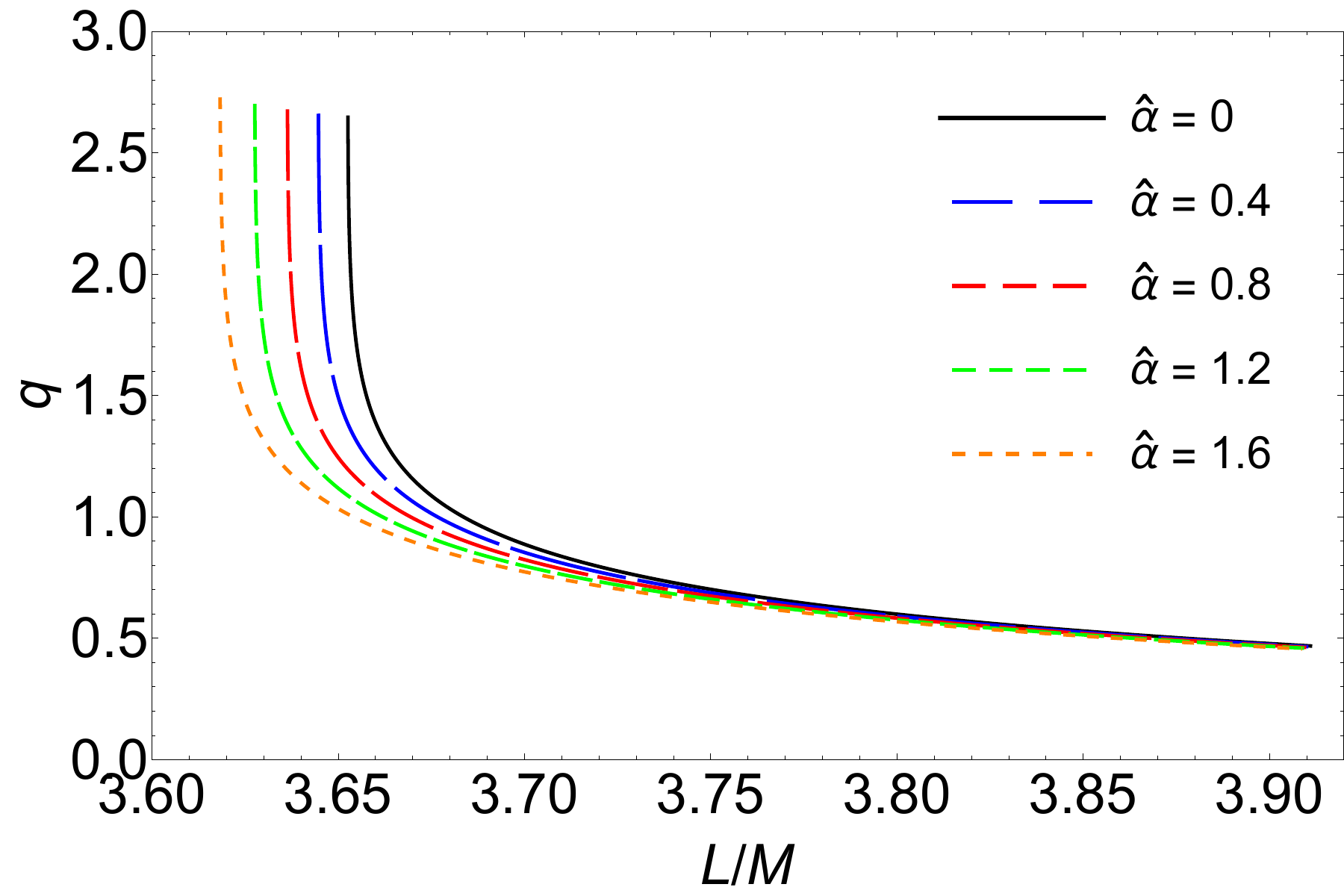}\label{plot-q-L}}
	\caption{(a) The rational number $q$ as a function of the energy of periodic orbits around a quantum-corrected black hole with different values of the parameter $\hat{\alpha}$. The orbital angular momentum is fixed as $L = (L_\tx{MBO} + L_\tx{ISCO})/2$. (b) The rational number $q$ as a function of the orbital angular momentum of periodic orbits around a quantum-corrected black hole with different values of the parameter $\hat{\alpha}$. The energy is fixed as $E = 0.96$.}
	\label{plot-q}
\end{figure}

The fundamental orbital frequencies of the periodic orbits are rationally related \cite{Levin:2008mq}. And for every periodic orbit indexed with a ($z$, $w$, $v$), one can define a rational
number as
\bqn\lb{q-define}
q \equiv \f{\omega_\phi}{\omega_r} - 1  = w + \f{v}{z},
\eqn 
where $\omega_\phi$ and $\omega_r$ are the radial frequency and the angular frequency of the periodic orbits, respectively. To eliminate degeneracy, the zoom number $w$ and the vertex number $v$ are required to be coprime. Combining the equations of motion for the particle \eqref{dott} and \eqref{dotphi}, Eq. \eqref{q-define} can be rewritten as
\bqn
q &=& \f{1}{\pi} \int^{r_2}_{r_1} \f{\dot{\phi}}{\dot{r}} dr -1 \nb\\
&=& \f{1}{\pi} \int^{r_2}_{r_1} \f{L}{r^2 \sqrt{E^2 - \lf( 1 - \f{2 M}{r} + \f{\alpha M^2}{r^4}  \rt) \lf( 1+\f{L^2}{r^2} \rt) } } dr -1,
\eqn 
where $r_1$ and $r_2$ are the radius of periapsis and apoapsis of the periodic orbits, respectively. Fixing the orbital angular momentum of the periodic orbits as $L = (L_\tx{MBO} + L_\tx{ISCO})/2$, we plot the rational number $q$ as a function of the energy with different values of the parameter $\hat{\alpha}$. And fixing the energy as $E = 0.96$, we plot the rational number $q$ as a function of the orbital angular momentum with different values of the parameter $\hat{\alpha}$. The results are shown in Fig. \ref{plot-q}. Figure \ref{plot-q-E} shows the rational number $q$ first increases slowly as the energy $E$ increases, then explodes as $E$ approaches its maximum value. Figure \ref{plot-q-L} shows the rational number $q$ explodes as the angular momentum $L$ approaches its minimal value, but decreases slowly as $L$ increases. One can also find in Fig. \ref{plot-q} that, with fixed rational number $q$, both the energy and the orbital angular momentum of the periodic orbits decrease with the parameter $\hat{\alpha}$. And both the maximum value of the energy and the minimal value of the orbital angular momentum of the periodic orbits decrease with the parameter $\hat{\alpha}$. Note that, in fact, the value of the parameter $\hat{\alpha}$ should be very small. However, we are qualitatively studying the impact of the parameter $\hat{\alpha}$ on periodic orbits and their gravitational waves in this work. Therefore, we choose relatively large values for the parameter $\hat{\alpha}$.  

For the periodic orbits indexed with different ($z$, $w$, $v$) around the quantum-corrected black hole with different values of the parameter $\hat{\alpha}$, we numerically calculate the energy $E$ of the periodic orbits with fixed orbital angular momentum $L = (L_\tx{MBO} + L_\tx{ISCO})/2$, and calculate the orbital angular momentum $L$ of the periodic orbits with fixed energy $E = 0.96 $, respectively. The numerical results are listed in Tables \ref{T-E} and \ref{T-L}. Our results indicate that both the energy and the orbital angular momentum of the test particle along the periodic orbits decrease with the parameter $\hat{\alpha}$.
\begin{table*}[t]
	\renewcommand\arraystretch{1.5} 
	\centering
	\begin{tabular}{p{1.0cm}p{1.8cm}p{1.8cm}p{1.8cm}p{1.8cm}p{1.8cm}p{1.8cm}p{1.8cm}p{1.8cm}}
		\hline
        \hline
		$\hat{\alpha}$ & $ E_{(1,1,0)} $ & $ E_{(1,2,0)} $ &$ E_{(2,1,1)} $ & $ E_{(2,2,1)} $ & $ E_{(3,1,2)} $ & $ E_{(3,2,2)} $ & $ E_{(4,1,3)} $ & $ E_{(4,2,3)} $\\
	  \hline
		$ 0 $ & $ 0.965425 $ & $ 0.968383 $ & $ 0.968026 $ & $ 0.968438 $ & $ 0.968225 $ & $ 0.968438 $ & $ 0.968285 $ & $ 0.968440 $ \\
        \hline 
        $ 0.4 $ & $ 0.965015$ & $ 0.968135 $ & $ 0.967750 $ & $ 0.968190 $ & $ 0.967963 $ & $ 0.968198 $ & $ 0.968029 $ & $ 0.968200 $ \\
        \hline 
        $ 0.8 $ & $ 0.964561 $ & $ 0.967873 $ & $ 0.967457 $ & $ 0.967936 $ & $ 0.967685 $ & $ 0.967942 $ & $ 0.967757 $ & $ 0.967943 $ \\
        \hline 
        $ 1.2 $ & $ 0.964062 $ & $ 0.967591 $ & $ 0.967135 $ & $ 0.967662 $ & $ 0.967377 $ & $ 0.967668 $ & $ 0.967462 $ & $ 0.967669 $ \\
        \hline 
        $ 1.6 $ & $ 0.963508 $ & $ 0.967285 $ & $ 0.966784 $ & $ 0.967367 $ & $ 0.967056 $ & $ 0.967375 $ & $ 0.967140 $ & $ 0.967377 $ \\
		\hline 
        \hline
	\end{tabular}
	\caption{The energy $E$ for the periodic orbits with different values of ($z$, $w$, $v$) and the quantum parameter $\hat{\alpha}$. The orbital angular momentum is fixed as $ L = (L_{\tx{MBO}} + L_\tx{{ISCO}})/2 $. }
	\label{T-E}
\end{table*}
\begin{table*}[t]
	\renewcommand\arraystretch{1.5} 
	\centering
	\begin{tabular}{p{1.0cm}p{1.8cm}p{1.8cm}p{1.8cm}p{1.8cm}p{1.8cm}p{1.8cm}p{1.8cm}p{1.8cm}}
		\hline
        \hline
		$\hat{\alpha}$ & $ L_{(1,1,0)} /M $ & $ L_{(1,2,0)} /M $ &$ L_{(2,1,1)} /M $ & $ L_{(2,2,1)} /M $ & $ L_{(3,1,2)} /M $ & $ L_{(3,2,2)} /M $ & $ L_{(4,1,3)} /M $ & $ L_{(4,2,3)} /M $\\
	  \hline
		$ 0 $ & $ 3.683593 $ & $ 3.653406$ & $ 3.657596 $ & $ 3.652700 $ & $ 3.655335 $ & $ 3.652636 $ & $ 3.654621 $ & $ 3.652616 $ \\
        \hline 
        $ 0.4 $ & $ 3.676625 $ & $  3.645547 $ & $ 3.649921 $ & $ 3.644803$ & $ 3.647555 $ & $ 3.644728 $ & $ 3.646823 $ & $ 3.644706 $ \\
        \hline 
        $ 0.8 $ & $  3.669394 $ & $ 3.637291 $ & $ 3.641876 $ & $ 3.636495 $ & $ 3.639415 $ & $ 3.636421 $ & $ 3.638636 $ & $ 3.636397 $ \\
        \hline 
        $ 1.2 $ & $ 3.661871 $ & $ 3.628592 $ & $ 3.633423 $ & $ 3.627742 $ & $ 3.630843 $ & $ 3.627654 $ & $ 3.630026 $ & $ 3.627627 $ \\
        \hline 
        $ 1.6 $ & $ 3.654000 $ & $ 3.619382 $ & $ 3.624505 $ & $ 3.618451 $ & $ 3.621785 $ & $ 3.618361 $ & $ 3.620908 $ & $ 3.618332 $ \\
		\hline 
        \hline
	\end{tabular}
	\caption{The orbital angular momentum $L$ for the periodic orbits with different values of ($z$, $w$, $v$) and the quantum parameter $\hat{\alpha}$. The energy is fixed as $ E = 0.960000 $. }
	\label{T-L}
\end{table*}

\begin{figure}[!t]
	\centering
	\subfigure{\includegraphics[scale =0.25]{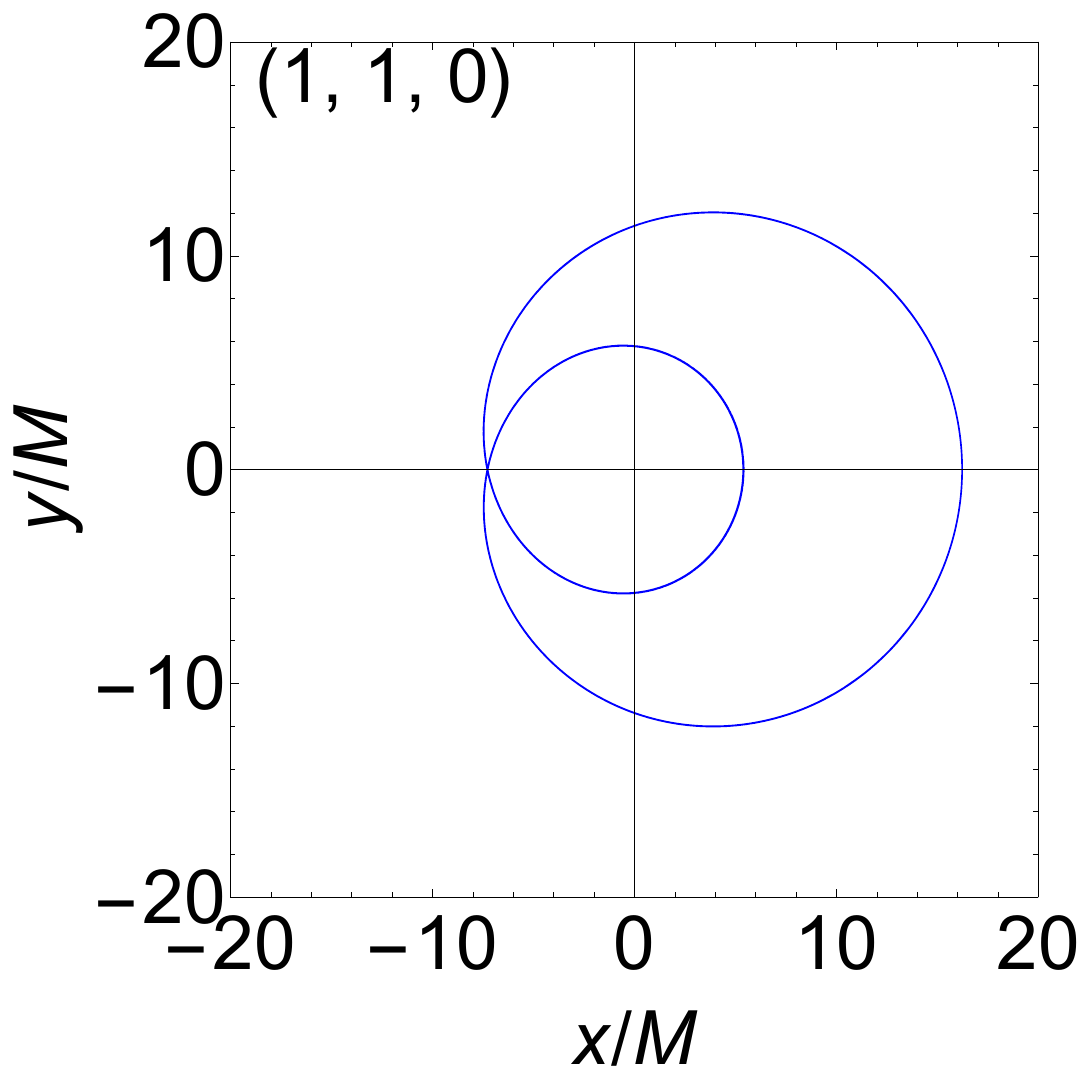}}
    \hspace{5mm}
	\subfigure{\includegraphics[scale =0.25]{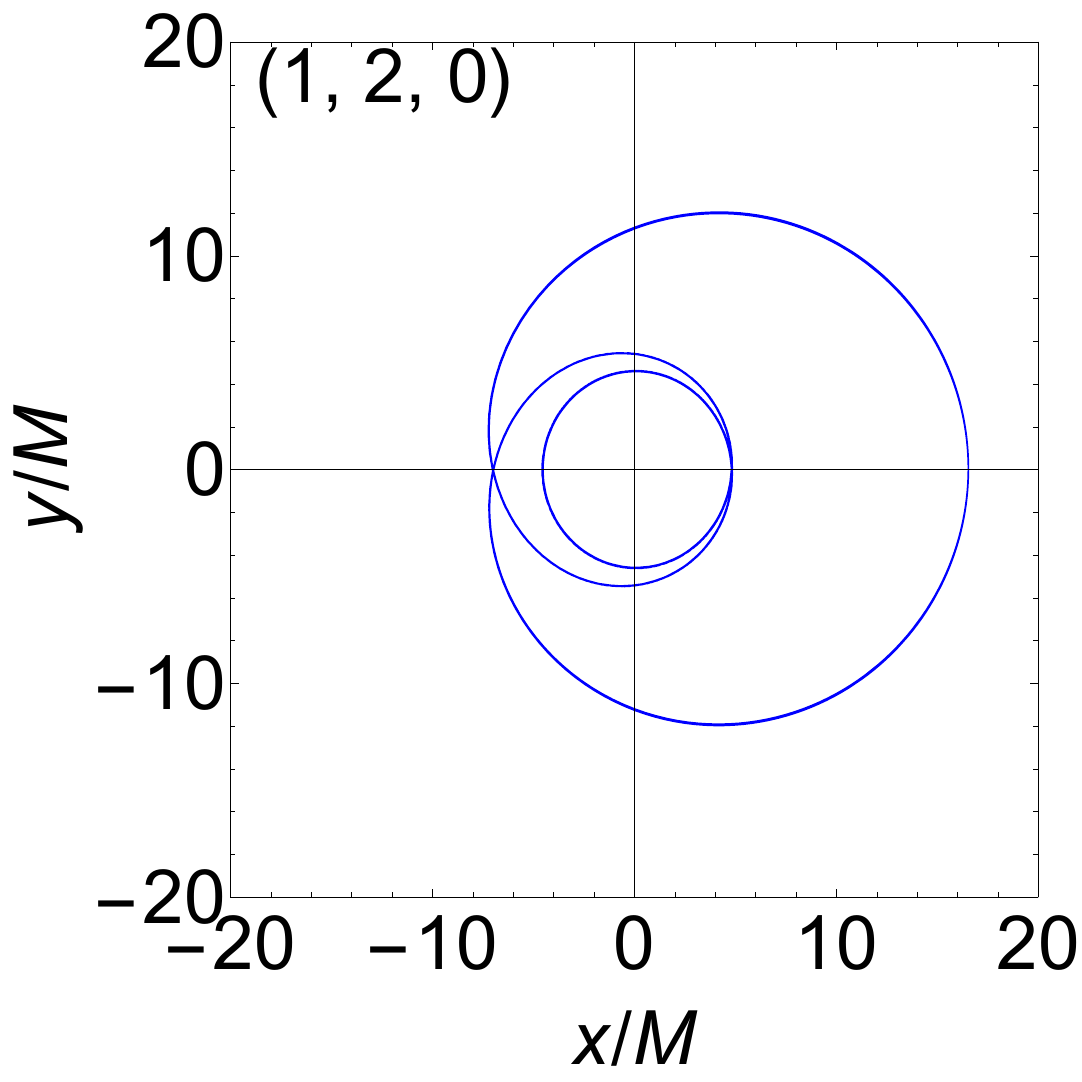}}
    \hspace{5mm}
    \subfigure{\includegraphics[scale =0.25]{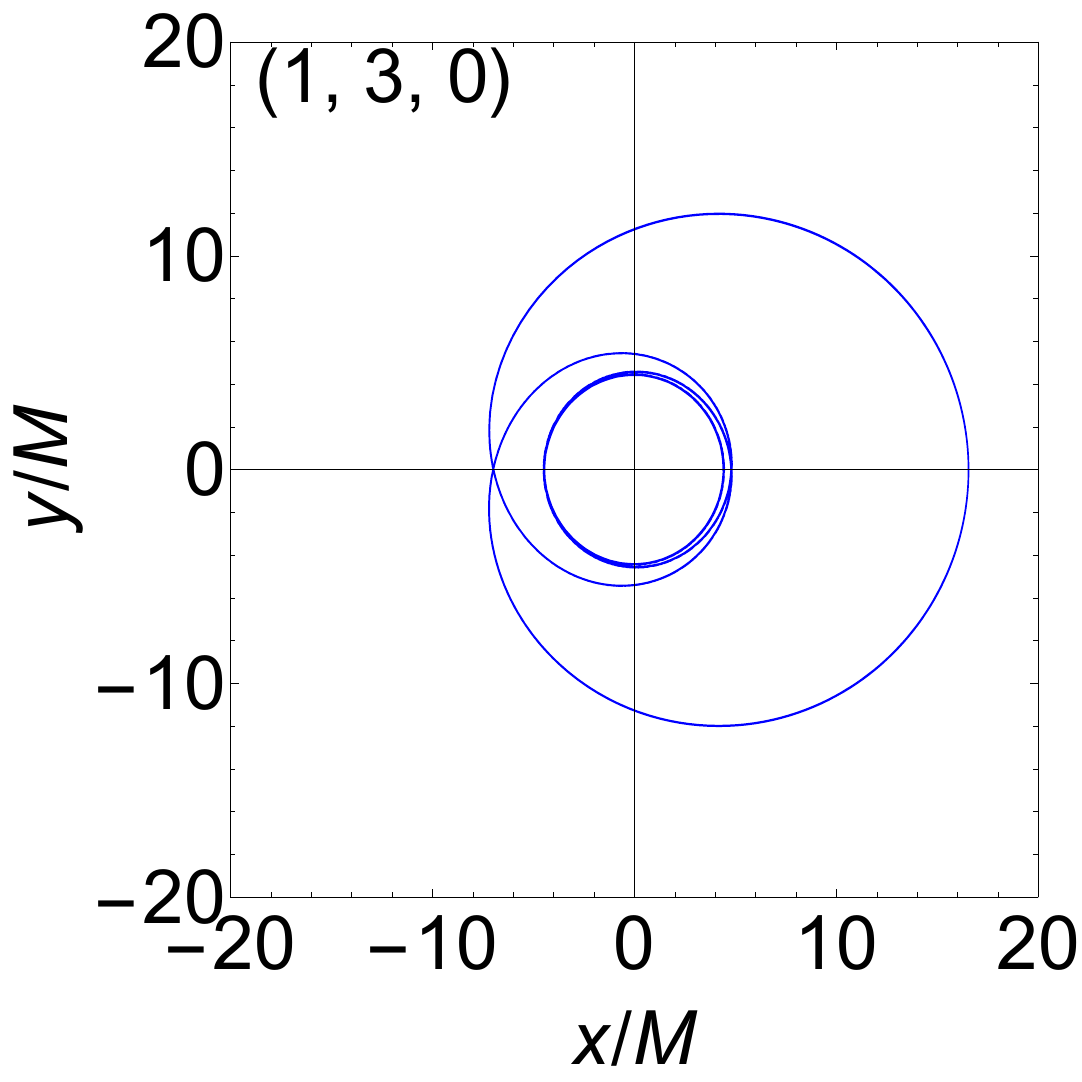}}
    \\
    \subfigure{\includegraphics[scale =0.25]{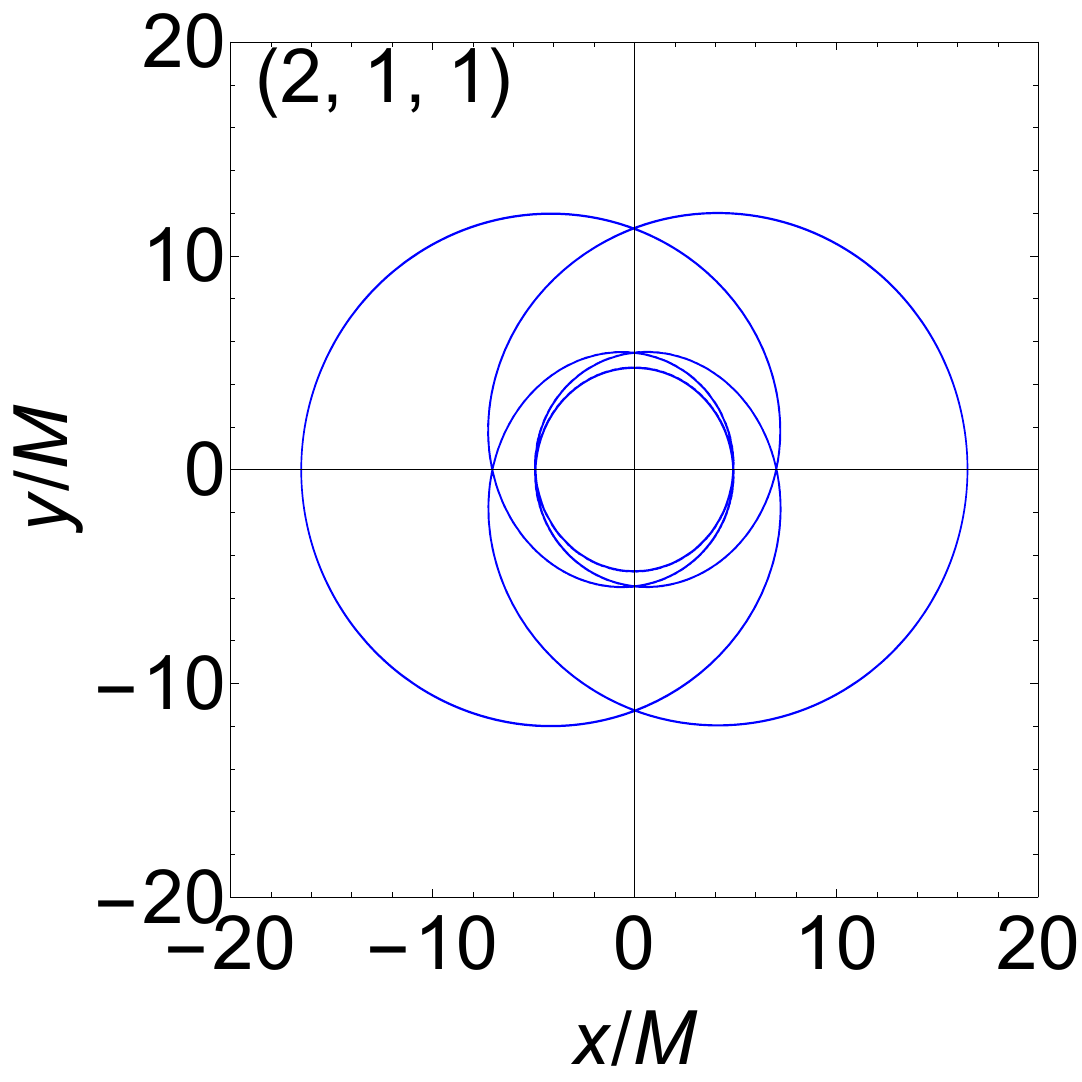}}
    \hspace{5mm}
	\subfigure{\includegraphics[scale =0.25]{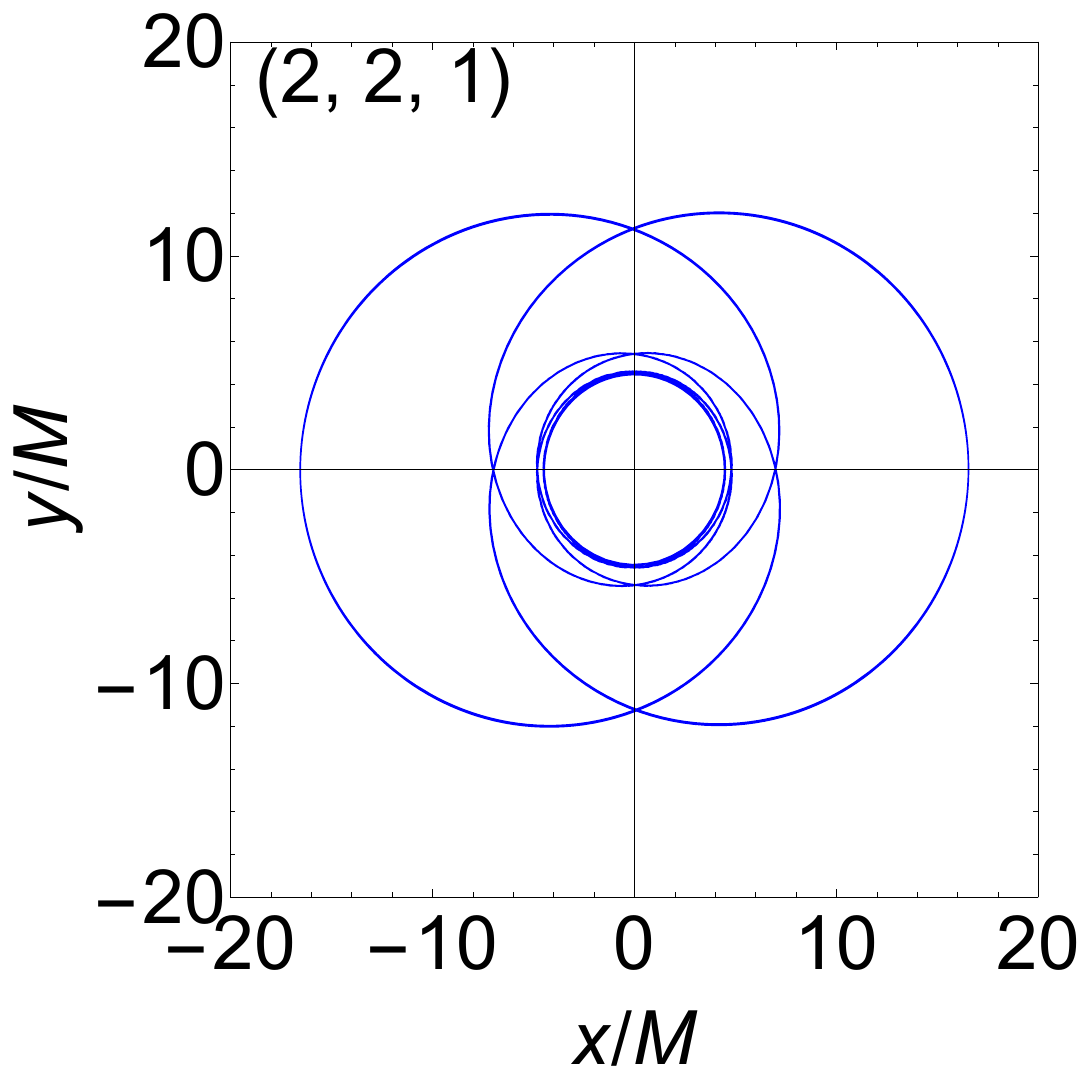}}
    \hspace{5mm}
    \subfigure{\includegraphics[scale =0.25]{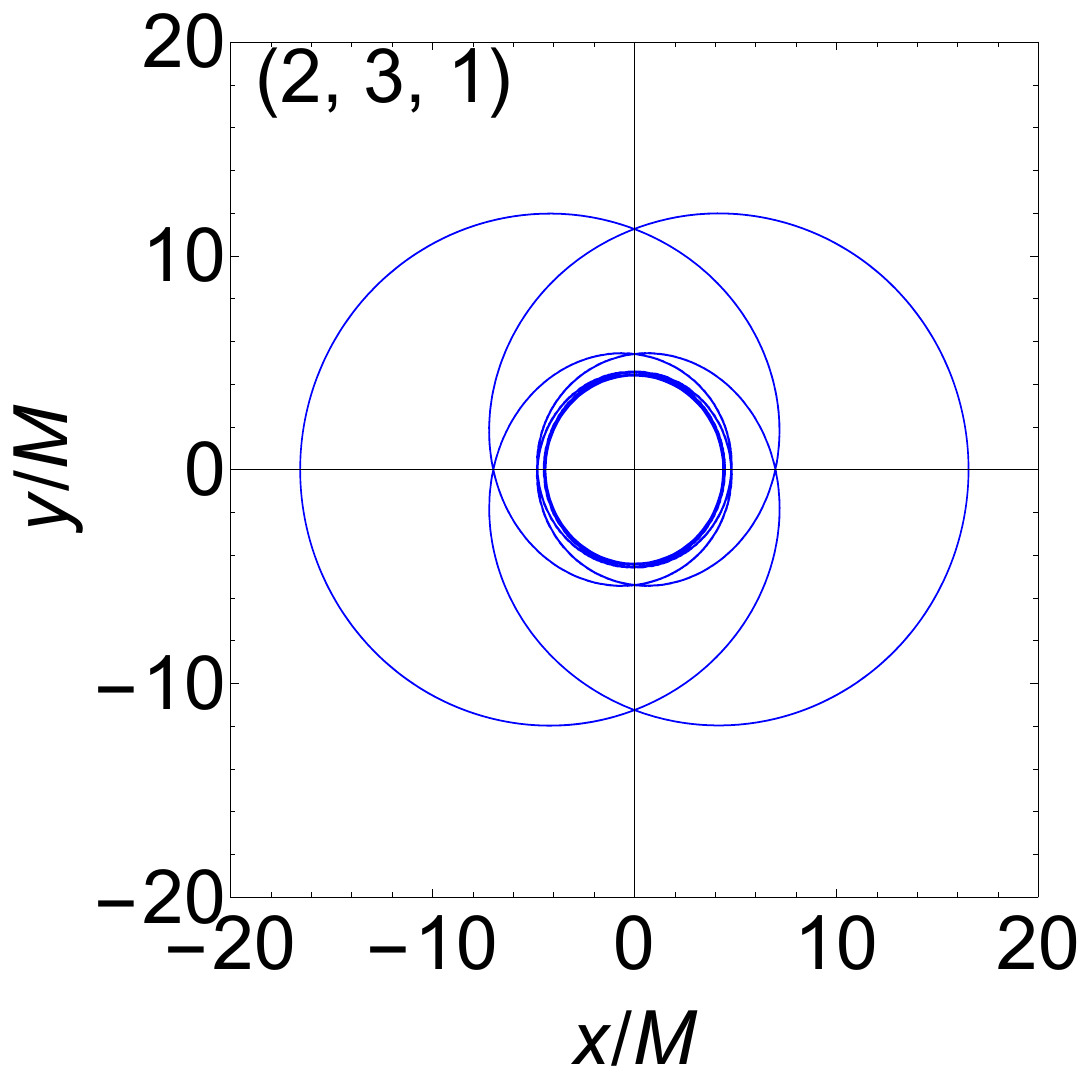}}
    \\
    \subfigure{\includegraphics[scale =0.25]{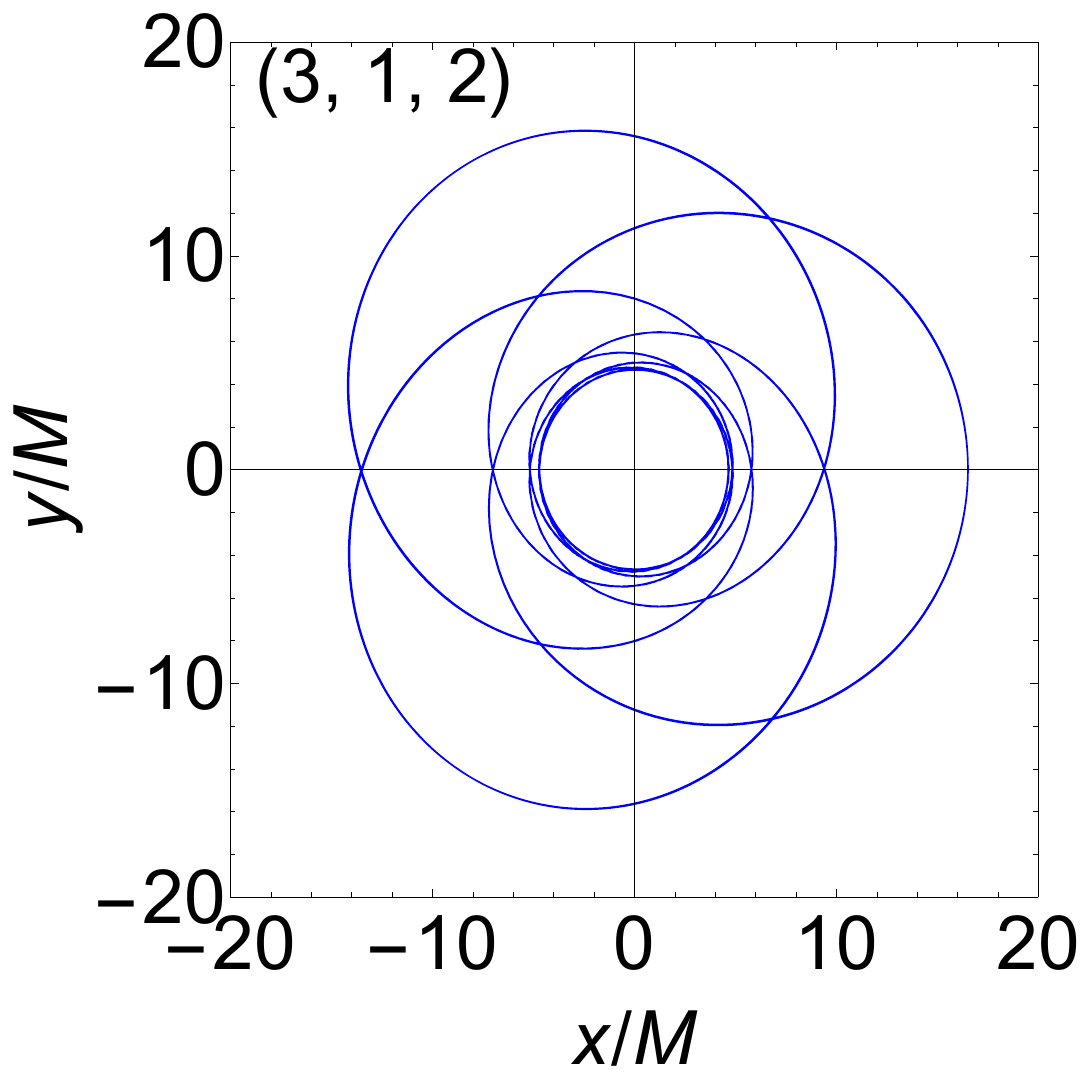}}
    \hspace{5mm}
	\subfigure{\includegraphics[scale =0.25]{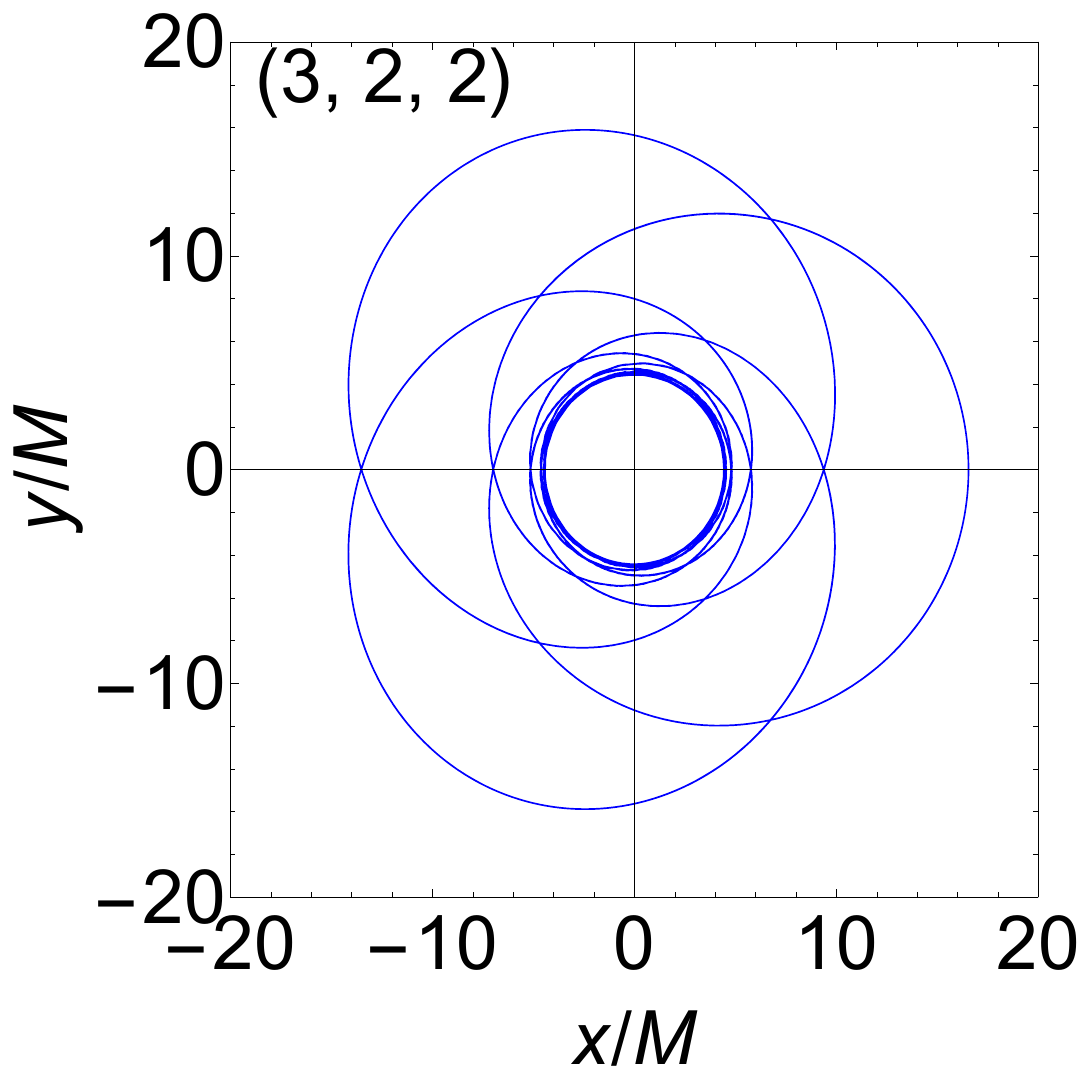}}
    \hspace{5mm}
    \subfigure{\includegraphics[scale =0.25]{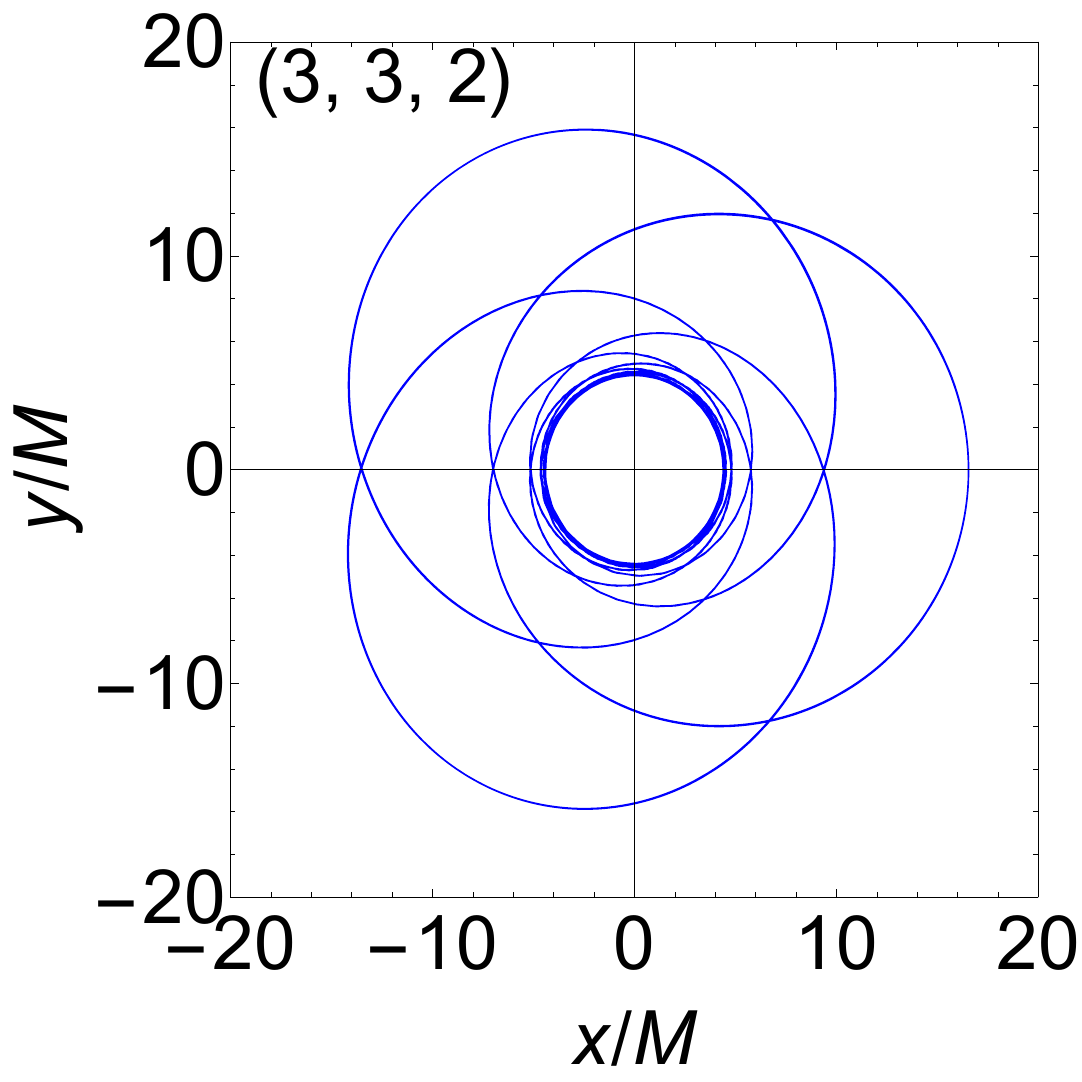}}
    \\
    \subfigure{\includegraphics[scale =0.25]{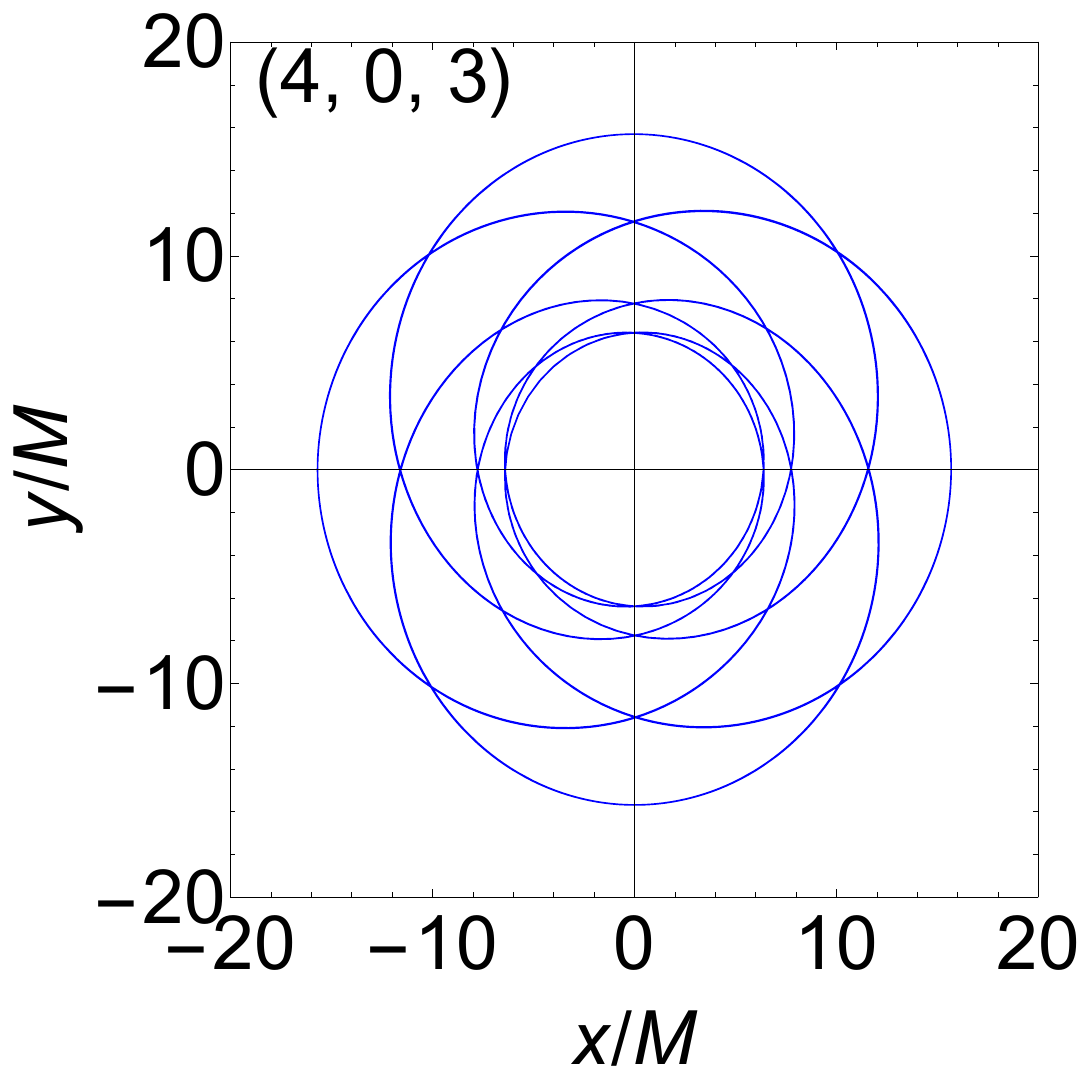}}
    \hspace{5mm}
	\subfigure{\includegraphics[scale =0.25]{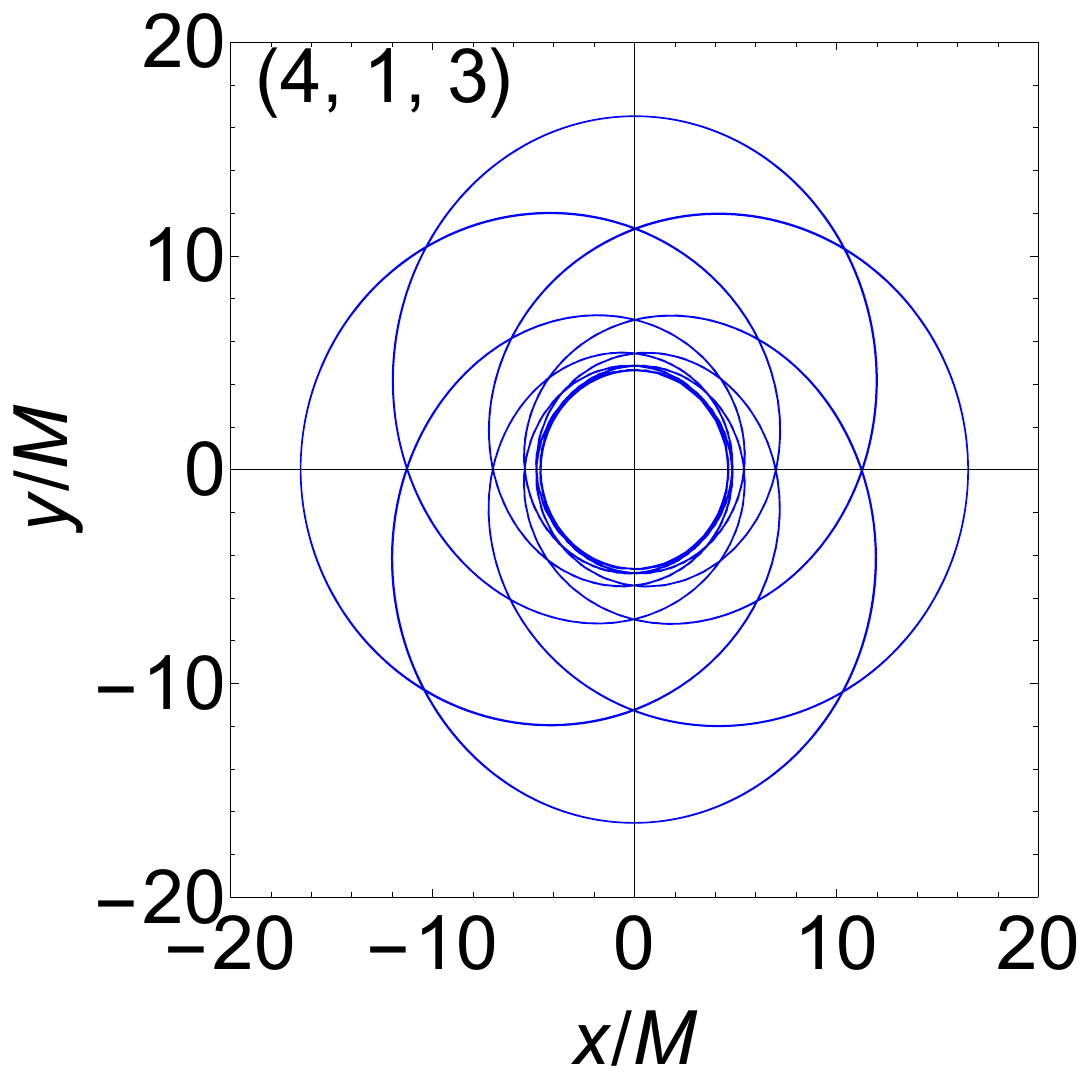}}
    \hspace{5mm}
    \subfigure{\includegraphics[scale =0.25]{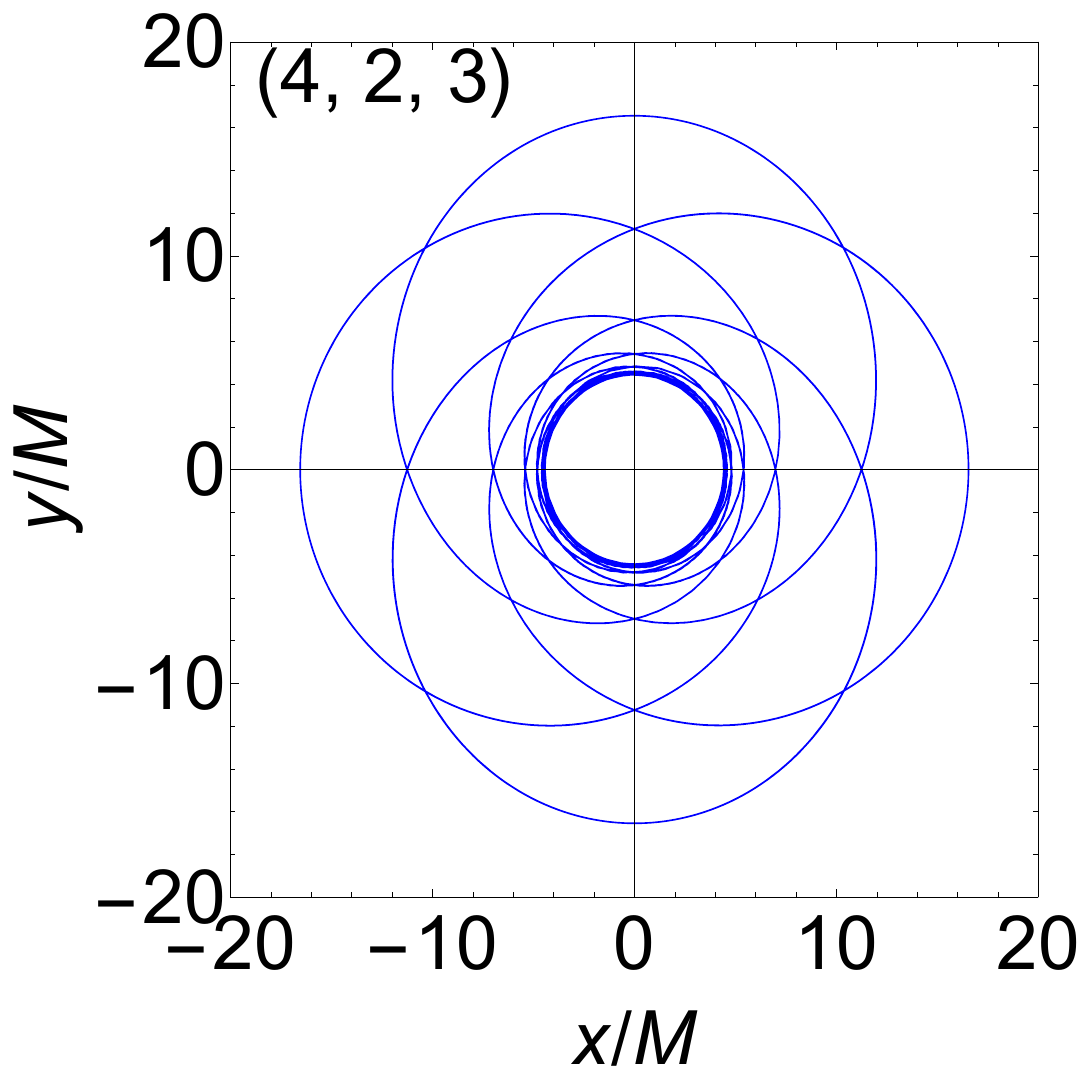}}
	\caption{Periodic orbits with different $(z, w, v)$ around the quantum-corrected black hole with $\hat{\alpha} = 0.8$ and $E = 0.96$.}
	\label{plot-orbit-a08-FE}
\end{figure}

\begin{figure}[!t]
	\centering
	\subfigure{\includegraphics[scale =0.25]{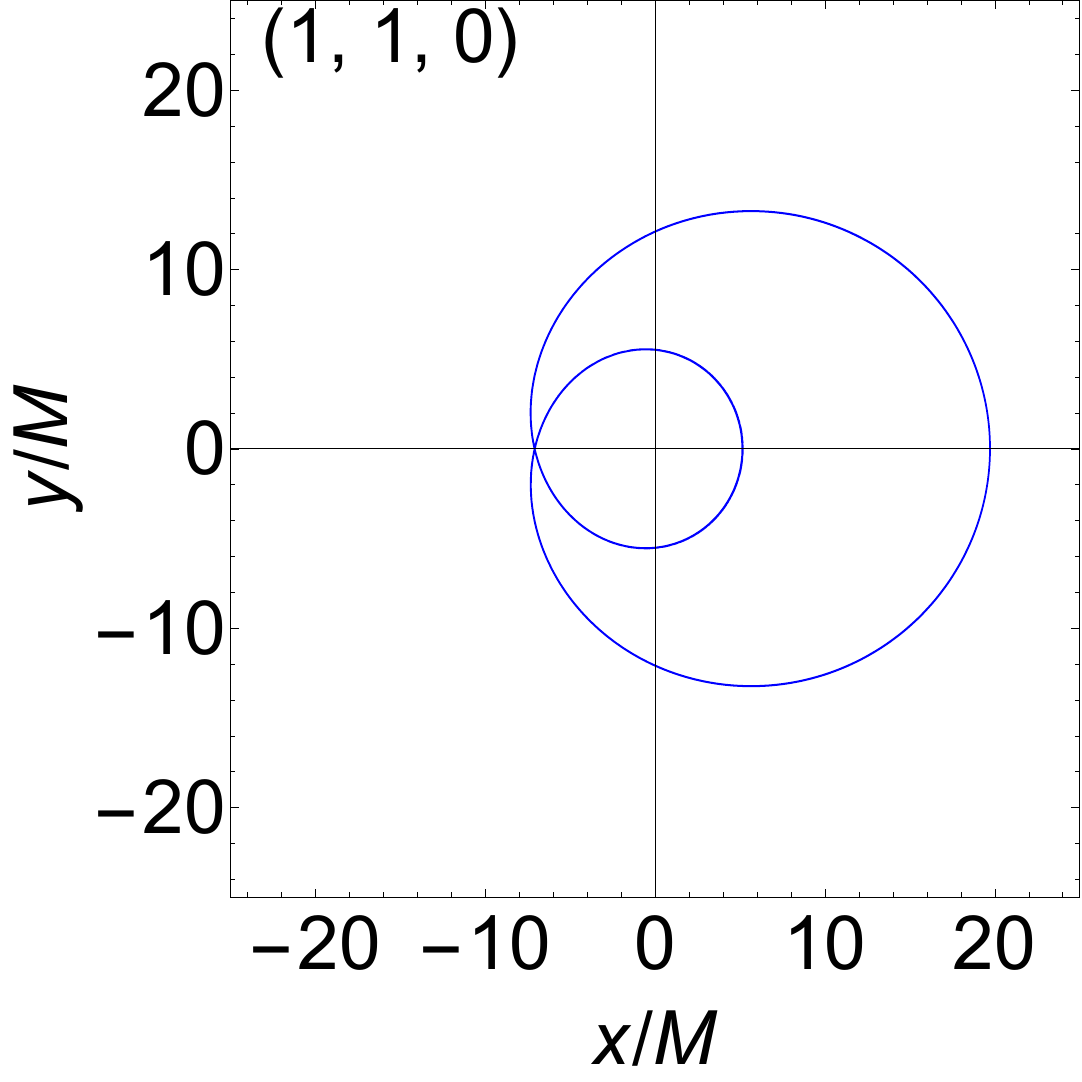}}
    \hspace{5mm}
	\subfigure{\includegraphics[scale =0.25]{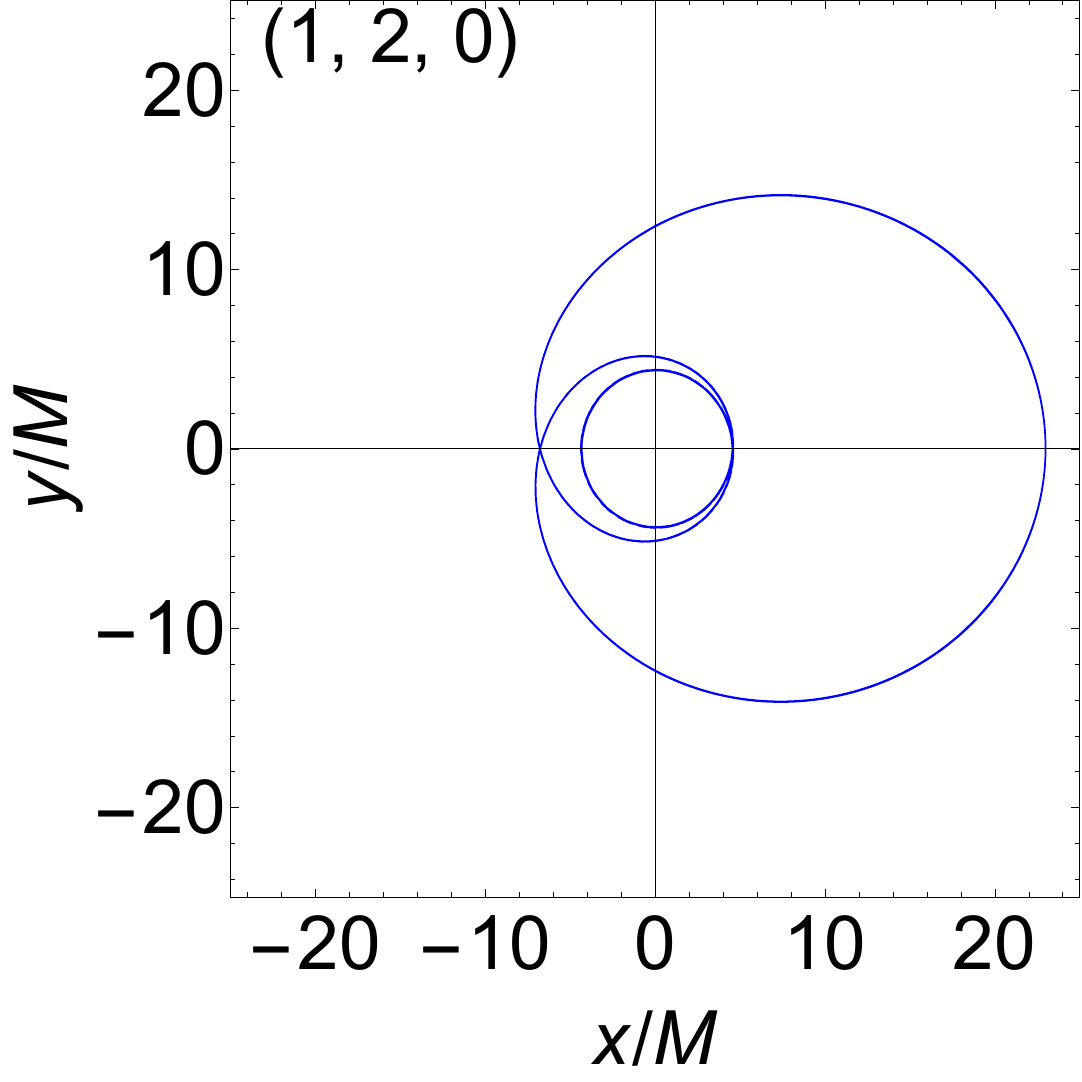}}
    \hspace{5mm}
    \subfigure{\includegraphics[scale =0.25]{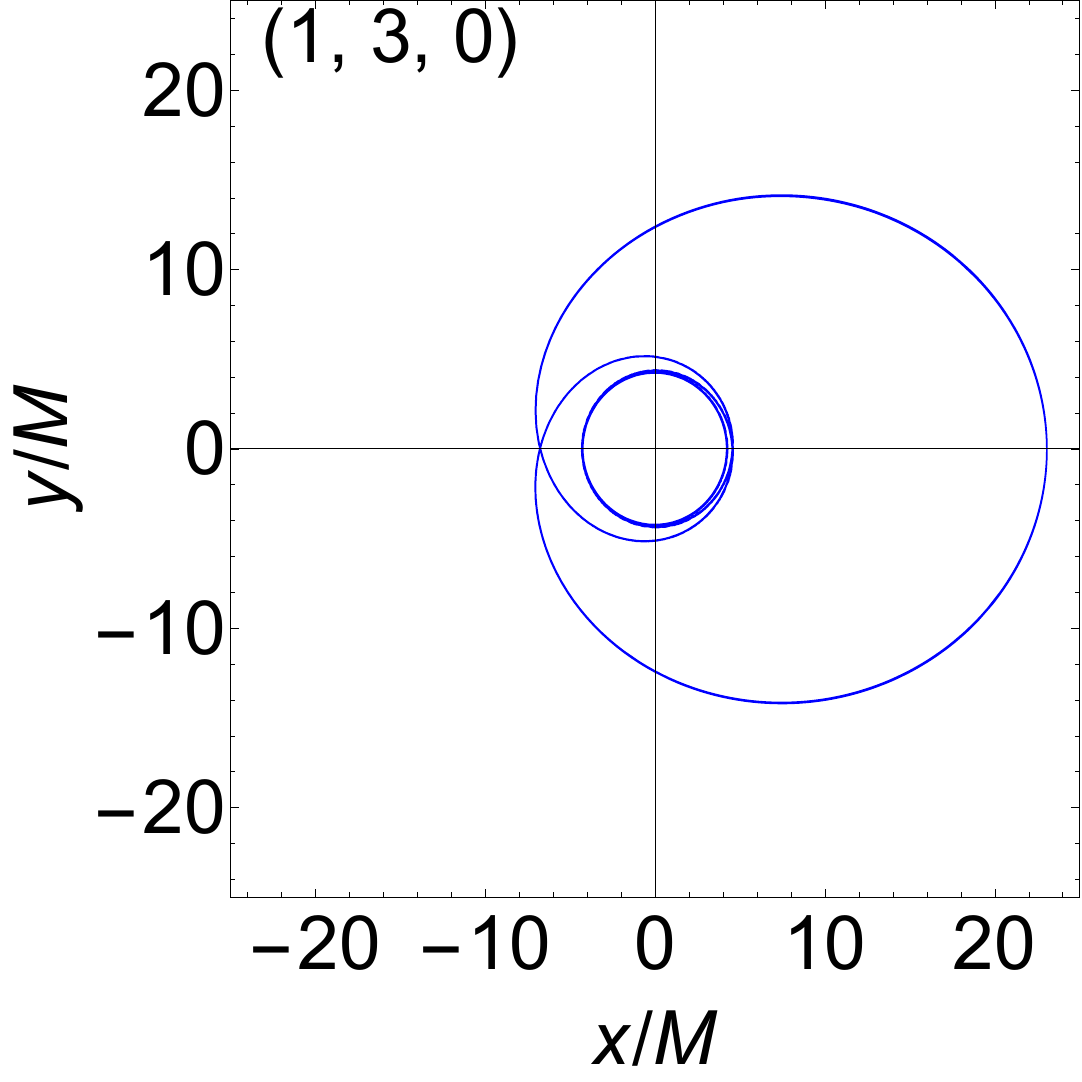}}
    \\
    \subfigure{\includegraphics[scale =0.25]{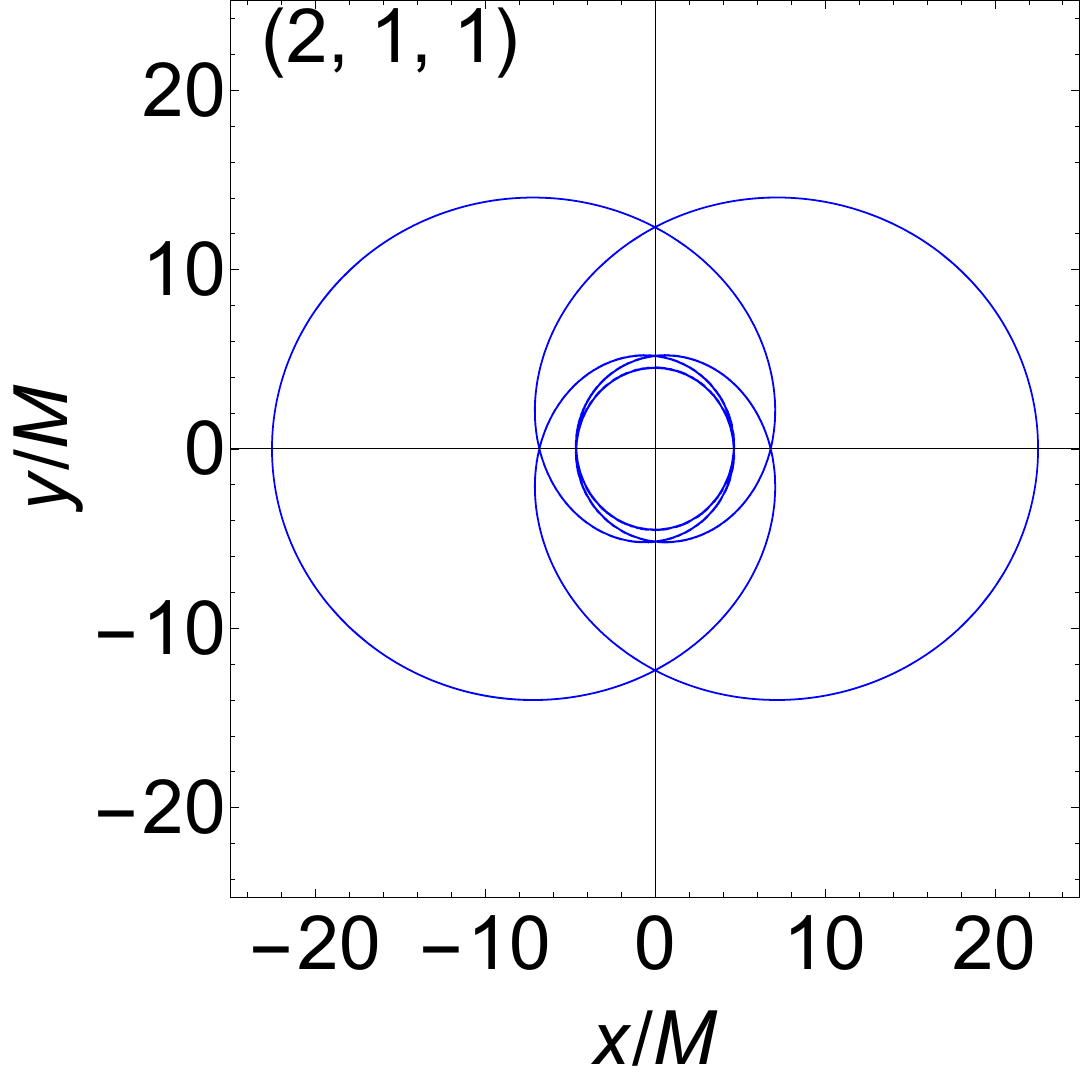}}
    \hspace{5mm}
	\subfigure{\includegraphics[scale =0.25]{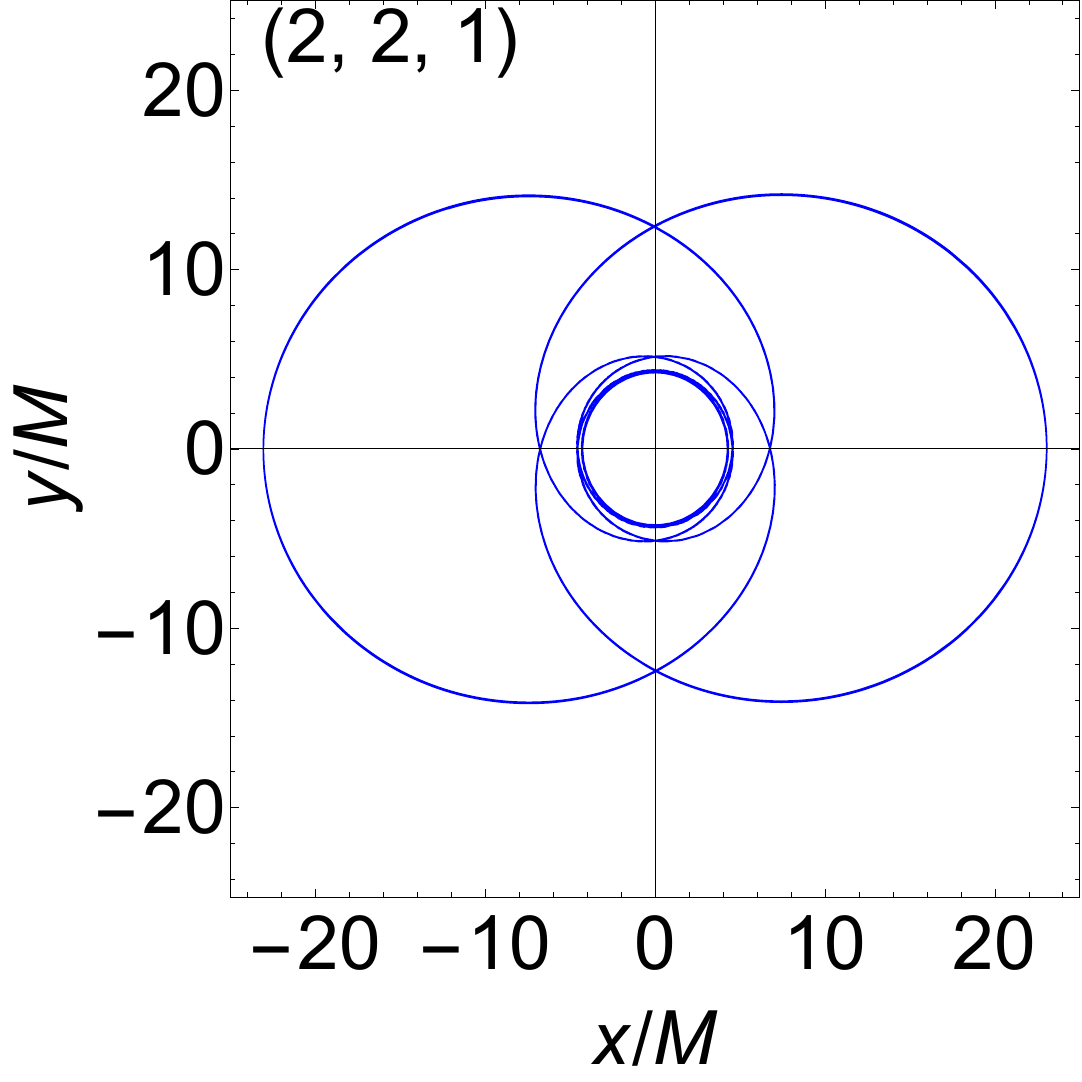}}
    \hspace{5mm}
    \subfigure{\includegraphics[scale =0.25]{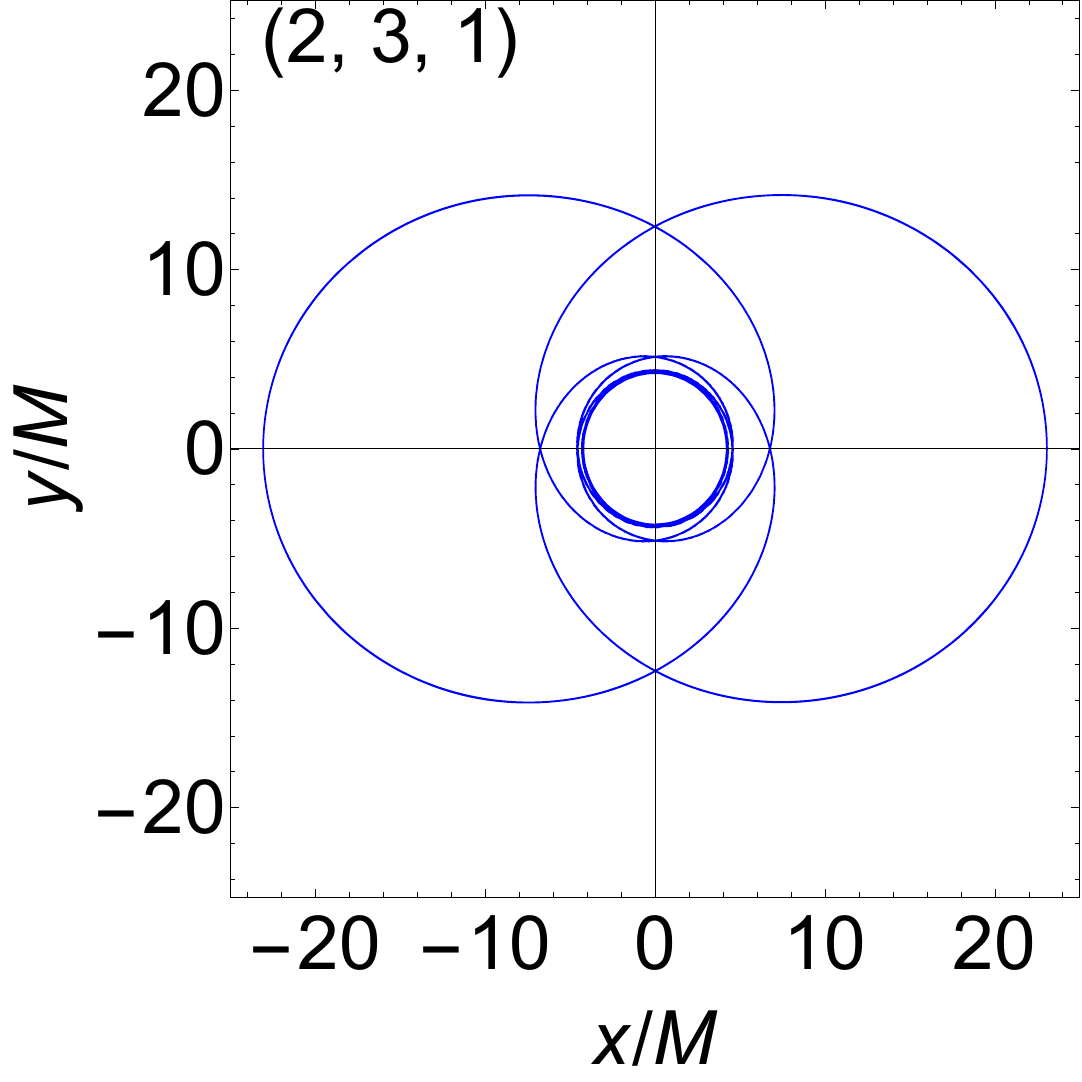}}
    \\
    \subfigure{\includegraphics[scale =0.25]{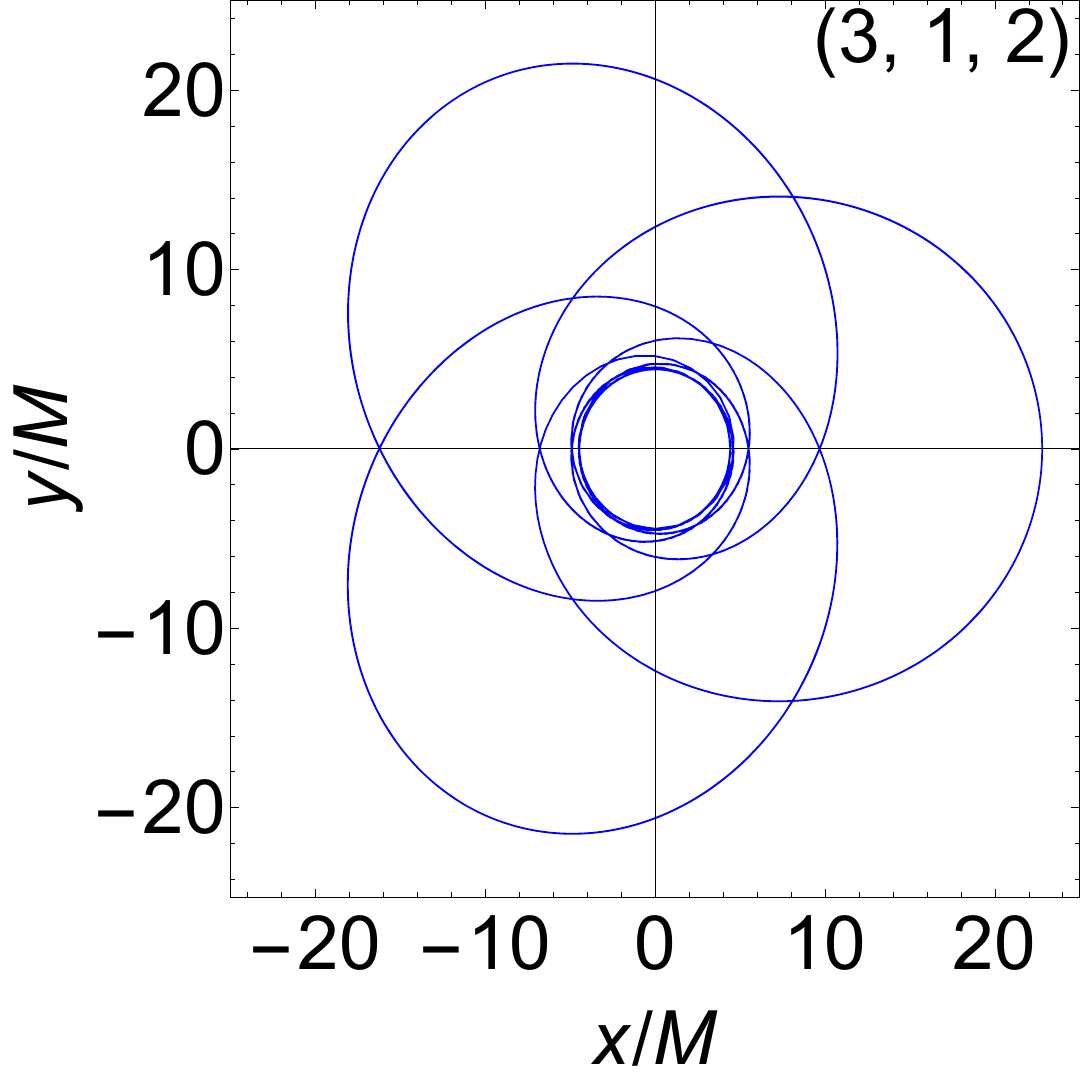}}
    \hspace{5mm}
	\subfigure{\includegraphics[scale =0.25]{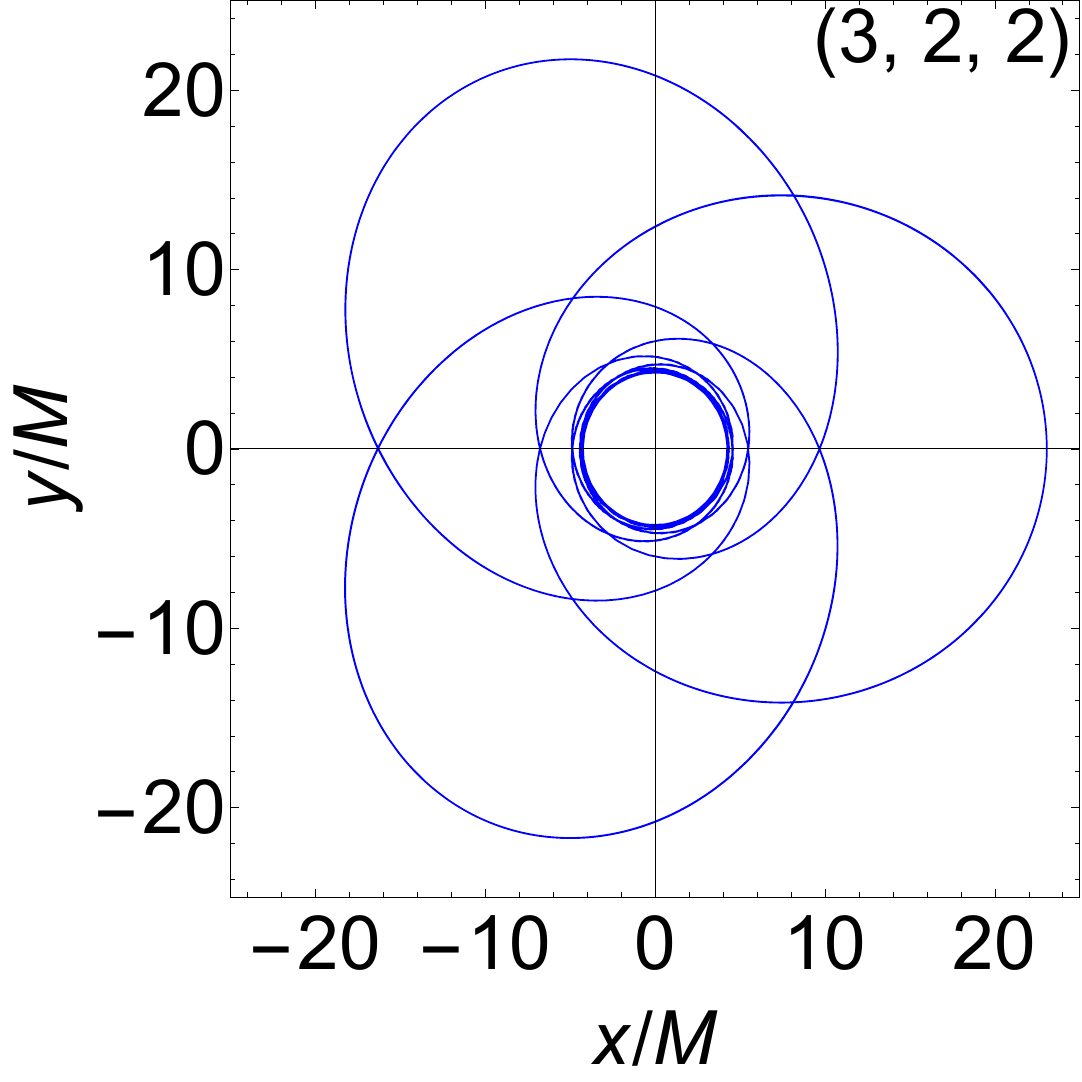}}
    \hspace{5mm}
    \subfigure{\includegraphics[scale =0.25]{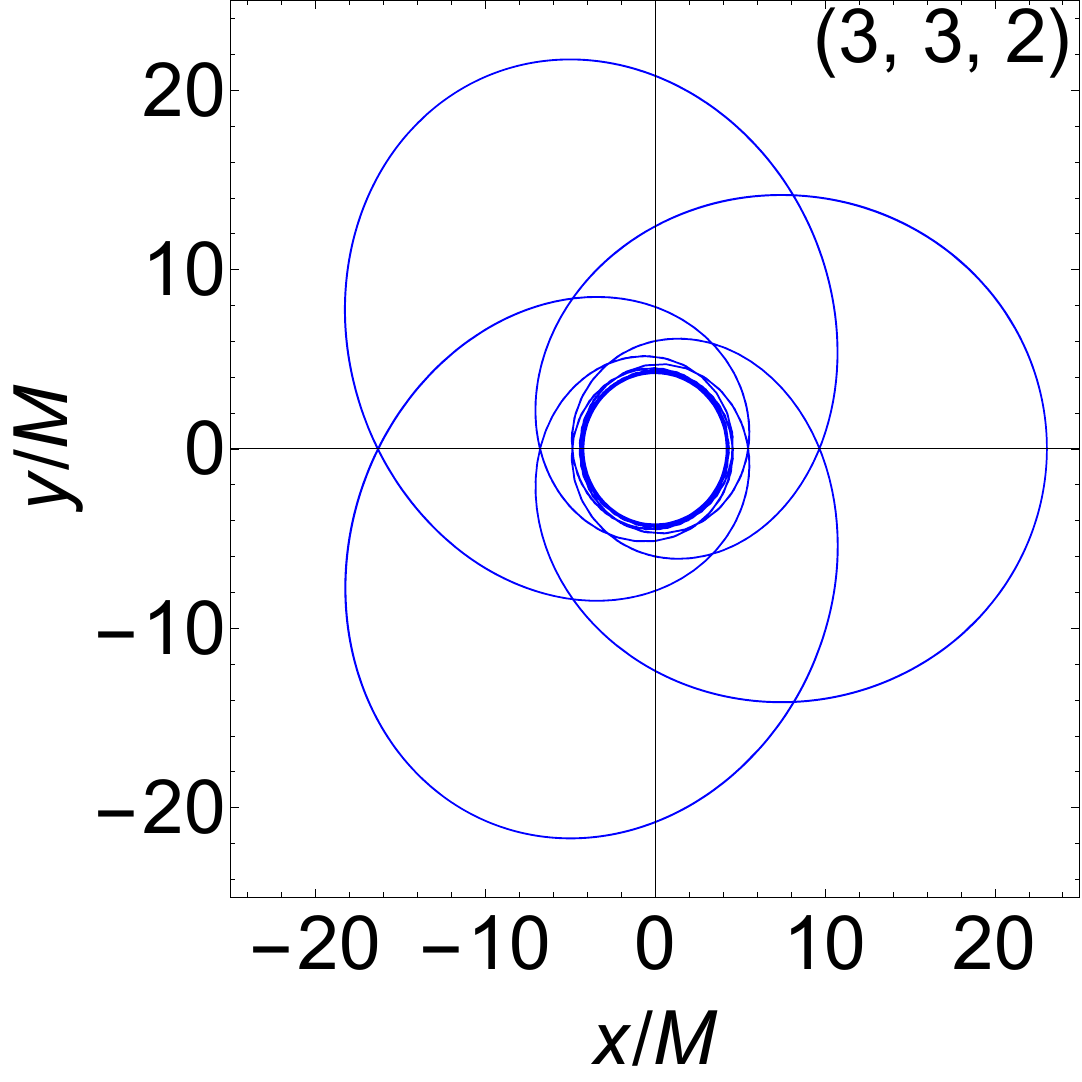}}
    \\
    \subfigure{\includegraphics[scale =0.25]{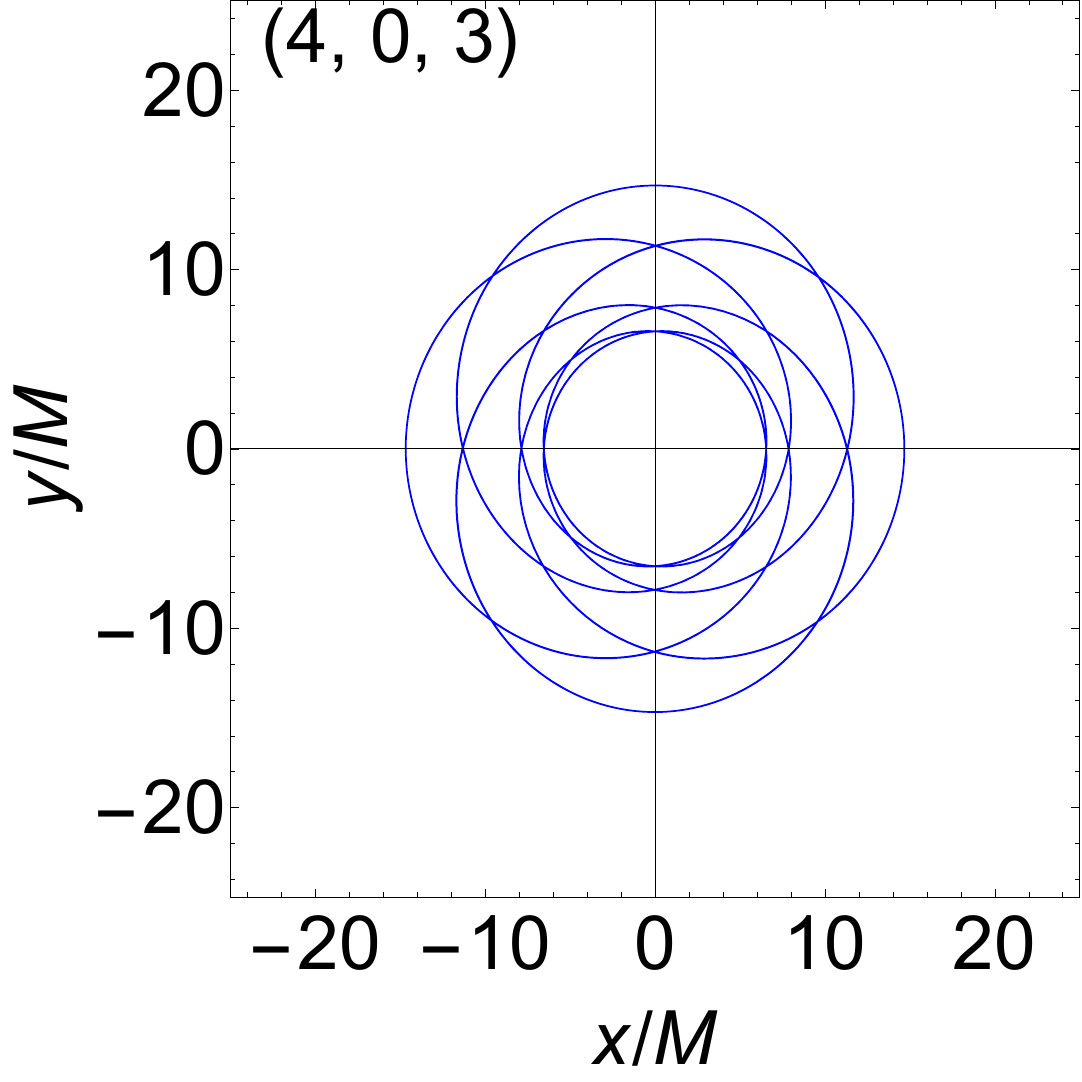}}
    \hspace{5mm}
	\subfigure{\includegraphics[scale =0.25]{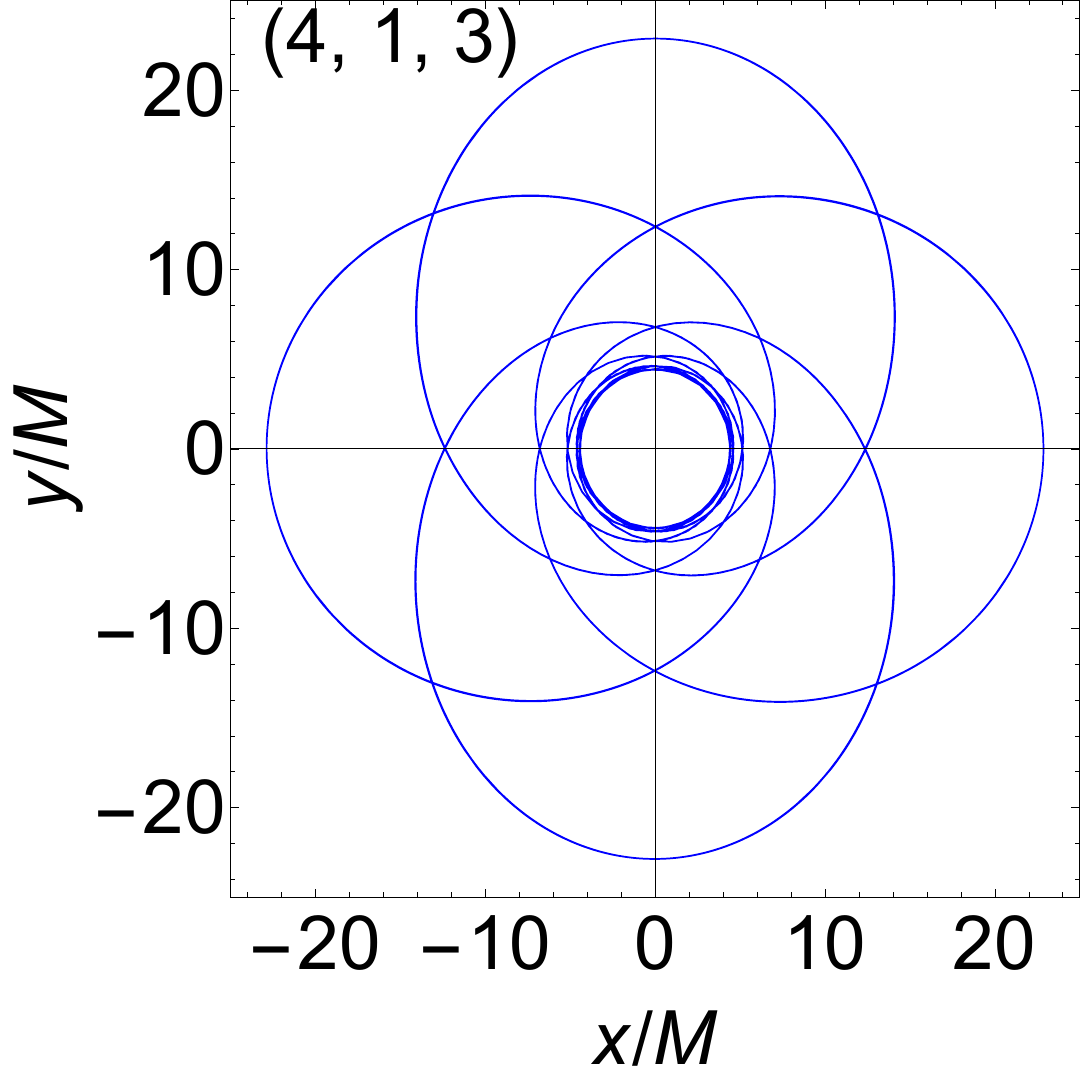}}
    \hspace{5mm}
    \subfigure{\includegraphics[scale =0.25]{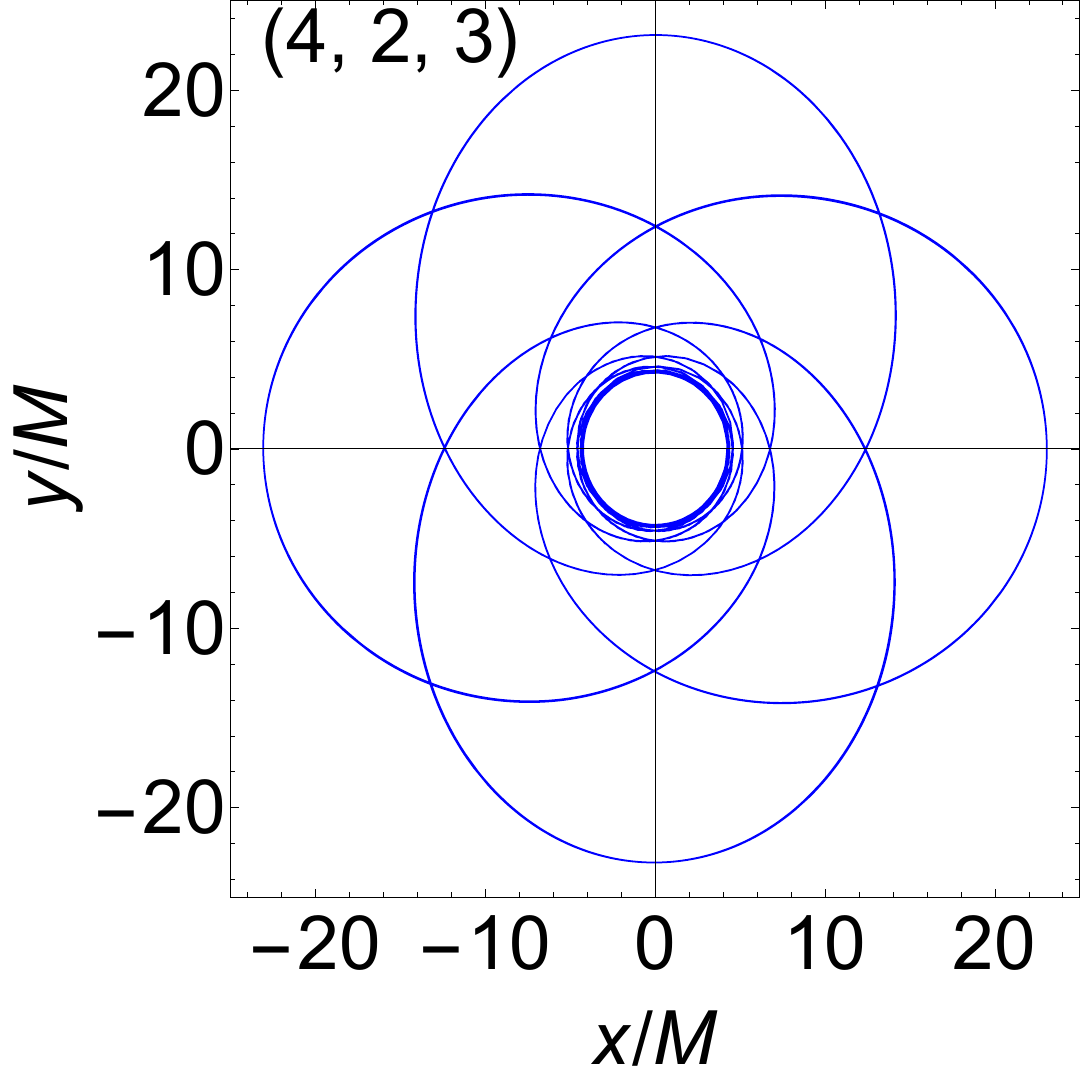}}
	\caption{Periodic orbits with different $(z, w, v)$ around the quantum-corrected black hole with $\hat{\alpha} = 0.8$ and $ L = (L_{\tx{MBO}} + L_\tx{{ISCO}})/2 $.}
	\label{plot-orbit-a08-FL}
\end{figure}

We plot the periodic orbits with different indexes $(z,~ w,~ v)$ around the quantum-corrected black hole with the parameter $\hat{\alpha} = 0.8$ in Figs. \ref{plot-orbit-a08-FE} and \ref{plot-orbit-a08-FL}, for which we fix the energy $ E = 0.96 $ or fix the orbital angular momentum $ L = (L_{\tx{MBO}} + L_\tx{{ISCO}})/2 $. One can see that the periodic orbit with a larger value of the zoom number $z$ has richer structural features, and the periodic orbit with a larger value of the whirl number $w$ revolves around the central black hole for more orbits between successive apoapses.

\section{Numerical Kludge gravitational waveforms from periodic orbits}\lb{sec4}
\renewcommand{\theequation}{4.\arabic{equation}}
\setcounter{equation}{0} 

A stellar-mass object along a periodic orbit around a supermassive quantum-corrected black hole forms an EMRI system. The gravitational waves radiated from this EMRI system may contain information about the periodic orbit and the quantum correction of the supermassive black hole. When computing gravitational waveforms from an EMRI, the adiabatic approximation method stands out as a prevalent technique \cite{Hughes:1999bq, Hughes:2001jr, Glampedakis:2002ya, Hughes:2005qb, Drasco:2005is, Gair:2005ih, Glampedakis:2005cf, Drasco:2005kz, Sundararajan:2007jg, Sundararajan:2008zm, Miller:2020bft, Isoyama:2021jjd}. Because the orbital parameters of the small object evolve on a much longer time scale than the orbital periods, one can set both the small object's energy and orbital angular momentum as constants and assume that the orbit is geodesic during a few periods of the orbit. In this section, we calculate the gravitational waveforms from the periodic orbits for one complete cycle. For this short duration, we can temporarily neglect the influence on the motion of the small object from the gravitational radiation.

The numerical Kludge waveform model has been proven to be effective in investigating gravitational waveforms from EMRIs \cite{Babak:2006uv}. Adopting the numerical kludge scheme, we first numerically solve the equations of motion \eqref{momentum-1} -\eqref{momentum-4} to obtain the periodic orbits of a small object around the supermassive quantum-corrected black hole. Then we use the quadrupole formula of gravitational radiation \cite{Thorne:1980ru, gravity-book} to generate the corresponding gravitational waveforms. For the small object with mass $m$, the symmetric and trace-free (STF) mass quadrupole is defined as \cite{Thorne:1980ru}
\bqn\lb{Iij}
I^{ij} = \left[ \int d^3x x^i x^j T^{tt}(t, x^i) \right]^{(\text{STF})},
\eqn 
where the $tt$-component of the stress-energy tensor for the small object with trajectory $Z^i(t)$ is
\bqn\lb{Ttt}
T^{tt}(t, x^i) = m \delta^{3} (x^i - Z^i(t)).
\eqn
Following the method in Ref.~\cite{Babak:2006uv}, we treat the Boyer-Lindquist coordinates as a fictitious spherical polar coordinate, then project the trajectory of the small object onto the following Cartesian coordinate
\bqn
x = r \sin \theta \cos \phi,~~~~y = r \sin \theta \sin \phi,~~~~z = r \cos \theta.
\eqn
Combining Eqs. \eqref{Iij} and \eqref{Ttt}, one can obtain the metric perturbations describing the gravitational waves as
\bqn\lb{metric perturbations}
h_{ij} = \f{2}{D_\tx{L}} \frac{d^2 I_{ij}}{dt^2} = \f{2 m}{D_\tx{L}} (a_i x_j + a_j x_i +2 v_i v_j),
\eqn 
where $D_\tx{L}$ is the luminosity distance from the EMRI system to the detector, $v_i$ and $a_i$ are the small object's spatial velocity and acceleration, respectively. We can construct a detector-adapted coordinate system $(X,~Y,~Z)$ with the origin coinciding with that of the original $(x,~y,~z)$ coordinate system, both centered on the supermassive black hole \cite{gravity-book}. The directions of the detector-adapted coordinate system in the original $(x,~y,~z)$ coordinates are  
\bqn
\mathbf{e}_X &=& [\cos \zeta,~-\sin\zeta, ~0], \lb{ex}\\
\mathbf{e}_Y &=& [\sin \iota \sin \zeta,~-\cos \iota \cos \zeta, ~- \sin \iota], \lb{ey}\\
\mathbf{e}_Z &=& [\sin \iota \sin \zeta,~-\sin \iota \cos \zeta,~ \cos \iota], \lb{ez}
\eqn
where $\iota$ is the inclination angle of the orbital plane of the smaller object to the $X-Y$ plane and $\zeta$ is the longitude of the pericenter measured in the orbital plane. After that, we can get the corresponding gravitational-wave polarizations from Eq. \eqref{metric perturbations} as follows
\bqn
h_{+} &=& \f{1}{2} (\mathbf{e}_X^i \mathbf{e}_X^j - \mathbf{e}_Y^i \mathbf{e}_Y^j) h_{ij}, \lb{waveform-11}\\
h_{\times} &=& \f{1}{2} (\mathbf{e}_X^i \mathbf{e}_Y^j + \mathbf{e}_Y^i \mathbf{e}_X^j) h_{ij}. \lb{waveform-12}
\eqn
Combining Eqs. \eqref{ex} and \eqref{ey}, the gravitational-wave polarizations \eqref{waveform-11} and \eqref{waveform-12} become
\bqn
h_{+} &=&  (h_{\zeta \zeta} - h_{\iota \iota})/2, \lb{waveform-21}\\
h_{\times} &=&  h_{\iota \zeta}, \lb{waveform-22}
\eqn
where the components $h_{\zeta \zeta}$, $h_{\iota \iota}$,  and $h_{\iota \zeta}$ are given by  \cite{Babak:2006uv}
\bqn
h_{\zeta \zeta} &=& h_{xx} \cos^2 \zeta - h_{xy}  \sin 2 \zeta + h_{yy} \sin^2 \zeta,\\
h_{\iota \iota} &=&  \cos^2 \iota [h_{xx} \sin^2 \zeta + h_{xy} \sin 2 \zeta + h_{yy} \cos^2 \zeta] + h_{zz} \sin^2 \iota - \sin 2 \iota [h_{xz} \sin \zeta + h_{yz} \cos \zeta], \\
h_{\iota \zeta} &=& \cos \iota \left[ 
 \f{1}{2} h_{xx} \sin 2 \zeta + h_{xy} \cos 2 \zeta - \f{1}{2} h_{yy} \sin 2 \zeta \right] + \sin \iota [h_{yz} \sin \zeta - h_{xz} \cos \zeta]. 
\eqn 

\begin{figure}[!t]
	\centering
	\subfigure[]{\includegraphics[scale =0.3]{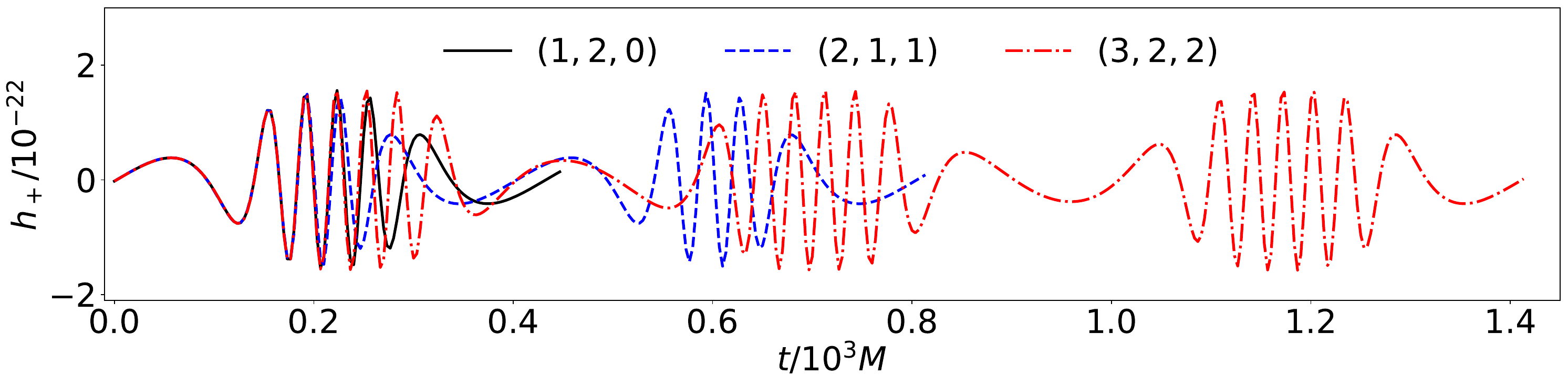}}
	\subfigure[]{\includegraphics[scale =0.3]{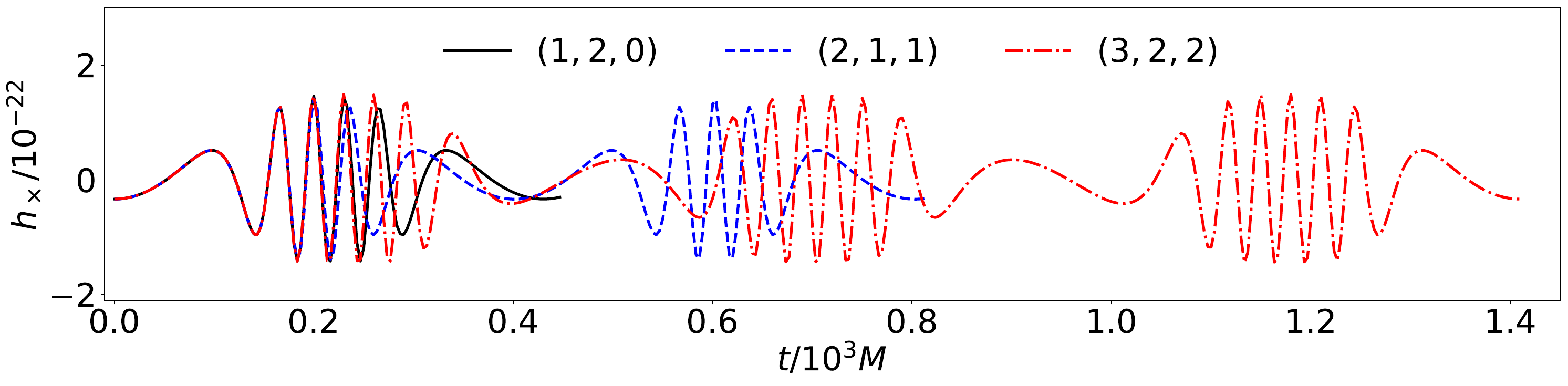}}
	\caption{Gravitational waveforms from a test object with $m = 10 M_\odot$ along different periodic orbits around a supermassive quantum-corrected black hole with $M = 10^6 M_\odot$ and $\hat{\alpha} = 0.8$. The energy is fixed as $E =0.96$.}
	\label{plot-waveform-a08}
\end{figure}

\begin{figure}[!t]
	\centering
	\subfigure[]{\includegraphics[scale =0.3]{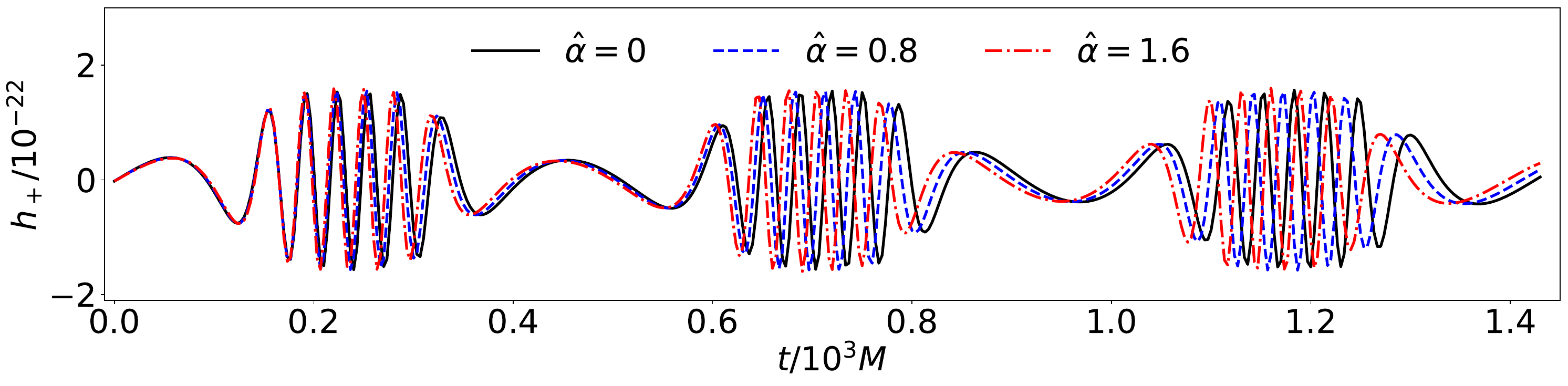}}
	\subfigure[]{\includegraphics[scale =0.3]{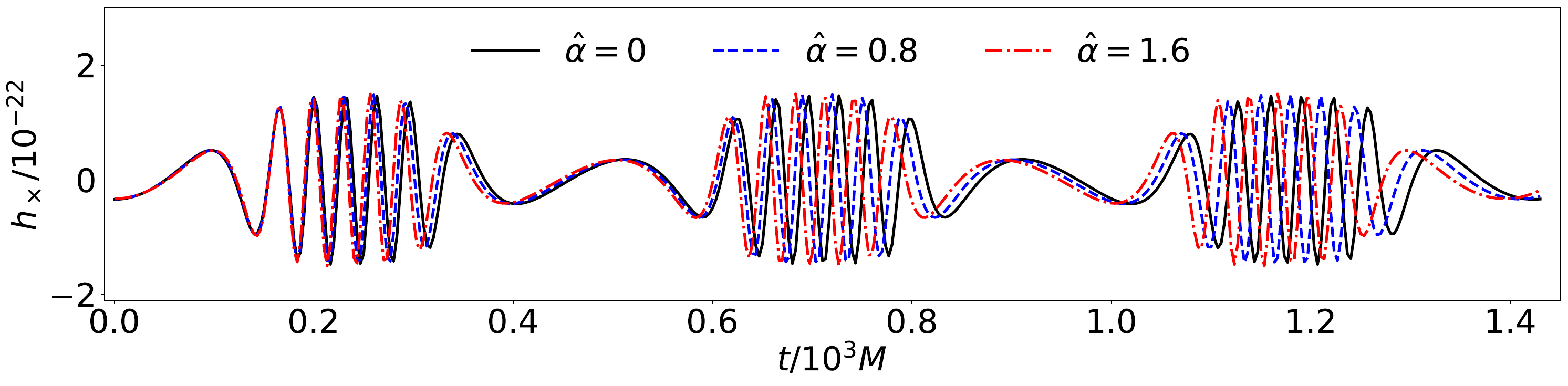}}
	\caption{Gravitational waveforms from a test object with $m = 10 M_\odot$ along the $(3,~2,~2)$ periodic orbits around a supermassive quantum-corrected black hole with $M = 10^6 M_\odot$ and different value of the parameter $\hat{\alpha}$. The energy is fixed as $E =0.96$.}
	\label{plot-waveform-322}
\end{figure}

\begin{figure}[!t]
	\centering
	\subfigure[]{\includegraphics[scale =0.3]{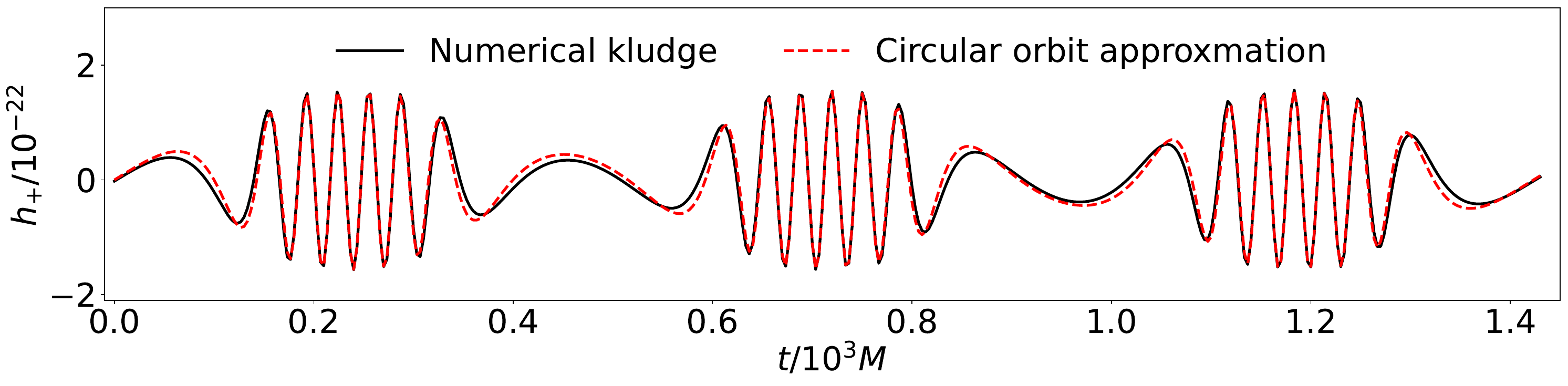}}
	\subfigure[]{\includegraphics[scale =0.3]{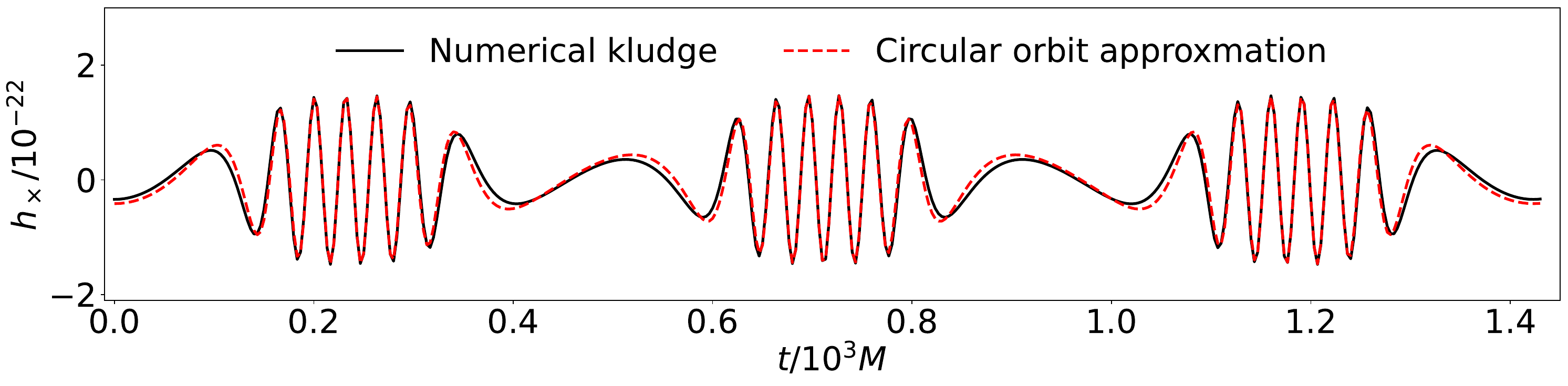}}
	\caption{Gravitational waveforms, calculated by two different methods, from a test object with $m = 10 M_\odot$ along the $(3,~ 2,~2)$ periodic orbits around a supermassive Schwarzschild black hole with $M = 10^6 M_\odot$ and $\hat{\alpha}=0$.}
	\label{plot-waveform-compare}
\end{figure}

Considering a small object with mass $m = 10 M_\odot$ along different periodic orbits around a supermassive quantum-corrected black hole with mass $M = 10^6 M_\odot$ at the luminosity distance $D_\tx{L} = 2\tx{Gpc}$, and setting the inclination angle $\iota = \pi/4$ and the longitude of pericenter $\zeta = \pi/4$, we calculate the polarizations \eqref{waveform-21} and \eqref{waveform-22} of the gravitational wave from the EMRI system with the parameter $\hat{\alpha} = 0.8$. We plot the numerical results in Fig.~\ref{plot-waveform-a08}. It shows that the gravitational waveforms have zoom and whirl phases in one complete period, which correspond to the zoom and whirl behaviors of the small object's periodic orbits. The gravitational waveforms from periodic orbits with larger zoom numbers $z$ exhibit richer substructures, which correspond to more leaves of a complete periodic orbit. Under the same setup, we also plot the gravitational waveforms from the EMRI system with fixed $(3,~2,~2)$ periodic orbits and different values of the parameter $\hat{\alpha}$ in Fig.~\ref{plot-waveform-322}. It can be found that the parameter $\hat{\alpha}$ slightly influences the amplitude of the waveforms and the phase of the waveforms advances as the parameter $\hat{\alpha}$ increases. Moreover, this phase advancement accumulates continuously with the evolution of the orbit and becomes more pronounced over time. Taking the $(3,~2,~2)$ periodic orbits around a supermassive Schwarzschild black hole as an example, we compare the gravitational waveforms obtained through the numerical kludge scheme with those derived from the circular orbit approximation \cite{Tu:2023xab, Li:2024tld}. We plot the results in Fig.~\ref{plot-waveform-compare}. It shows that the gravitational waveforms for periodic orbits obtained using the circular orbit approximation effectively capture the characteristics of the waveforms from periodic orbits. However, the gravitational waveforms obtained through the numerical kludge scheme are more realistic and accurate.

\begin{figure}[!t]
	\centering
	\subfigure[]{\includegraphics[scale =0.3]{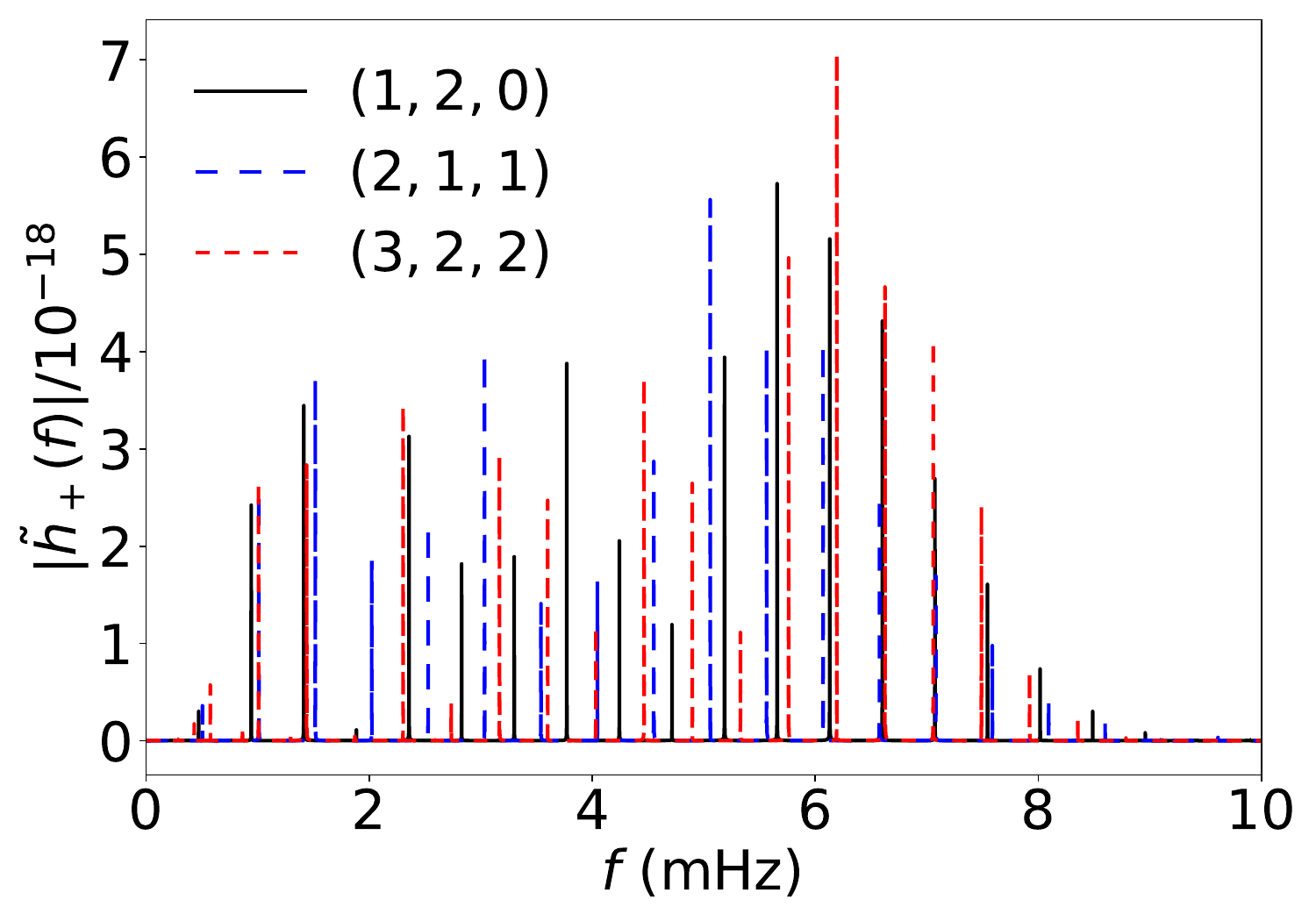}}
	\subfigure[]{\includegraphics[scale =0.3]{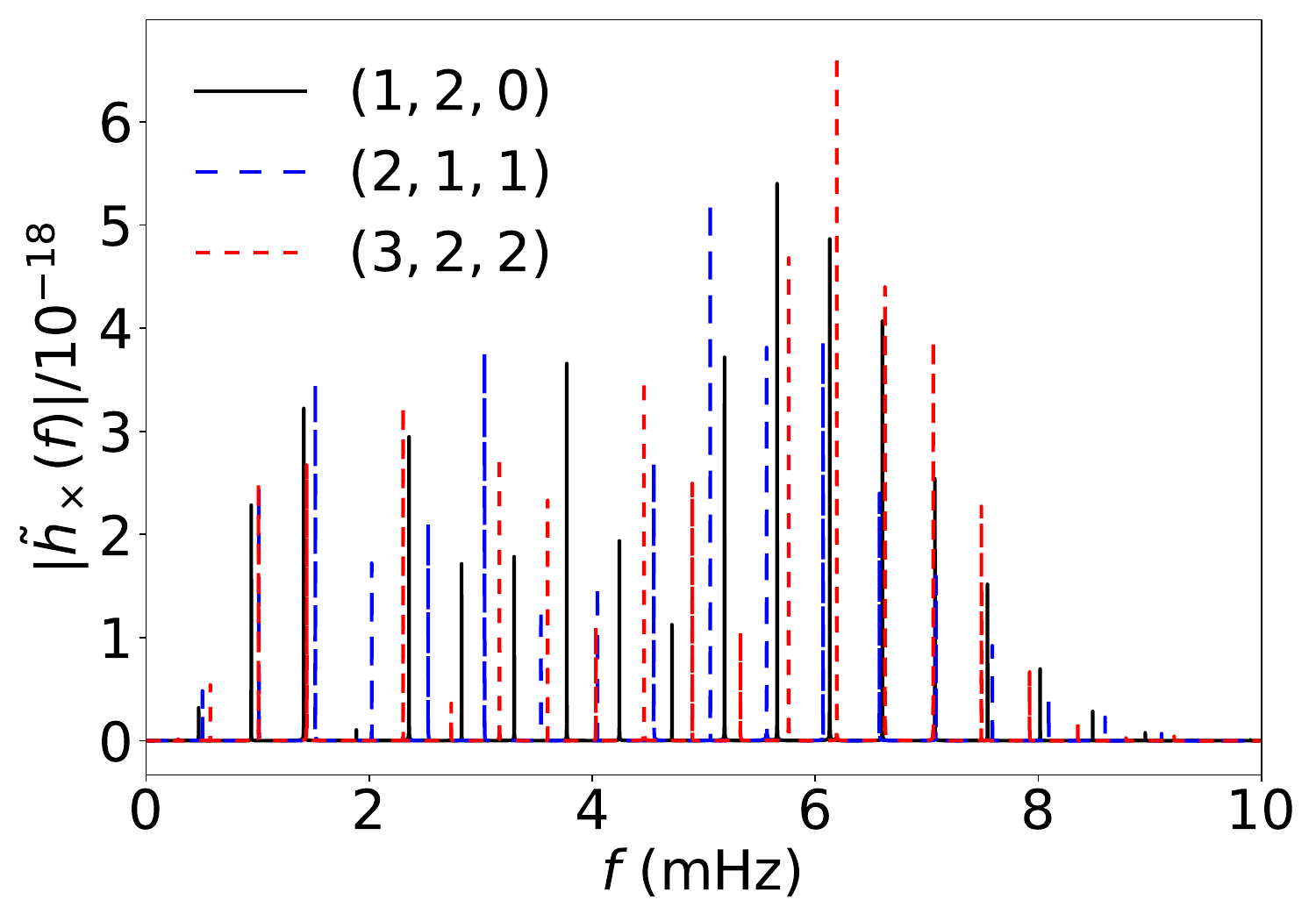}}
	\caption{Absolute values of the Fourier transforms of the gravitational waveforms from a test object with $m = 10 M_\odot$ along different periodic orbits around a supermassive quantum-corrected black hole with $M = 10^6 M_\odot$ and $\hat{\alpha} = 0.8$. The energy is fixed as $E = 0.96$.}
	\label{plot-FFT-a08}
\end{figure}

\begin{figure}[!t]
	\centering
	\subfigure[]{\includegraphics[scale =0.3]{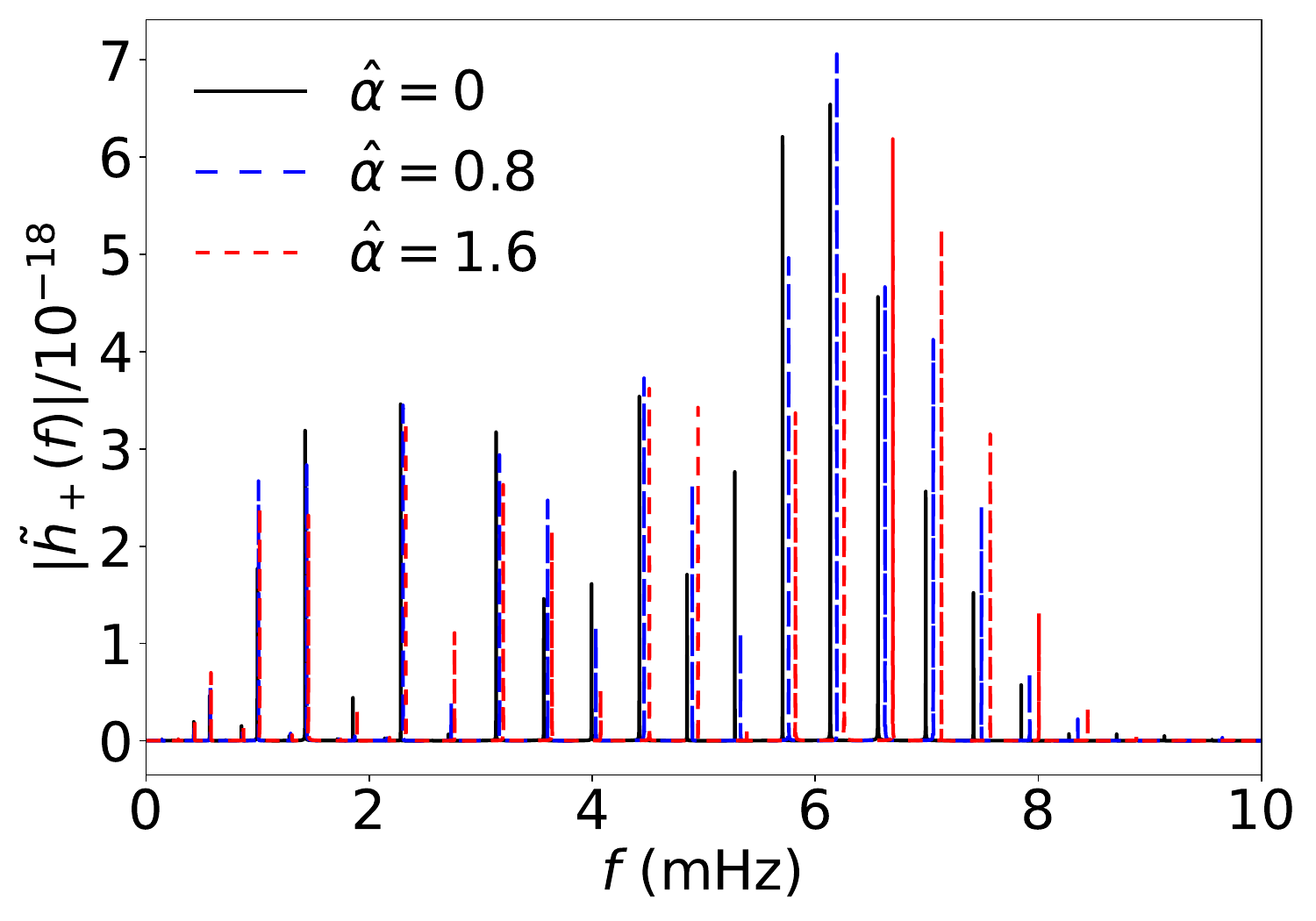}}
	\subfigure[]{\includegraphics[scale =0.3]{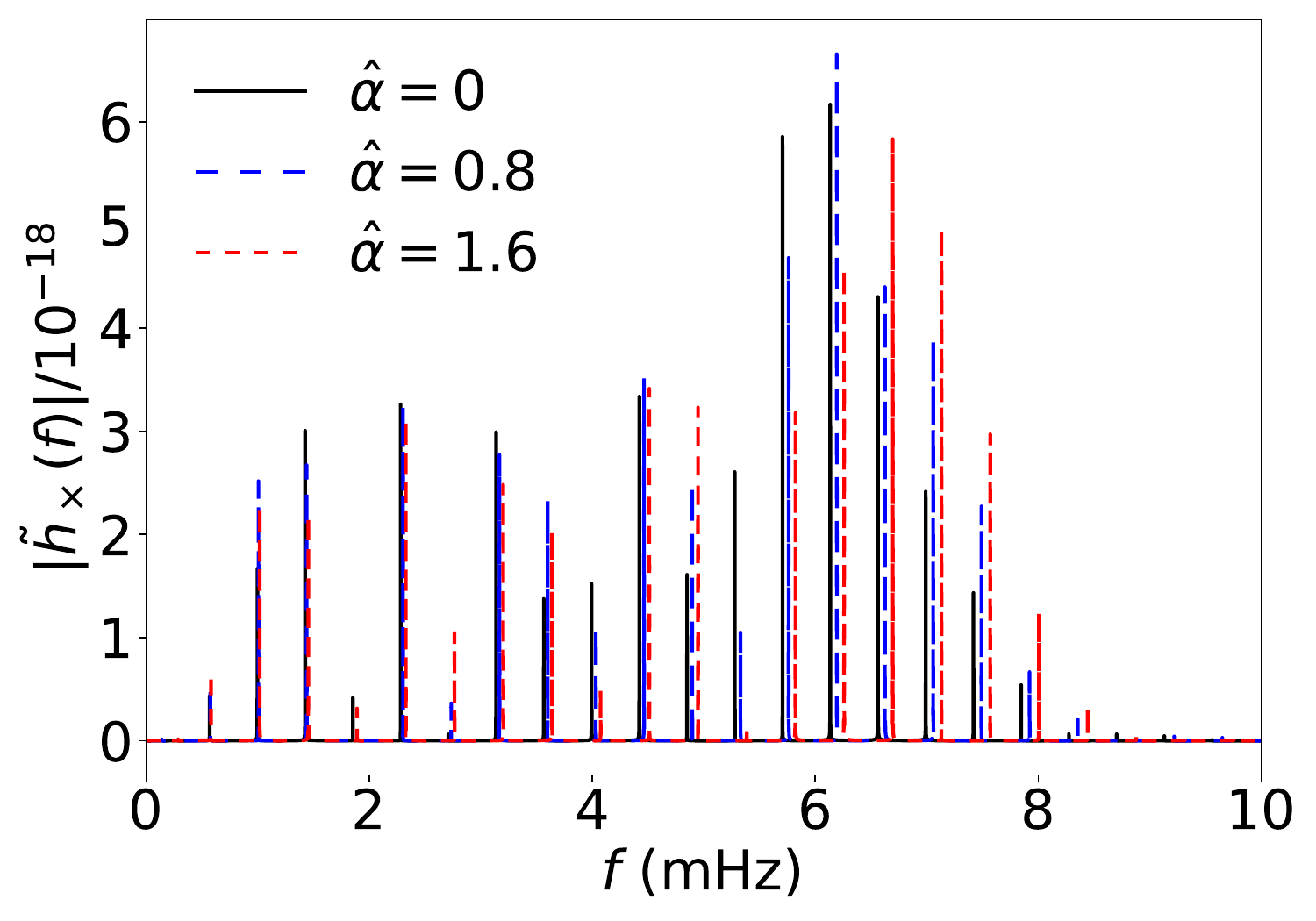}}
	\caption{Absolute values of the Fourier transforms of the gravitational waveforms from a test object with $m = 10 M_\odot$ along the $(3,~2,~2)$ periodic orbits around a supermassive quantum-corrected black hole with $M = 10^6 M_\odot$ and different value of the parameter $\hat{\alpha}$. The energy is fixed as $E =0.96$.}
	\label{plot-FFT-322}
\end{figure}

\begin{figure}[!t]
	\centering
	\subfigure[]{\includegraphics[scale =0.3]{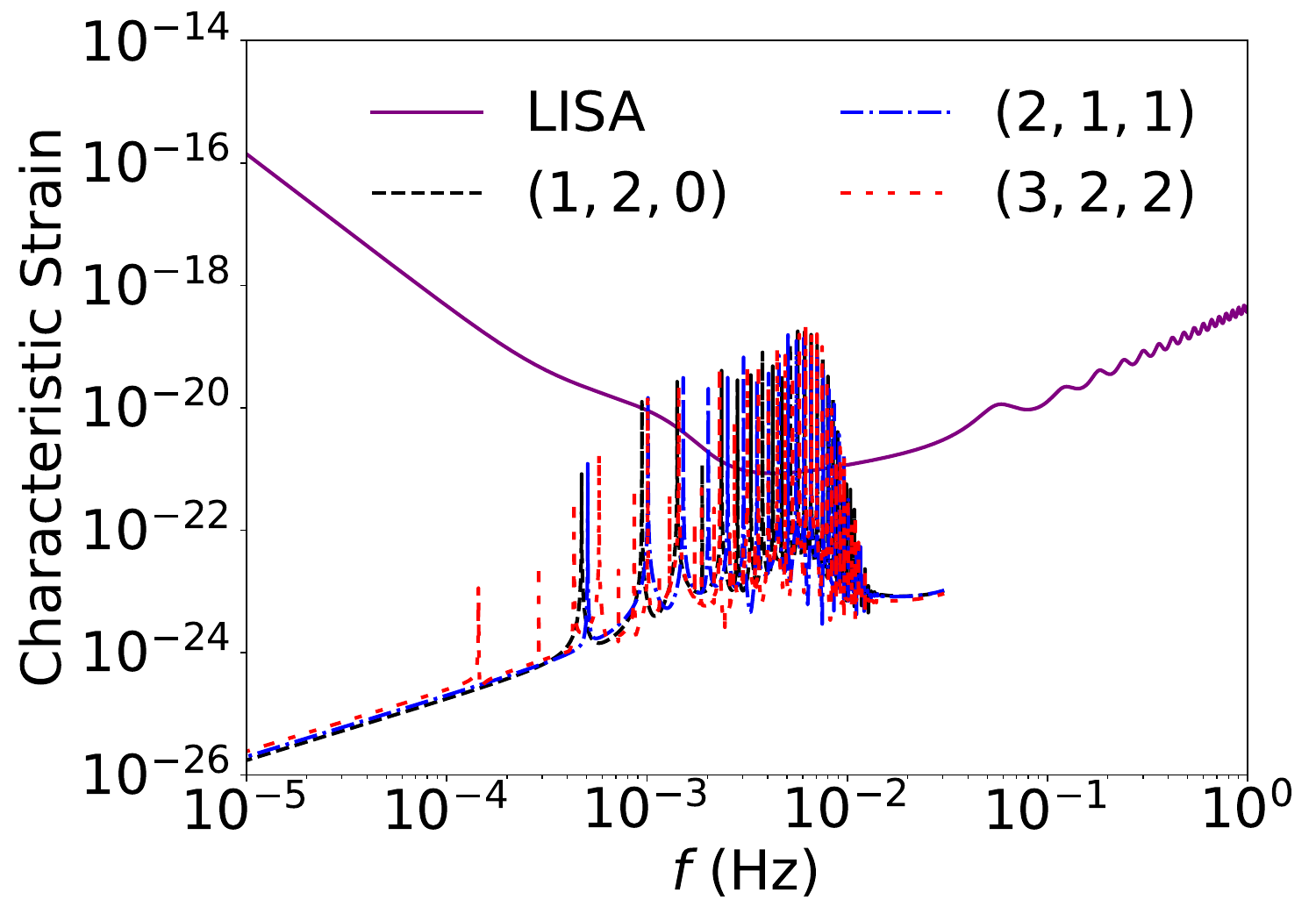}}
	\subfigure[]{\includegraphics[scale =0.3]{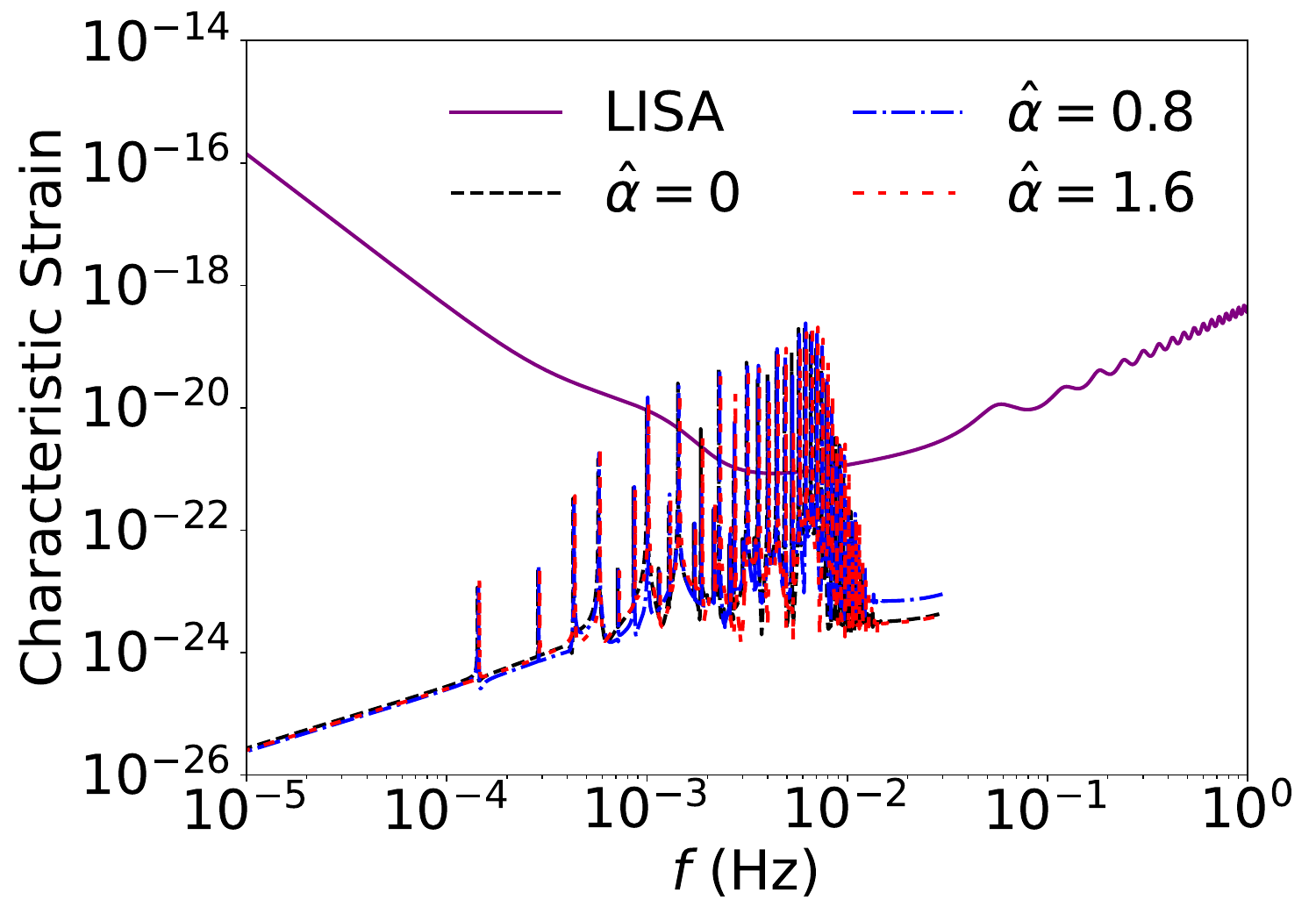}}
	\caption{Comparison of the characteristic strains of gravitational waveforms of periodic orbits in Figs.~\ref{plot-waveform-a08} and \ref{plot-waveform-322} with the sensitivity curve of LISA.}
	\label{plot-strain}
\end{figure}

We obtain the corresponding frequency spectra by performing discrete Fourier transforms on the time-domain gravitational waveforms in Figs. \ref{plot-waveform-a08} and \ref{plot-waveform-322}. We plot the absolute values of the frequency spectra $\tilde{h}_{+,\times}(f)$ in Figs. \ref{plot-FFT-a08} and \ref{plot-FFT-322}. One can find that the characteristic frequencies of gravitational waves emitted from EMRIs with periodic orbits typically lie within the mHz range, precisely within the detection capabilities of space-based gravitational wave detectors. Figure \ref{plot-FFT-a08} shows that the characteristic frequencies of the periodic orbits with different $(z,~w,~v)$, and Fig. \ref{plot-FFT-322} shows that the characteristic frequency of the $(3,~2,~2)$ periodic orbits with different values of $\hat{\alpha}$. To assess the detectability of gravitational waves of periodic orbits, we calculate the corresponding characteristic strain by
\bqn
h_\tx{c}(f) = 2 f \left(|\tilde{h}_{+}(f)|^2 + |\tilde{h}_{\times}(f)|^2 \right)^{1/2},
\eqn
and compare it with the sensitivity curve of LISA \cite{Robson:2018ifk}. We plot the results in Fig.~\ref{plot-strain}. It is shown that parts of the characteristic strains, with different $(w,~v,~z)$ or with different values of the parameter $\hat{\alpha}$, are above the sensitivity curve of LISA, which means the gravitational waves emitted from EMRIs with periodic orbits and quantum correction are potentially detectable by LISA.

\section{Conclusions and Discussions}\lb{sec5}

In this work, we investigated the geodesic motion of a massive test particle around a quantum-corrected black hole. To determine the radial motion of the test particle, we derived the effective potential of motion. We found that the height of the barrier of the effective potential increases with the dimensionless quantum-corrected parameter $\hat{\alpha}$. With different conditions, we numerically obtained the relations between the parameter $\hat{\alpha}$ and the properties of the MBO and ISCO, respectively. We found that both the radius and the orbital angular momentum of the MBO decrease with the parameter $\hat{\alpha}$. All of the radius, the orbital angular momentum, and the energy of the ISCO decrease with the parameter $\hat{\alpha}$. Subsequently, we explored the allowed parameter space of the orbital angular momentum and the energy for the bound orbits around the quantum-corrected black hole. We observed that the bound orbit, characterized by a constant orbital angular momentum and a large parameter $\hat{\alpha}$, possesses an elevated upper energy limit.

Next, we studied the influences of the parameter $\hat{\alpha}$ on the periodic orbits of a particle around the quantum-corrected black hole. We numerically studied the relation between the rational number $q$ and the energy or the orbital angular momentum of periodic orbits. We found that the energy with fixed rational number $q$ decreases with the parameter $\hat{\alpha}$ and the orbital angular momentum with fixed rational number $q$ decreases with the parameter $\hat{\alpha}$. Both the maximum value of the energy and the minimal value of the orbital angular momentum of the periodic orbits decrease with the parameter $\hat{\alpha}$. For different ($z$, $w$, $v$) periodic orbits, we numerically calculated the energy and the orbital angular momentum, respectively. Our findings indicate that both the energy and the orbital angular momentum of the periodic orbits with fixed ($z$, $w$, $v$) decrease with the parameter $\hat{\alpha}$. 

Finally, we considered an EMRI system composed of a test object with mass $m = 10 M_\odot$ along the periodic orbits around a supermassive quantum-corrected black hole with mass $M = 10^6 M_\odot$. As the first application of the numerical kludge scheme to explore the gravitational waveforms from periodic orbits, we studied the corresponding gravitational waveforms by setting the luminosity distance from the EMRI system to the detector $D_\tx{L} = 2 \;\tx{Gpc}$, the inclination angle $\iota = \pi/4$, and the longitude of pericenter $\zeta = \pi/4$. We found that the gravitational waveforms have zoom and whirl phases in one complete period, and the phase of waveforms advances as the parameter $\hat{\alpha}$ increases. This phase advancement accumulates continuously with the evolution of the orbit and becomes more pronounced over time. To evaluate the detectability of gravitational waves from EMRIs with periodic orbits, we performed discrete Fourier transforms on the time-domain gravitational waveforms and obtained the corresponding frequency spectra. Our findings revealed that the characteristic frequencies of gravitational waves emitted from EMRIs with periodic orbits typically lie within the detection capabilities of space-based
gravitational wave detectors. We calculated the characteristic strains of gravitational waves from the frequency spectra. We found that parts of the characteristic strains, with different $(w,~v,~z)$ or with different values of the parameter $\hat{\alpha}$, exceed the sensitivity curve of LISA. This implies that future space-based gravitational wave detectors may detect gravitational waves emitted from EMRIs with periodic orbits.

It is imperative to emphasize that in this work, we only employed the simplest modeling approach when studying gravitational waves from EMRIs. Resonances in EMRIs play a crucial role in shaping gravitational wave signals, with a significant impact on their phenomenolog~\cite{Apostolatos:2009vu, Lukes-Gerakopoulos:2010ipp, Contopoulos:2011dz, Lukes-Gerakopoulos:2014dpa,  Zelenka:2019nyp, Destounis:2020kss, Lukes-Gerakopoulos:2021ybx, Destounis:2021mqv, Destounis:2023gpw, Destounis:2023cim, Destounis:2021rko,Mukherjee:2022dju,Destounis:2023khj,Strateny:2023edo,Eleni:2024fgs}. These resonances merit further investigation to better understand their effects. For long periods of orbits, we must consider the impact of gravitational radiation on the orbital parameters of small celestial bodies. In realistic astrophysical situations, the supermassive black holes in the center of galaxies are rotating. We should investigate the motion of the test particle around the rotating black holes with quantum correction. On the other hand, the supermassive black holes are believed to be enveloped by dark matter~\cite{Cardoso:2021wlq}. So one should consider the environment's influence on the small object's orbit and propagation of gravitational waves~\cite{Li:2021pxf, Rahman:2023sof, Cardoso:2022whc, Destounis:2022obl}. Another interesting question is whether gravitational waves from EMRIs can be used to determine whether the supermassive object at the center of a galaxy is a black hole \cite{Zhang:2021ojz}. Recently, Ref. \cite{Liu:2024qci} studied gravitational waves for eccentric EMRIs in self-dual spacetime. It is highly worthwhile to explore using gravitational waves from EMRIs to constrain the parameters of modified theories of gravity. We will address these issues in our future work.

\section*{Acknowledgements}

This work was supported by the National Key Research and Development Program of China (Grant No. 2021YFC2203003), the National Natural Science Foundation of China (Grants No.~12475056, No.~12105126, No. 12275238, and No. 12247101), the 111 Project (Grant No. B20063), Gansu Province's Top Leading Talent Support Plan, the Zhejiang Provincial Natural Science Foundation of China under Grant No. LR21A050001 and No. LY20A050002, and the Fundamental Research Funds for the Provincial Universities of Zhejiang in China under Grant No. RF-A2019015.


\begin{thebibliography}{399}

\bibitem{LIGOScientific:2016aoc}
B.~P.~Abbott \textit{et al.} [LIGO Scientific and Virgo], Observation of Gravitational Waves from a Binary Black Hole Merger, \href{\doibase 10.1103/PhysRevLett.116.061102}{Phys. Rev. Lett. \textbf{116}, 6, 061102 (2016)}, arXiv:1602.03837 [gr-qc].

\bibitem{LIGOScientific:2016vlm}
B.~P.~Abbott \textit{et al.} [LIGO Scientific and Virgo], Properties of the Binary Black Hole Merger GW150914,
\href{\doibase 10.1103/PhysRevLett.116.241102}{Phys. Rev. Lett. \textbf{116}, 24, 241102 (2016)},
arXiv:1602.03840 [gr-qc].

\bibitem{LIGOScientific:2018mvr}
B.~P.~Abbott \textit{et al.} [LIGO Scientific and Virgo], GWTC-1: A Gravitational-Wave Transient Catalog of Compact Binary Mergers Observed by LIGO and Virgo during the First and Second Observing Runs, \href{\doibase 10.1103/PhysRevX.9.031040}{Phys. Rev. X \textbf{9}, 3, 031040 (2019)}, arXiv:1811.12907 [astro-ph.HE].

\bibitem{LIGOScientific:2020ibl}
R.~Abbott \textit{et al.} [LIGO Scientific and Virgo], GWTC-2: Compact Binary Coalescences Observed by LIGO and Virgo During the First Half of the Third Observing Run, \href{\doibase 10.1103/PhysRevX.11.021053}{Phys. Rev. X \textbf{11}, 021053 (2021)}, arXiv:2010.14527 [gr-qc].

\bibitem{LIGOScientific:2021usb}
R.~Abbott \textit{et al.} [LIGO Scientific and VIRGO], GWTC-2.1: Deep extended catalog of compact binary coalescences observed by LIGO and Virgo during the first half of the third observing run, \href{\doibase 10.1103/PhysRevD.109.022001}{
Phys. Rev. D \textbf{109}, 2, 022001 (2024)}, arXiv:2108.01045 [gr-qc].

\bibitem{KAGRA:2021vkt}
R.~Abbott \textit{et al.} [KAGRA, VIRGO and LIGO Scientific], GWTC-3: Compact Binary Coalescences Observed by LIGO and Virgo during the Second Part of the Third Observing Run, \href{https://doi.org/10.1103/PhysRevX.13.04103}{Phys. Rev. X \textbf{13}, 4, 041039 (2023)},
arXiv:2111.03606 [gr-qc].

\bibitem{Hughes:2000ssa}
S.~A.~Hughes, Gravitational waves from extreme mass ratio inspirals: Challenges in mapping the space-time of massive, compact objects, \href{\doibase 10.1088/0264-9381/18/19/314}{Class. Quant. Grav. \textbf{18}, 4067-4074 (2001)},
arXiv:gr-qc/0008058.

\bibitem{Glampedakis:2005hs}
K.~Glampedakis, Extreme mass ratio inspirals: LISA's unique probe of black hole gravity, \href{\doibase 10.1088/0264-9381/22/15/004}{Class. Quant. Grav. \textbf{22}, S605-S659 (2005)}, arXiv:gr-qc/0509024.

\bibitem{LISA:2017pwj}
P.~Amaro-Seoane \textit{et al.} [LISA],
Laser Interferometer Space Antenna, \href{https://arxiv.org/abs/1702.00786}{arXiv:1702.00786 [astro-ph.IM]}.

\bibitem{Hu:2017mde}
W.~R.~Hu and Y.~L.~Wu,
The Taiji Program in Space for gravitational wave physics and the nature of gravity,
\href{\doibase 10.1093/nsr/nwx116}{Natl. Sci. Rev. \textbf{4}, 5, 685-686 (2017)}.

\bibitem{TianQin:2015yph}
J.~Luo \textit{et al.} [TianQin], TianQin: a space-borne gravitational wave detector, \href{\doibase 10.1088/0264-9381/33/3/035010}{Class. Quant. Grav. \textbf{33}, 3, 035010 (2016)},
arXiv:1512.02076 [astro-ph.IM].

\bibitem{Musha:2017usi}
M.~Musha [DECIGO Working group], Space gravitational wave detector DECIGO/pre-DECIGO, \href{\doibase 10.1117/12.2296050}{Proc. SPIE Int. Soc. Opt. Eng. \textbf{10562}, 105623T (2017)}.

\bibitem{Barausse:2020rsu}
E.~Barausse, E.~Berti, T.~Hertog, S.~A.~Hughes, P.~Jetzer, P.~Pani, T.~P.~Sotiriou, N.~Tamanini, H.~Witek, and K.~Yagi, \textit{et al.}, Prospects for Fundamental Physics with LISA, \href{\doibase 10.1007/s10714-020-02691-1}{Gen. Rel. Grav. \textbf{52}, 8, 81 (2020)},
arXiv:2001.09793 [gr-qc].

\bibitem{LISA:2022yao}
P.~A.~Seoane \textit{et al.} [LISA],
Astrophysics with the Laser Interferometer Space Antenna,
\href{\doibase 10.1007/s41114-022-00041-y}{Living Rev. Rel. \textbf{26}, 1, 2 (2023)},
arXiv:2203.06016 [gr-qc].

\bibitem{Wang:2021srv}
L.~F.~Wang, S.~J.~Jin, J.~F.~Zhang, and X.~Zhang,
Forecast for cosmological parameter estimation with gravitational-wave standard sirens from the LISA-Taiji network, \href{\doibase 10.1007/s11433-021-1736-6}{
Sci. China Phys. Mech. Astron. \textbf{65}, 1, 210411 (2022)}, arXiv:2101.11882 [gr-qc].

\bibitem{Gao:2022hho}
Q.~Gao, Constraint on the mass of graviton with gravitational waves, \href{\doibase 10.1007/s11433-022-1971-9}{Sci. China Phys. Mech. Astron. \textbf{66}, 2, 220411 (2023)}, arXiv:2206.02140 [gr-qc].

\bibitem{Jin:2023sfc}
S.~J.~Jin, Y.~Z.~Zhang, J.~Y.~Song, J.~F.~Zhang, and X.~Zhang, Taiji-TianQin-LISA network: Precisely measuring the Hubble constant using both bright and dark sirens, \href{doi:10.1007/s11433-023-2276-1}{Sci. China Phys. Mech. Astron. \textbf{67}, 2, 220412 (2024)}, arXiv:2305.19714 [astro-ph.CO].

\bibitem{Zhong:2023pjz}
X.~Zhong, W.~Han, Z.~Luo, and Y.~Wu, Exploring the nature of black hole and gravity with an imminent merging binary of supermassive black holes,
\href{\doibase 10.1007/s11433-022-2028-7}{Sci. China Phys. Mech. Astron. \textbf{66}, 3, 230411 (2023)}, arXiv:2305.04478 [gr-qc].

\bibitem{GWSTQLISA}
Torres-Orjuela Alejandro, Huang Shun-Jia, Liang Zheng-Cheng, Liu Shuai, Wang Hai-Tian, Ye Chang-Qing, Hu Yi-Ming, and Mei Jianwei, Detection of astrophysical gravitational wave sources by TianQin and LISA, \href{https://doi.org/10.1007/s11433-023-2308-x}{Sci. China Phys. Mech. Astron. \textbf{67}, 5, 259511 (2024)}.

\bibitem{dirty}
Ye Jiang and Wen-Biao Han, General formalism for dirty extreme-mass-ratio inspirals, \href{https://doi.org/10.1007/s11433-024-2366-5}{Sci. China Phys. Mech. Astron. \textbf{67}, 7, 270411  (2024)}.

\bibitem{Ghosh:2024arw}
R.~Ghosh and K.~Chakravarti,
Parameterized Non-circular Deviation from the Kerr Paradigm and Its Observational Signatures: Extreme Mass Ratio Inspirals and Lense-Thirring Effect,
\href{https://arxiv.org/abs/2406.02454}{arXiv:2406.02454 [gr-qc]}.

\bibitem{Babak:2017tow}
S.~Babak, J.~Gair, A.~Sesana, E.~Barausse, C.~F.~Sopuerta, C.~P.~L.~Berry, E.~Berti, P.~Amaro-Seoane, A.~Petiteau and A.~Klein, Science with the space-based interferometer LISA. V: Extreme mass-ratio inspirals,
\href{\doibase 10.1103/PhysRevD.95.103012}{Phys. Rev. D \textbf{95}, 10, 103012 (2017)}, arXiv:1703.09722 [gr-qc].

\bibitem{Levin:2008mq}
J.~Levin and G.~Perez-Giz, A Periodic Table for Black Hole Orbits, \href{\doibase 10.1103/PhysRevD.77.103005}{
Phys. Rev. D \textbf{77}, 103005 (2008)}, arXiv:0802.0459 [gr-qc].

\bibitem{Levin:2008ci}
J.~Levin and B.~Grossman, Dynamics of Black Hole Pairs. I. Periodic Tables,
\href{\doibase 10.1103/PhysRevD.79.043016}{Phys. Rev. D \textbf{79}, 043016 (2009)}, arXiv:0809.3838 [gr-qc].

\bibitem{Grossman:2008yk}
R.~Grossman and J.~Levin, Dynamics of Black Hole Pairs II: Spherical Orbits and the Homoclinic Limit of Zoom-Whirliness,
\href{\doibase 10.1103/PhysRevD.79.043017}{Phys. Rev. D \textbf{79}, 043017 (2009)}, arXiv:0811.3798 [gr-qc].

\bibitem{Flanagan:2010cd}
E.~E.~Flanagan and T.~Hinderer,
Transient resonances in the inspirals of point particles into black holes, \href{https://journals.aps.org/prl/abstract/10.1103/PhysRevLett.109.071102}{
Phys. Rev. Lett. \textbf{109}, 071102 (2012)},
arXiv:1009.4923 [gr-qc].

\bibitem{Berry:2016bit}
C.~P.~L.~Berry, R.~H.~Cole, P.~Ca\~nizares, and J.~R.~Gair, Importance of transient resonances in extreme-mass-ratio inspirals,
\href{https://journals.aps.org/prd/abstract/10.1103/PhysRevD.94.124042}{Phys. Rev. D \textbf{94}, 12, 124042 (2016)},
arXiv:1608.08951 [gr-qc].

\bibitem{Speri:2021psr}
L.~Speri and J.~R.~Gair, Assessing the impact of transient orbital resonances,
\href{https://journals.aps.org/prd/abstract/10.1103/PhysRevD.103.124032}{Phys. Rev. D \textbf{103}, 12, 124032 (2021)},
arXiv:2103.06306 [gr-qc].

\bibitem{Misra:2010pu}
V.~Misra and J.~Levin, Rational Orbits around Charged Black Holes,
\href{\doibase 10.1103/PhysRevD.82.083001}{Phys. Rev. D \textbf{82}, 083001 (2010)},
arXiv:1007.2699 [gr-qc].

\bibitem{Babar:2017gsg}
G.~Z.~Babar, A.~Z.~Babar, and Y.~K.~Lim,
Periodic orbits around a spherically symmetric naked singularity,
\href{\doibase 10.1103/PhysRevD.96.084052}{Phys. Rev. D \textbf{96}, 8, 084052 (2017)},
arXiv:1710.09581 [gr-qc].

\bibitem{Liu:2018vea}
C.~Liu, C.~Ding, and J.~Jing,
Periodic orbits around Kerr Sen black holes, \href{\doibase 10.1088/0253-6102/71/12/1461}{Commun. Theor. Phys. \textbf{71}, 12, 1461 (2019)},
arXiv:1804.05883 [gr-qc].

\bibitem{Lin:2021noq}
H.~Y.~Lin and X.~M.~Deng, Rational orbits around 4 $\mathcal D$ Einstein\textendash{}Lovelock black holes, \href{\doibase 10.1016/j.dark.2020.100745}{Phys. Dark Univ. \textbf{31}, 100745 (2021)}.

\bibitem{Wei:2019zdf}
S.~W.~Wei, J.~Yang, and Y.~X.~Liu,
Geodesics and periodic orbits in Kehagias-Sfetsos black holes in deformed Ho$\check{\text{r}}$ava-Lifshitz gravity,
\href{\doibase 10.1103/PhysRevD.99.104016}{Phys. Rev. D \textbf{99}, 10, 104016 (2019)},
arXiv:1904.03129 [gr-qc].

\bibitem{Deng:2020yfm}
X.~M.~Deng, Geodesics and periodic orbits around quantum-corrected black holes, \href{\doibase 10.1016/j.dark.2020.100629}{Phys. Dark Univ. \textbf{30}, 100629 (2020)}.

\bibitem{Azreg-Ainou:2020bfl}
M.~Azreg-A\"\i{}nou, Z.~Chen, B.~Deng, M.~Jamil, T.~Zhu, Q.~Wu, and Y.~K.~Lim,
Orbital mechanics and quasiperiodic oscillation resonances of black holes in Einstein-\AE{}ther theory,
\href{\doibase 10.1103/PhysRevD.102.044028}{Phys. Rev. D \textbf{102}, 4, 044028 (2020)},
arXiv:2004.02602 [gr-qc].

\bibitem{Wang:2022tfo}
R.~Wang, F.~Gao, and H.~Chen,
Periodic orbits around a static spherically symmetric black hole surrounded by quintessence,
\href{\doibase 10.1016/j.aop.2022.169167}{Annals Phys. \textbf{447}, 1, 169167 (2022)}.

\bibitem{Lin:2023rmo}
H.~Y.~Lin and X.~M.~Deng,
Precessing and periodic orbits around hairy black holes in Horndeski\textquoteright{}s Theory, \href{\doibase 10.1140/epjc/s10052-023-11487-x}{Eur. Phys. J. C \textbf{83}, 4, 311 (2023)}.

\bibitem{Zhang:2022psr}
J.~Zhang and Y.~Xie,
Probing a self-complete and Generalized-Uncertainty-Principle black hole with precessing and periodic motion,
\href{\doibase 10.1007/s10509-022-04046-5}{Astrophys. Space Sci. \textbf{367}, 2, 17 (2022)}.

\bibitem{Gao:2021arw}
B.~Gao and X.~M.~Deng,
Bound orbits around modified Hayward black holes,
\href{\doibase 10.1142/S0217732321502370}{Mod. Phys. Lett. A \textbf{36}, 33, 2150237 (2021)}.

\bibitem{Zhou:2020zys}
T.~Y.~Zhou and Y.~Xie, Precessing and periodic motions around a black-bounce/traversable wormhole,
\href{\doibase 10.1140/epjc/s10052-020-08661-w}{Eur. Phys. J. C \textbf{80}, 11, 1070 (2020)}.

\bibitem{Deng:2020hxw}
X.~M.~Deng, Periodic orbits around brane-world black holes,
\href{\doibase 10.1140/epjc/s10052-020-8067-7}{Eur. Phys. J. C \textbf{80}, 6, 489 (2020)}.

\bibitem{Tu:2023xab}
Z.~Y.~Tu, T.~Zhu, and A.~Wang,
Periodic orbits and their gravitational wave radiations in a polymer black hole in loop quantum gravity,
\href{\doibase 10.1103/PhysRevD.108.024035}{Phys. Rev. D \textbf{108}, 2, 2 (2023)},
arXiv:2304.14160 [gr-qc].

\bibitem{Li:2024tld}
Y.~Z.~Li and X.~M.~Kuang,
Precessing and periodic timelike orbits and their potential applications in Einsteinian cubic gravity,
\href{https://arxiv.org/abs/2401.16071}{arXiv:2401.16071 [gr-qc]}.

\bibitem{Rovelli:1997yv}
C.~Rovelli, Loop quantum gravity,
\href{\doibase 10.12942/lrr-1998-1}{Living Rev. Rel. \textbf{1}, 1 (1998)}.

\bibitem{Lewandowski:2022zce}
J.~Lewandowski, Y.~Ma, J.~Yang, and C.~Zhang, Quantum Oppenheimer-Snyder and Swiss Cheese Models, \href{\doibase 10.1103/PhysRevLett.130.101501}{Phys. Rev. Lett. \textbf{130}, 10, 101501 (2023)}, arXiv:2210.02253 [gr-qc].

\bibitem{Ye:2023qks}
J.~P.~Ye, Z.~Q.~He, A.~X.~Zhou, Z.~Y.~Huang, and J.~H.~Huang,
Shadows and photon rings of a quantum black hole, \href{\doibase 10.1016/j.physletb.2024.138566}{Phys. Lett. B \textbf{851}, 138566 (2024)},
arXiv:2312.17724 [gr-qc].

\bibitem{Yang:2022btw}
J.~Yang, C.~Zhang, and Y.~Ma,
Shadow and stability of quantum-corrected black holes,
\href{\doibase 10.1140/epjc/s10052-023-11800-8}{Eur. Phys. J. C \textbf{83}, 7, 619 (2023)},
arXiv:2211.04263 [gr-qc].

\bibitem{Zhang:2023okw}
C.~Zhang, Y.~Ma, and J.~Yang, Black hole image encoding quantum gravity information,
\href{\doibase 10.1103/PhysRevD.108.104004}{Phys. Rev. D \textbf{108}, 10, 104004 (2023)},
arXiv:2302.02800 [gr-qc].

\bibitem{Gong:2023ghh}
H.~Gong, S.~Li, D.~Zhang, G.~Fu, and J.~P.~Wu,
Quasinormal modes of quantum-corrected black holes,
\href{\doibase 10.1103/PhysRevD.110.044040}{Phys. Rev. D \textbf{110}, 4, 044040 (2024)},
arXiv:2312.17639 [gr-qc].

\bibitem{Cao:2024oud}
L.~M.~Cao, J.~N.~Chen, L.~B.~Wu, L.~Xie, and Y.~S.~Zhou,
The pseudospectrum and spectrum (in)stability of quantum corrected Schwarzschild black hole,
\href{\doibase 10.1007/s11433-024-2435-5}{Sci. China Phys. Mech. Astron. \textbf{67}, 10, 100412 (2024)},
arXiv:2401.09907 [gr-qc].

\bibitem{Shao:2023qlt}
C.~Y.~Shao, C.~Zhang, W.~Zhang, and C.~G.~Shao, Scalar fields around a loop quantum gravity black hole in de Sitter spacetime: Quasinormal modes, late-time tails and strong cosmic censorship,
\href{\doibase 10.1103/PhysRevD.109.064012}{Phys. Rev. D \textbf{109}, 6, 064012 (2024)},
arXiv:2309.04962 [gr-qc].

\bibitem{Zhao:2024elr}
L.~Zhao, M.~Tang, and Z.~Xu,
The lensing effect of quantum-corrected black hole and parameter constraints from EHT observations, \href{\doibase 10.1140/epjc/s10052-024-13342-z}{Eur. Phys. J. C \textbf{84}, 9, 971 (2024)},
arXiv:2403.18606 [gr-qc].

\bibitem{You:2024jeu}
L.~You, Y.~H.~Feng, R.~B.~Wang, X.~R.~Hu, and J.~B.~Deng,
Decoding Quantum Gravity Information with Black Hole Accretion Disk,
\href{\doibase 10.3390/universe10100393}{Universe \textbf{10}, 10, 393 (2024)},
arXiv:2404.01418 [gr-qc].

\bibitem{Chandrasekhar:1985kt}
S.~Chandrasekhar, The mathematical theory of black holes, Oxford University Press, 1985.

\bibitem{Misner:1973prb}
C.~W.~Misner, K.~S.~Thorne and J.~A.~Wheeler, Gravitation, W. H. Freeman, 1973, ISBN 978-0-7167-0344-0, 978-0-691-17779-3.

\bibitem{Babak:2006uv}
S.~Babak, H.~Fang, J.~R.~Gair, K.~Glampedakis, and S.~A.~Hughes, `Kludge' gravitational waveforms for a test-body orbiting a Kerr black hole,
\href{https://journals.aps.org/prd/abstract/10.1103/PhysRevD.75.024005}{Phys. Rev. D \textbf{75}, 024005 (2007)},
arXiv:gr-qc/0607007.

\bibitem{Robson:2018ifk}
T.~Robson, N.~J.~Cornish, and C.~Liu, The construction and use of LISA sensitivity curves,
\href{\doibase 10.1088/1361-6382/ab1101}{Class. Quant. Grav. \textbf{36}, 10, 105011 (2019)}, arXiv:1803.01944 [astro-ph.HE].

\bibitem{Hughes:1999bq}
S.~A.~Hughes, The Evolution of circular, nonequatorial orbits of Kerr black holes due to gravitational wave emission, \href{https://journals.aps.org/prd/abstract/10.1103/PhysRevD.61.084004}{Phys. Rev. D \textbf{61}, 8, 084004 (2000)}, arXiv:gr-qc/9910091.

\bibitem{Hughes:2001jr}
S.~A.~Hughes, Evolution of circular, nonequatorial orbits of Kerr black holes due to gravitational wave emission. II. Inspiral trajectories and gravitational wave forms,
\href{https://journals.aps.org/prd/abstract/10.1103/PhysRevD.64.064004}{Phys. Rev. D \textbf{64}, 064004 (2001)}, arXiv:gr-qc/0104041.

\bibitem{Glampedakis:2002ya}
K.~Glampedakis and D.~Kennefick, Zoom and whirl: Eccentric equatorial orbits around spinning black holes and their evolution under gravitational radiation reaction,
\href{https://journals.aps.org/prd/abstract/10.1103/PhysRevD.66.044002}{Phys. Rev. D \textbf{66}, 044002 (2002)}, arXiv:gr-qc/0203086.
	
\bibitem{Hughes:2005qb}
S.~A.~Hughes, S.~Drasco, E.~E.~Flanagan, and J.~Franklin, Gravitational radiation reaction and inspiral waveforms in the adiabatic limit,
\href{https://journals.aps.org/prl/abstract/10.1103/PhysRevLett.94.221101}{Phys. Rev. Lett. \textbf{94}, 221101 (2005)}, arXiv:gr-qc/0504015.
	
\bibitem{Drasco:2005is}
S.~Drasco, E.~E.~Flanagan, and S.~A.~Hughes, Computing inspirals in Kerr in the adiabatic regime. I. The Scalar case, \href{https://iopscience.iop.org/article/10.1088/0264-9381/22/15/011}{Class. Quant. Grav. \textbf{22}, S801-846 (2005)}, arXiv:gr-qc/0505075.

\bibitem{Gair:2005ih}
J.~R.~Gair and K.~Glampedakis,
Improved approximate inspirals of test-bodies into Kerr black holes, \href{https://journals.aps.org/prd/abstract/10.1103/PhysRevD.73.064037}{Phys. Rev. D \textbf{73}, 064037 (2006)}, arXiv:gr-qc/0510129.

\bibitem{Glampedakis:2005cf}
K.~Glampedakis and S.~Babak,
Mapping spacetimes with LISA: Inspiral of a test-body in a `quasi-Kerr' field,
\href{https://iopscience.iop.org/article/10.1088/0264-9381/23/12/013}{Class. Quant. Grav. \textbf{23}, 4167-4188 (2006)}, arXiv:gr-qc/0510057.

\bibitem{Drasco:2005kz}
S.~Drasco and S.~A.~Hughes, Gravitational wave snapshots of generic extreme mass ratio inspirals,
\href{https://journals.aps.org/prd/abstract/10.1103/PhysRevD.73.024027}{Phys. Rev. D \textbf{73}, 2, 024027 (2006)}, arXiv:gr-qc/0509101.

\bibitem{Sundararajan:2007jg}
P.~A.~Sundararajan, G.~Khanna, and S.~A.~Hughes, Towards adiabatic waveforms for inspiral into Kerr black holes. I. A New model of the source for the time domain perturbation equation,
\href{https://journals.aps.org/prd/abstract/10.1103/PhysRevD.76.104005}{Phys. Rev. D \textbf{76}, 104005 (2007)}, arXiv:gr-qc/0703028.

\bibitem{Sundararajan:2008zm}
P.~A.~Sundararajan, G.~Khanna, S.~A.~Hughes, and S.~Drasco, Towards adiabatic waveforms for inspiral into Kerr black holes: II. Dynamical sources and generic orbits,
\href{https://journals.aps.org/prd/abstract/10.1103/PhysRevD.78.024022}{Phys. Rev. D \textbf{78}, 024022 (2008)}, arXiv:0803.0317 [gr-qc].

\bibitem{Miller:2020bft}
J.~Miller and A.~Pound, Two-timescale evolution of extreme-mass-ratio inspirals: waveform generation scheme for quasicircular orbits in Schwarzschild spacetime,
\href{https://journals.aps.org/prd/abstract/10.1103/PhysRevD.103.064048}{Phys. Rev. D \textbf{103}, 6, 064048 (2021)}, arXiv:2006.11263 [gr-qc].

\bibitem{Isoyama:2021jjd}
S.~Isoyama, R.~Fujita, A.~J.~K.~Chua, H.~Nakano, A.~Pound, and N.~Sago, Adiabatic Waveforms from Extreme-Mass-Ratio Inspirals: An Analytical Approach,
\href{https://journals.aps.org/prl/abstract/10.1103/PhysRevLett.128.231101}{Phys. Rev. Lett. \textbf{128}, 23, 231101 (2022)}, arXiv:2111.05288 [gr-qc].

\bibitem{Thorne:1980ru}
K.~S.~Thorne, Multipole Expansions of Gravitational Radiation,
\href{\doibase 10.1103/RevModPhys.52.299}{Rev. Mod. Phys. \textbf{52}, 299-339 (1980)}.

\bibitem{gravity-book}
E. Poisson and C. M. Will, Gravity: Newtonian, Post-Newtonian, Relativistic (Cambridge University Press,
Cambridge, England, 2014).

\bibitem{Apostolatos:2009vu}
T.~A.~Apostolatos, G.~Lukes-Gerakopoulos, and G.~Contopoulos, How to Observe a Non-Kerr Spacetime Using Gravitational Waves,
\href{https://journals.aps.org/prl/abstract/10.1103/PhysRevLett.103.111101}{Phys. Rev. Lett. \textbf{103}, 111101 (2009)}, arXiv:0906.0093 [gr-qc].

\bibitem{Lukes-Gerakopoulos:2010ipp}
G.~Lukes-Gerakopoulos, T.~A.~Apostolatos, and G.~Contopoulos, Observable signature of a background deviating from the Kerr metric,
\href{https://journals.aps.org/prd/abstract/10.1103/PhysRevD.81.124005}{Phys. Rev. D \textbf{81}, 124005 (2010)}, arXiv:1003.3120 [gr-qc].

\bibitem{Contopoulos:2011dz}
G.~Contopoulos, G.~Lukes-Gerakopoulos, and T.~A.~Apostolatos,
Orbits in a non-Kerr Dynamical System,
\href{https://www.worldscientific.com/doi/abs/10.1142/S0218127411029768}{Int. J. Bifurc. Chaos \textbf{21}, 2261-2277 (2011)}, arXiv:1108.5057 [gr-qc].

\bibitem{Lukes-Gerakopoulos:2014dpa}
G.~Lukes-Gerakopoulos, G.~Contopoulos, and T.~A.~Apostolatos, Non-Linear Effects in Non-Kerr spacetimes,
\href{https://link.springer.com/chapter/10.1007/978-3-319-06761-2_16}{Springer Proc. Phys. \textbf{157}, 129-136 (2014)}, arXiv:1408.4697 [gr-qc].

\bibitem{Zelenka:2019nyp}
O.~Zelenka, G.~Lukes-Gerakopoulos, V.~Witzany, and O.~Kop\'a\v{c}ek,
Growth of resonances and chaos for a spinning test particle in the Schwarzschild background,
\href{https://journals.aps.org/prd/abstract/10.1103/PhysRevD.101.024037}{Phys. Rev. D \textbf{101}, 2, 024037 (2020)},
arXiv:1911.00414 [gr-qc].

\bibitem{Destounis:2020kss}
K.~Destounis, A.~G.~Suvorov, and K.~D.~Kokkotas, Testing spacetime symmetry through gravitational waves from extreme-mass-ratio inspirals,
\href{https://journals.aps.org/prd/abstract/10.1103/PhysRevD.102.064041}{Phys. Rev. D \textbf{102}, 6, 064041 (2020)}, arXiv:2009.00028 [gr-qc].

\bibitem{Lukes-Gerakopoulos:2021ybx}
G.~Lukes-Gerakopoulos and V.~Witzany, Non-linear effects in EMRI dynamics and their imprints on gravitational waves,
\href{https://arxiv.org/abs/2103.06724}{arXiv:2103.06724 [gr-qc]}.

\bibitem{Destounis:2021mqv}
K.~Destounis, A.~G.~Suvorov, and K.~D.~Kokkotas, Gravitational-wave glitches in chaotic extreme-mass-ratio inspirals, \href{https://journals.aps.org/prl/abstract/10.1103/PhysRevLett.126.141102}{Phys. Rev. Lett. \textbf{126}, 14, 141102 (2021)},
arXiv:2103.05643 [gr-qc].

\bibitem{Destounis:2021rko}
K.~Destounis and K.~D.~Kokkotas, Gravitational-wave glitches: Resonant islands and frequency jumps in nonintegrable extreme-mass-ratio inspirals,
\href{https://journals.aps.org/prd/abstract/10.1103/PhysRevD.104.064023}{Phys. Rev. D \textbf{104}, 6, 064023 (2021)},
arXiv:2108.02782 [gr-qc].

\bibitem{Mukherjee:2022dju}
S.~Mukherjee, O.~Kopacek, and G.~Lukes-Gerakopoulos,
Resonance crossing of a charged body in a magnetized Kerr background: An analog of extreme mass ratio inspiral,
\href{https://journals.aps.org/prd/abstract/10.1103/PhysRevD.107.064005}{Phys. Rev. D \textbf{107}, 6, 064005 (2023)},
arXiv:2206.10302 [gr-qc].

\bibitem{Destounis:2023gpw}
K.~Destounis, G.~Huez, and K.~D.~Kokkotas,
Geodesics and gravitational waves in chaotic extreme-mass-ratio inspirals: the curious case of Zipoy-Voorhees black-hole mimickers,
\href{https://link.springer.com/article/10.1007/s10714-023-03119-2}{Gen. Rel. Grav. \textbf{55}, 6, 71 (2023)},
arXiv:2301.11483 [gr-qc].

\bibitem{Destounis:2023cim}
K.~Destounis and K.~D.~Kokkotas,
Slowly-rotating compact objects: the nonintegrability of Hartle\textendash{}Thorne particle geodesics,
\href{https://link.springer.com/article/10.1007/s10714-023-03170-z}{Gen. Rel. Grav. \textbf{55}, 11, 123 (2023)},
arXiv:2305.18522 [gr-qc].

\bibitem{Destounis:2023khj}
K.~Destounis, F.~Angeloni, M.~Vaglio, and P.~Pani, Extreme-mass-ratio inspirals into rotating boson stars: Nonintegrability, chaos, and transient resonances,
\href{https://journals.aps.org/prd/abstract/10.1103/PhysRevD.108.084062}{Phys. Rev. D \textbf{108}, 8, 8 (2023)}, arXiv:2305.05691 [gr-qc].

\bibitem{Strateny:2023edo}
M.~Straten\'y and G.~Lukes-Gerakopoulos,
Growth of orbital resonances around a black hole surrounded by matter,
\href{https://arxiv.org/abs/2311.08818}{arXiv:2311.08818 [gr-qc]}.

\bibitem{Eleni:2024fgs}
A.~Eleni, K.~Destounis, T.~A.~Apostolatos, and K.~D.~Kokkotas,
Resonant excitation of eccentricity in spherical extreme-mass-ratio inspirals,
\href{https://arxiv.org/abs/2408.02004}{arXiv:2408.02004 [gr-qc]}.

\bibitem{Cardoso:2021wlq}
V.~Cardoso, K.~Destounis, F.~Duque, R.~P.~Macedo, and A.~Maselli, Black holes in galaxies: Environmental impact on gravitational-wave generation and propagation, \href{https://journals.aps.org/prd/abstract/10.1103/PhysRevD.105.L061501}{
Phys. Rev. D \textbf{105}, 6, L061501 (2022)}, arXiv:2109.00005 [gr-qc].

\bibitem{Li:2021pxf}
G.~L.~Li, Y.~Tang, and Y.~L.~Wu,
Probing dark matter spikes via gravitational waves of extreme-mass-ratio inspirals,
\href{\doibase 10.1007/s11433-022-1930-9}{Sci. China Phys. Mech. Astron. \textbf{65}, 10, 100412 (2022)}, arXiv:2112.14041 [astro-ph.CO].

\bibitem{Rahman:2023sof}
M.~Rahman, S.~Kumar, and A.~Bhattacharyya, Probing astrophysical environment with eccentric extreme mass-ratio inspirals, \href{\doibase 10.1088/1475-7516/2024/01/035}{JCAP \textbf{01}, 035 (2024)}, arXiv:2306.14971 [gr-qc].

\bibitem{Cardoso:2022whc}
V.~Cardoso, K.~Destounis, F.~Duque, R.~Panosso Macedo, and A.~Maselli, Gravitational Waves from Extreme-Mass-Ratio Systems in Astrophysical Environments, \href{https://journals.aps.org/prl/abstract/10.1103/PhysRevLett.129.241103}{Phys. Rev. Lett. \textbf{129}, 24, 241103 (2022)},
arXiv:2210.01133 [gr-qc].

\bibitem{Destounis:2022obl}
K.~Destounis, A.~Kulathingal, K.~D.~Kokkotas, and G.~O.~Papadopoulos, Gravitational-wave imprints of compact and galactic-scale environments in extreme-mass-ratio binaries, \href{https://journals.aps.org/prd/abstract/10.1103/PhysRevD.107.084027}{
Phys. Rev. D \textbf{107}, 8, 084027 (2023)},
arXiv:2210.09357 [gr-qc].

\bibitem{Zhang:2021ojz}
Y.~P.~Zhang, Y.~B.~Zeng, Y.~Q.~Wang, S.~W.~Wei, P.~A.~Seoane, and Y.~X.~Liu, Gravitational radiation pulses from Extreme-Mass-Ratio-Inspiral system with a supermassive boson star, \href{https://arxiv.org/abs/2108.13170}{arXiv:2108.13170 [gr-qc]}.

\bibitem{Liu:2024qci}
Y.~Liu and X.~Zhang,
Gravitational waves for eccentric extreme mass ratio inspirals of self-dual spacetime,
\href{\doibase:10.1088/1475-7516/2024/10/056}{JCAP \textbf{10}, 056 (2024)},
arXiv:2404.08454 [gr-qc].



\end{thebibliography}
\end{document}